\documentclass[floatfix,a4paper,tightenlines,amsmath,amssymb,nofootinbib,notitlepage,showkeys,prd,twocolumn,10pt,aps,superscriptaddress]{revtex4-2}

\usepackage[utf8]{inputenc}
\usepackage{amsmath}
\usepackage{amssymb}
\usepackage{graphicx}
\usepackage{hyperref}
\usepackage[export]{adjustbox}
\usepackage[usenames,dvipsnames,svgnames,table]{xcolor}

\newcommand{\eref}[1]{Eq.~(\ref{#1})}
\newcommand{\fref}[1]{Fig.~\ref{#1}}
\newcommand{\tref}[1]{Tab.~\ref{#1}}
\newcommand{\nnnl}{\nonumber\\}	

\allowdisplaybreaks 

\begin{document}

\title{The analytic structure of three-point functions from contour deformations}

\author{Markus Q.~Huber}
\email{markus.huber@physik.jlug.de}
\affiliation{Institut f\"ur Theoretische Physik, Justus-Liebig-Universit\"at Giessen, 35392 Giessen, Germany}

\author{Wolfgang Kern}
\email{w.kern@uni-graz.at}
\affiliation{Institute of Physics, University of Graz, NAWI Graz, Universit\"atsplatz 5, 8010 Graz, Austria}
\affiliation{Institute of Mathematics and Scientific Computing, University of Graz, NAWI Graz, Heinrichstraße 36, 8010 Graz, Austria}

\author{Reinhard~Alkofer}
\email{reinhard.alkofer@uni-graz.at}
\affiliation{Institute of Physics, University of Graz, NAWI Graz, Universit\"atsplatz 5, 8010 Graz, Austria}

\date{\today}

\begin{abstract}
We explore the analytic structure of three-point functions using contour deformations.
This method allows continuing calculations analytically from the spacelike to the timelike regime.
We first elucidate the case of two-point functions with explicit explanations how to deform the integration contour \textit{and} the cuts in the integrand to obtain the known cut structure of the integral.
This is then applied to one-loop three-point integrals.
We explicate individual conditions of the corresponding Landau analysis in terms of contour deformations.
In particular, the emergence and position of singular points in the complex integration plane are relevant to determine the physical thresholds.
As an exploratory demonstration of this method's numerical implementation we apply it to a coupled system of functional equations for the propagator and the three-point vertex of $\phi^3$ theory.
We demonstrate that  under generic circumstances the three-point vertex function displays cuts which can be determined from modified Landau conditions.
\end{abstract}



\maketitle

\section{Introduction}

The spectrum of a quantum field theory is encoded in the spectral properties of appropriate correlation functions.
Functional methods like Dyson-Schwinger equations (DSEs) or the functional renormalization group provide a continuum approach for its calculation.
In the context of quantum chromodynamics (QCD), they are tools which are by now well developed for spacelike momenta.
Timelike momenta, on the other hand, still pose a challenge.
Nevertheless, such methods are successfully used, for example, in the case when suitable simplifying models can be employed, see, e.g., \cite{Cloet:2013jya,Eichmann:2016yit,Eichmann:2020oqt} for reviews on mesons, baryons and tetraquarks.
Given the progress in calculating elementary correlation functions in advanced truncation schemes for spacelike momenta at a quantitative level, e.g., \cite{Williams:2015cvx,Cyrol:2016tym,Cyrol:2017ewj,Huber:2018ned,Huber:2020keu,Gao:2021wun,Pawlowski:2022oyq}, it is of course enticing to envisage taking this to the timelike regime as well and perform bound state calculations from first principles.
The potential of this was recently illustrated for glueballs \cite{Huber:2020ngt,Huber:2021yfy}.

For propagators, various techniques for calculating on the timelike side are already established, e.g., the contour deformation method \cite{Maris:1995ns,Alkofer:2003jj,Eichmann:2007nn,Windisch:2012zd,Windisch:2012sz,Strauss:2012dg,Windisch:2013dxa,Eichmann:2019dts,Fischer:2020xnb}, the shell method \cite{Fischer:2008sp}, use of the Cauchy-Riemann equations \cite{GimenoSegovia:2008sx}, the covariant spectator theory framework \cite{Biernat:2018khd}, the Cauchy method \cite{Fischer:2005en,Krassnigg:2009gd}, or spectral representations including the Nakanishi integral representation \cite{Nakanishi:1963zz,Nakanishi:1969ph,Nakanish:i1971gtf,Sauli:2001mb,Sauli:2002tk,Sauli:2006ba,Jia:2017niz,Solis:2019fzm,Frederico:2019noo,Horak:2020eng,Mezrag:2020iuo,Horak:2021pfr,Horak:2022myj,Duarte:2022yur,Horak:2022aza}.
Also indirect methods have been explored.
This includes fits with trial functions, the Bayesian spectral reconstruction method, machine learning methods, Gaussian processes, the Tikhonov regularization, or Pad\'e approximants in various forms, see, for instance, \cite{Cucchieri:2011ig,Dudal:2013yva,Cucchieri:2016jwg,Siringo:2016jrc,Cyrol:2018xeq,Tripolt:2018xeo,Dudal:2019gvn,Binosi:2019ecz,Li:2019hyv,Falcao:2020vyr,Horak:2021syv,Lechien:2022ieg,Falcao:2022gxt,Boito:2022rad,Horak:2023xfb}.
Moreover, purely analytical treatments can provide additional constraints \cite{Lowdon:2017uqe,Lowdon:2018mbn,Cyrol:2018xeq,Hayashi:2018giz,Kondo:2019rpa,Hayashi:2020few,Hayashi:2021nnj,Hayashi:2021jju}.

For three-point (and higher $n$-point) functions, which are necessary ingredients for bound state calculations as well, the situation is more complicated due to the presence of three (or more) independent kinematic variables instead of only one.
However, for the application to the bound state spectrum of QCD they are essential to go beyond the widely used rainbow-ladder truncation with its inherent limitations.
Up to now, Bethe-Salpeter equations and transition form factors were investigated, e.g., with spectral representations \cite{Carbonell:2010zw,Frederico:2013vga,dePaula:2017ikc,Ydrefors:2019jvu,Sauli:2001we,Kusaka:1997xd,Karmanov:2005nv,Karmanov:2005yg,Frederico:2015ufa,dePaula:2016oct,AlvarengaNogueira:2019zcs,Gutierrez:2016ixt,Moita:2022lfu} or contour deformations \cite{Weil:2017knt,Williams:2018adr,Miramontes:2019mco,Eichmann:2019dts,Miramontes:2021xgn,Alkofer:2022hln,Eichmann:2021vnj}, see also \cite{Leitao:2017esb,Leitao:2017mlx}.
The shell method was used in Ref.~\cite{Williams:2015cvx} to calculate the quark-gluon vertex.

Here we push further in this direction and explore the contour deformation method (CDM) for the calculation of three-point functions.
In Ref.~\cite{Maris:1995ns} it was applied to the special case of the fermion propagator of QED making use of a special momentum routing.
Later, it was continuously developed.
For example, it was employed to explain the Landau conditions for two-point integrals \cite{Windisch:2013dxa}.
The so-called ray method is a special realization using a pre-defined grid consisting of rays in the complex plane \cite{Fischer:2020xnb}.
The versatility of the method was demonstrated, for instance, by the application to scattering amplitudes \cite{Eichmann:2019dts}.
Here we consider general three-point functions.
Because of the more complicated kinematics, this is more convoluted than in the case of two-point functions and we start out with simplified kinematics before we discuss the general case.
As a direct application, we illuminate how the Landau conditions \cite{Landau:1959fi,Collins:2020euz}, which describe when a singularity in the form of a branch point or pole in the external momenta arises, are realized within the CDM.
Understanding the origin of nonanalyticities of the integral is mandatory to be able to perform the integrals numerically for timelike momenta.
In particular, it becomes clear, as discussed in Sec.~\ref{sec:nonperturbative}, that for the cases considered here, the positions of poles are not required to be known exactly to take them into account in the deformation of the integration contours of numeric calculations.
For the convenience of the reader, we shortly summarize the Landau conditions in Sec.~\ref{sec:Landau_conds}, as one of the main points of this work is their derivation within the CDM.
As an example, we study $\phi^3$ theory in three dimensions numerically.
En route, we also point out some details for two-point integrals which were not necessarily clear from the available literature.
A concise summary of the presented analytic results can be found in Ref.~\cite{Huber:2023uzd}.

We will shortly introduce $\phi^3$ theory and its equations of motion in the next section.
The Landau conditions are explained in Sec.~\ref{sec:Landau_conds}.
We then discuss the numerical methods to solve the equations for complex momenta in Sec.~\ref{sec:integration} and present the results in Sec.~\ref{sec:results}.
We conclude with a summary.
In the appendices we provide details for the Landau analysis of three-point functions and the generalization to three different masses.

\section{Setup: $\phi^3$ theory and its equations of motion}
\label{sec:phi3_theory}

\begin{figure}
\hskip-2cm\includegraphics[width=0.4\textwidth]{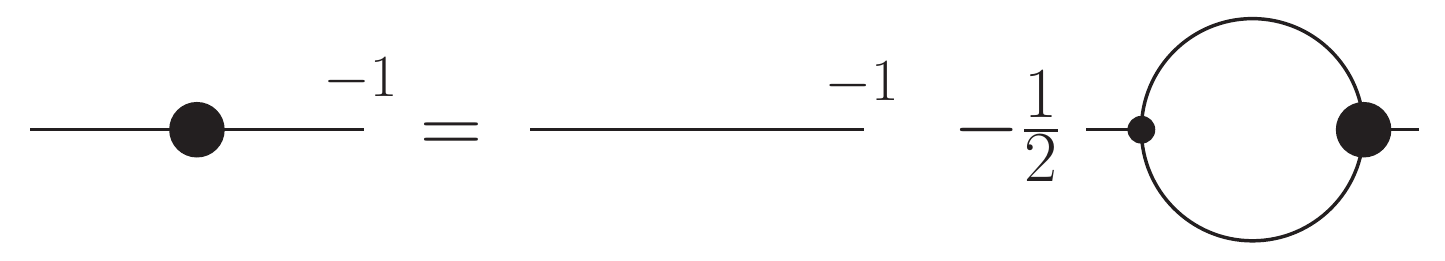}\\
\vskip3mm
\vskip3mm
\hskip-2cm\includegraphics[width=0.4\textwidth]{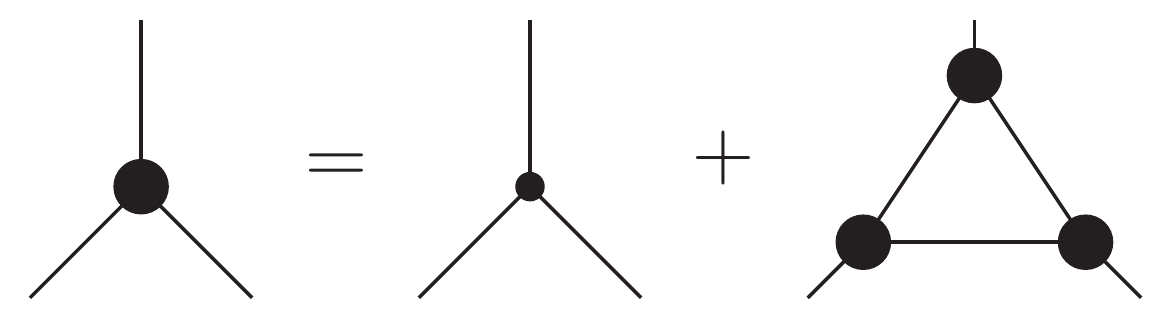}
\caption{The equations of motion for the propagator (top) and the vertex (bottom) from the three-loop truncated 3PI effective action of $\phi^3$ theory.
The former agrees with the one-particle irreducible Dyson-Schwinger equation.
The big blobs denote dressed vertices, the small ones bare vertices.
All internal propagators are dressed.}
\label{fig:DSEs}
\end{figure}

It is often useful to explore properties of a theory or new techniques using simple toy models.
Here we choose $\phi^3$ theory to this end.
The theory contains a scalar field with a cubic interaction and is defined by the Lagrangian density
\begin{equation}
\label{eq:Lagrangian}
\mathcal{L}=\frac{1}{2}\partial_\mu \phi  \partial^\mu \phi - \frac{1}{2} m^2 \phi^2+\frac{g}{3!}\phi^3.
\end{equation}
We are interested in the analytic properties of the theory.
Since the most relevant contribution for the vertex is given by a triangle diagram, see \fref{fig:DSEs}, this provides useful insights into the methodology also for other theories, among them QCD, where such diagrams appear.

The analytic analysis can be done in $d$ dimensions.
Only for the numeric part in Sec. \ref{sec:results} we choose specifically $d=3$ because the theory is finite then and we do not need to renormalize.
For Yang-Mills theory, some aspects of the analytic structure of propagators in lower dimensions were studied using lattice, e.g., \cite{Maas:2007uv,Maas:2011se,Maas:2015nva} or functional methods, e.g., \cite{Huber:2007kc,Dudal:2008xd,Dudal:2008rm,Huber:2012zj,Huber:2016tvc,Corell:2018yil,Huber:2018ned}.

Finally, we need to comment on the fact that the theory is unstable.
Because of the form of the potential, any state can decay into a state with a lower energy leading to an infinite cascade.
For the analysis of the analytic structure, this is irrelevant.
Perturbation theory can be applied formally \cite{Collins:2008re} and is known to five-loop order \cite{Borinsky:2021jdb}.
It should also be kept in mind that the theory can be made physically meaningful by embedding it into another theory.
As a simple example, consider adding a quartic interaction term which renders the potential bounded from below.
$\phi^3$ theory is also closely related to other model theories like the Wick-Cutkosky model \cite{Wick:1954eu} with its three-point interaction of two scalar fields.

In the following, we work with Euclidean metric, viz., a Wick rotation $p_0\rightarrow i\,p_4$ from Minkowksi space was performed.
The propagator reads then
\begin{align}
\label{eq:pert_prop}
D(p^2)= \frac{1}{p^2+m^2+\Sigma(p^2)}.
\end{align}
$m$ is the bare mass which is shifted by the selfenergy $\Sigma(p^2)$.
At one-loop order of perturbation theory, it is (in three dimensions)
\begin{align}\label{eq:prop_pert}
\Sigma(p^2)&=
-\frac{g^2}{8 \pi} \frac{\text{arccot}\left(\frac{2m}{\sqrt{p^2}}\right)}{\sqrt{p^2}}.
\end{align}
The pole is shifted from $-m^2$ by the selfenergy and a cut opens at $p^2=-4m^2$.

For nonperturbative calculations we consider the equations of motion for the propagator and the vertex from the three-particle irreducible (3PI) effective action truncated to three loops shown in \fref{fig:DSEs} \cite{Berges:2004pu,Carrington:2010qq}.\footnote{For conciseness, we also call the equations of motion of $n$PI effective actions Dyson-Schwinger equations, although they are conceptually somewhat different and usually refer to the equations of motion from the 1PI effective action.
When we want to distinguish equations of motion from different effective actions, we refer to them as $n$PI-DSEs.}
The resulting equation for the propagator is identical to its 1PI-DSE, see, e.g., Refs.~\cite{Berges:2004pu,Alkofer:2008nt,Carrington:2010qq,Huber:2011qr,Huber:2018ned,Huber:2019dkb} for details on the equations' derivations and useful tools.
For the vertex, we work with the 3PI equation to avoid the four-point function contained in the 1PI-DSE.
(NB: The respective 1PI-DSE would have (i) the triangle diagram with one vertex bare instead of all three dressed
 and (ii) an additional swordfish diagram with a dressed four-point function.)
The equation is also manifestly symmetric in the external legs in contrast to the corresponding 1PI-DSE.
In addition, we know that quantitatively the 3PI-DSE is preferable over a one-loop truncated 1PI-DSE in the case of the three-gluon vertex \cite{Huber:2020keu}, see Refs.~\cite{Huber:2016tvc,Huber:2018ned} for more discussions and comparisons between equations of motion from the 1PI and 3PI effective actions.
However, the technical aspects of the analysis below can be repeated identically for the vertex 1PI-DSE as well.

\begin{figure*}
\includegraphics[width=0.3\textwidth]{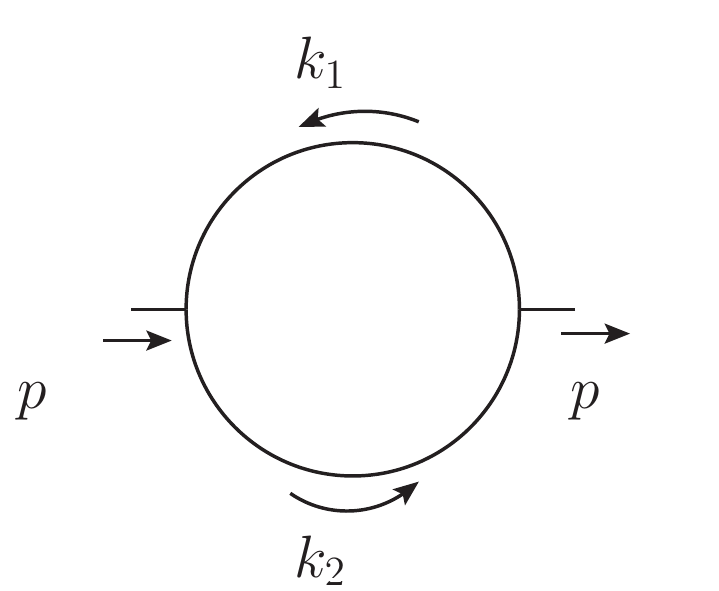}\hfill
\includegraphics[width=0.3\textwidth]{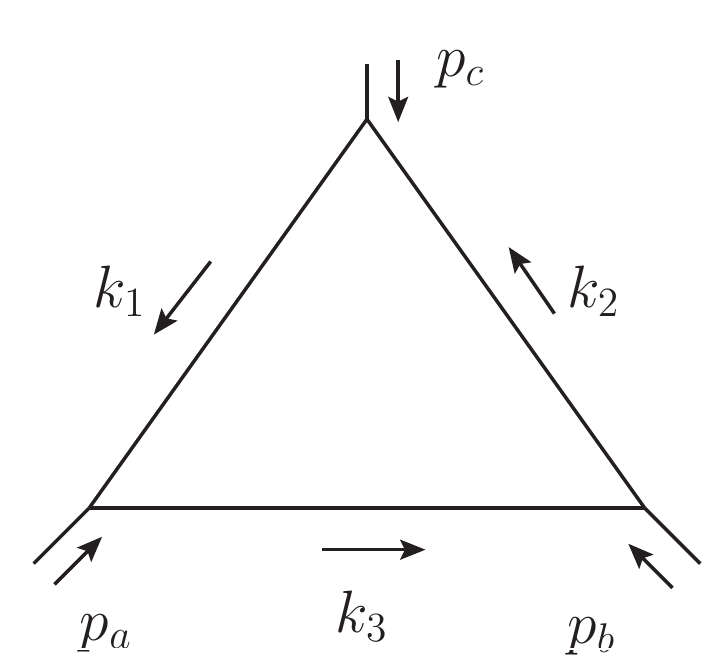}\hfill
\includegraphics[width=0.3\textwidth]{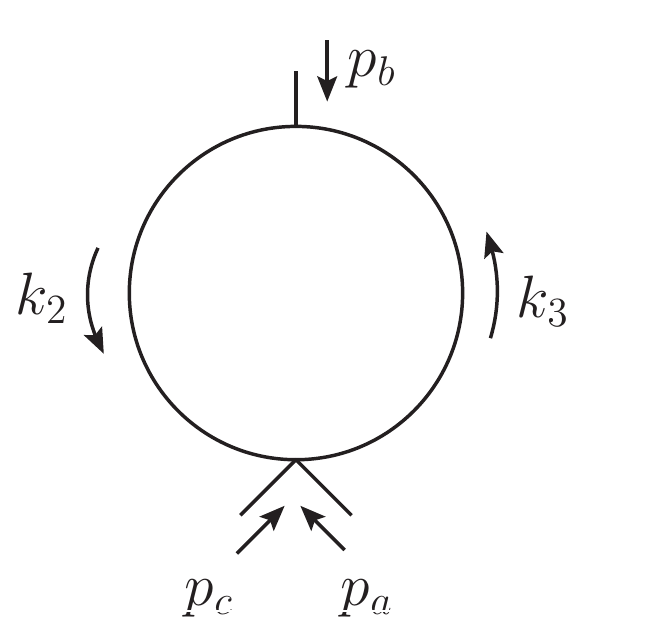}
\caption{Momentum routing for the propagator's one-loop selfenergy (left), the triangle diagram (center) and the swordfish diagram (right).}
\label{fig:LC_routing}
\end{figure*}

We parametrize the propagator $D(p^2)$ and the vertex $\Gamma(p_a,p_b,p_c)$ as
\begin{align}
D(p^2)&:=\frac{Z(p^2)}{p^2+m^2} \\
\Gamma(p_a,p_b,p_c)&:=g\, \overline{\Gamma}(p_a^2,p_b^2,p_c^2).
\end{align}
The equations of motion read
\begin{subequations}
\label{eq:DSEs}
\begin{align}
\label{eq:DSE_prop}
D(p^2)^{-1}&=p^2+m^2\nnnl 
&-\frac{g^2}{2} \int_q  D\left(k^2\right) D\left(q^2\right) \overline\Gamma \left(k^2,q^2,p^2\right),\\
\label{eq:DSE_vert}
\overline\Gamma(p_a^2,p_b^2,p_c^2)&=1+g^2\int_q D(k_a^2) D(k_b^2) D(q^2) \times \nonumber \\ & \quad \overline\Gamma \left(p_a^2, q^2, k_a^2 \right) \overline\Gamma \left(p_b^2,q^2,k_b^2 \right) \overline\Gamma \left(p_c^2,k_a^2,k_b^2 \right),
\end{align}
\end{subequations}
with $\int_q=\int d^dq/(2\pi)^d$ and
\begin{align}
k=q-p,& \qquad k_a=q-p_a,& \qquad k_b=q+p_b.
\end{align}

\section{Landau conditions}
\label{sec:Landau_conds}

We shortly summarize the main points of Landau's analysis of the analytic structure of Feynman diagrams leading to the \textit{Landau conditions} \cite{Landau:1959fi}, see also \cite{Bjorken:1967qfa}.
The starting point is the generic expression of a Feynman diagram $F(p_1,\ldots,p_n)$ with $n$ legs, each with a momentum $p_i$, $I$ internal propagators and $L$ loops:
\begin{align}
\label{eq:F}
F(p_1, \dots, p_n)=\int \prod \limits_{l=1}^{L} \frac{d^dq_l}{(2\pi)^d} \prod \limits_{i=1}^{I} \frac{E(\{p_j\},\{q_j\})}{\left(k_i^2 +m^2 \right)}. 
\end{align}
The dimension $d$ can remain general.
The internal momenta $k_i$ are linear combinations of the external momenta $p_i$ and the loop momenta $q_i$.
In general, the numerator $E(\{p_j\},\{q_j\})$ can depend on internal and external momenta.
For $\phi^3$ theory it is 1 in the perturbative case.
Landau's original analysis was indeed for perturbative diagrams, but under certain conditions, to be discussed in Sec.~\ref{sec:nonperturbative}, the analysis can be extended to nonperturbative diagrams.
Eq.~(\ref{eq:F}) can be rewritten using Feynman parametrization: 
\begin{align}
\label{eq:F_Feynman}
&F(p_1, \dots, p_n)\propto\nnnl
&\int \prod \limits_{l=1}^{L} \frac{d^dq_l}{(2\pi)^d} \int_0^1 \prod \limits_{i=1}^{I} d\alpha_i \frac{\delta \left(1-\sum \limits_{i=1}^{I} \alpha_i \right)}{\left( \sum \limits_{i=1}^{I} \alpha_i (k_i^2+m^2) \right)^{I}}.
\end{align}

In short, the Landau conditions are
\begin{enumerate}
 \item For each propagator $i=1,\ldots,I$:\\
 \begin{enumerate}
 \item $k_i^2=-m^2$\newline~\newline
 \textit{or}
 \item $\alpha_i=0$.
 \end{enumerate}
 \item For each loop $l=1,\ldots,L$: $\sum \limits_{i\, \in \text{ loop } l} \alpha_i k_i =0$.
\end{enumerate}
We assumed that in the loops the internal momenta $k_i$ are chosen such that the loop momentum enters with positive sign and prefactor 1, viz., $k_i =q_l + \dots $.
If this is not the case, additional minus signs appear in the second condition.

The first condition corresponds to two different cases.
If the on-shell condition is fulfilled, the propagators lead to a vanishing denominator and thus a pole in the integrand.
If, on the other hand, for an internal propagator labeled by $i$ the condition $\alpha_i=0$ applies, this line will not contribute to the diagram.
Hence, we can contract that line to a point and analyze the resulting diagram, which is called a contracted diagram.
This can be illustrated by the triangle diagram that is reduced to a swordfish diagram when one $\alpha_i=0$, see App.~\ref{sec:Landau_cond_3p} and \fref{fig:LC_routing}.
As a consequence, not only the original diagram but also all contracted variants have to be analyzed and it needs to be determined which one creates the leading singularity.
The second Landau condition ensures that the momenta are parallel. Hence, the surfaces containing a potential singularity approach the integration path parallel, and the contour cannot be deformed to avoid the singularity.

To find the singularities of a given diagram, we have to solve the Landau equations for nonvanishing $\alpha_i$ and consider also all possibilities of $\alpha_i=0$.
Finally, it has to be determined which of the resulting singularities is leading.

It is convenient to contract the second condition with the vectors $k_j$ to obtain a more compact representation:
\begin{align}
\label{eq:LandauCond}
Q\cdot \vec{\alpha}=0, \qquad (Q)_{ij}=k_i \cdot k_j, \qquad \vec{\alpha}=\left( \begin{array}{c} \alpha_1 \\ \vdots \\ \alpha_n \end{array} \right).
\end{align}
A solution to this equation automatically fulfills the second Landau condition.
From \eref{eq:F_Feynman}, one can see that $0\leq\alpha_i\leq1$ with $\alpha_1+\alpha_2+\dots+\alpha_n=1$.
Solutions with other values of the $\alpha_i$ correspond to singularities on nonphysical sheets which we do not take into account here.

As an illustrative example we show how the Landau conditions can be used to obtain the branch point of the one-loop selfenergy diagram depicted in \fref{fig:LC_routing}.
Although it is immaterial, we can think of $k_1$ and $k_2$ as $k_1=q-p$ and $k_2=q$.
The matrix $Q$ is given by
\begin{align}
Q=\left( \begin{array}{cc} k_1^2 & k_1\cdot k_2 \\ k_1\cdot k_2  & k_2^2 \end{array} \right).
\end{align}
Because of momentum conservation at the vertices, the internal momenta $k_1$ and $k_2$ are related by $k_2-k_1=p$ so that their scalar product yields
\begin{align}
k_1\cdot k_2=\frac{1}{2}(k_1^2 +k_2^2 - p^2).
\end{align}
From the first Landau condition, we get for nonvanishing $\alpha_i$ that $k_1^2=k_2^2=-m^2$.
Hence, 
\begin{align}
Q=\left( \begin{array}{cc} -m^2 & -m^2-\frac{p^2}{2} \\ -m^2-\frac{p^2}{2}  & -m^2 \end{array} \right).
\label{eq:LC_Q_prop}
\end{align}
Setting the determinant of $Q$ to zero, we obtain
\begin{align}
p^2 \left( m^2+\frac{p^2}{4} \right) =0.
\end{align}
There are two solutions $p_{1,2}^2$ to this equation.
We still have to check which of them corresponds to a physical threshold.
We plug the solution $p_1^2=0$ into \eref{eq:LandauCond}:
\begin{align}
Q \cdot \vec{\alpha} = \left( \begin{array}{cc} -m^2 & -m^2 \\ -m^2  & -m^2 \end{array}  \right) \left( \begin{array}{c} \alpha_1 \\ \alpha_2 \end{array} \right) =\vec{0}.
\end{align}
The only nontrivial solution is
\begin{align}
\alpha_1=-\alpha_2,
\end{align}
which cannot be fulfilled by two positive numbers.
Thus, for $p_1^2=0$ we would need to deform the integration contour for the Feynman parameter integration, and we cannot determine if this singularity is on the first Riemann sheet, see, e.g., Ref.~\cite{Zwicky:2016lka} for more details.

The case $p_2^2=-4m^2$ leads to 
\begin{align}
\label{eq:Landau_prop_physical}
Q \cdot \vec{\alpha} = \left( \begin{array}{cc} -m^2 & m^2 \\ m^2  & -m^2 \end{array}  \right) \left( \begin{array}{c} \alpha_1 \\ \alpha_2 \end{array} \right) =\vec{0}.
\end{align}
The solution is
\begin{align}
\alpha_1=\alpha_2.
\end{align}
From $\alpha_1 +\alpha_2 =1$ we obtain then
\begin{align}
\alpha_1=\alpha_2=\frac{1}{2}.
\end{align}
Thus,  $p^2=-4m^2$ is a physical singularity of the selfenergy diagram.

For the triangle diagram the same analysis can be repeated.
The procedure is exactly the same as before.
We refer to App.~\ref{sec:Landau_cond_3p} for details and continue here with the solution.
For the external momenta $p_a, p_b$ and $p_c$, the position of the branch point is determined by
\begin{align}
\label{eq:Landau_cond_vertex}
p_a^2 \,p_b^2 \,p_c^2 = m^2 \left(p_a^4 + p_b^4 +p_c^4 -2(p_a^2\,p_b^2+p_a^2\,p_c^2+p_b^2\,p_c^2)\right).
\end{align}

The analysis in App.~\ref{sec:Landau_cond_3p} shows that the physical thresholds are given by
\begin{align}\label{eq:Landau_sol_3p_loc}
\left. \begin{array}{l}
p_c^2=\frac{2m^2(p_a^2+p_b^2) + p_a^2 p_b^2 + \sqrt{p_a^2(4m^2+p_a^2)} \sqrt{p_b^2(4m^2+p_b^2)}}{2m^2} \\  \qquad \text{for} \quad -4m^2\leq p_a^2,p_b^2 \leq 0, \quad \text{and} \quad p_a^2+p_b^2\leq-4m^2  \\ ~\\
p_a^2=-4m^2 \qquad \text{for} \quad p_b^2+p_c^2\geq-4m^2\\~\\
p_b^2=-4m^2 \qquad \text{for} \quad p_c^2+p_a^2\geq-4m^2\\~\\
p_c^2=-4m^2 \qquad \text{for} \quad p_a^2+p_b^2\geq-4m^2\,.\\
\end{array} \right. 
\end{align}
The two distinct regions stem from the full triangle diagram and the contracted triangle diagram, viz., the swordfish diagram, see \fref{fig:LC_routing} and App.~\ref{sec:Landau_cond_3p}.

Finally, we state the Landau condition for the specific kinematic configuration $p_a^2=p_b^2=p^2$ for which $p_c^2=2p^2(1+\cos\theta)$ with $\theta$ the angle between $p_a$ and $p_b$.
As derived in App.~\ref{sec:Landau_cond_3p}, the branch point $p_B^2$ is then
\begin{align}
p_B^2=\left \lbrace \begin{array}{l l}
-4m^2\sin^2\frac{\theta}{2} & \text{for} ~ \pi/2\leq \theta \leq \pi  \, , \\
\frac{-m^2}{\cos^2\frac{\theta}{2}} & \text{for} \quad  0\leq \theta \leq \pi/2 \,.
\end{array} \right.
\label{eq:Br_point_vert_sym}
\end{align}
As an alternative to an explicit calculation, one can directly plug the given kinematics into \eref{eq:Landau_sol_3p_loc} to arrive at this result. From the condition $p_a^2+p_b^2=2p^2<-4m^2$ it follows that $p^2=-2m^2$ separates the two solutions.
The condition in terms of the angle $\theta$ follows from setting $p^2=-2m^2$ in $p_c^2=2p^2(1+\cos\theta)=-4m^2$.
Graphically, one can visualize this by setting $p_a^2=p_b^2$ in \fref{fig:threshold_3p}.

\section{Integration for complex momenta}
\label{sec:integration}

\begin{figure*}
\centering
\includegraphics[width=0.48\textwidth]{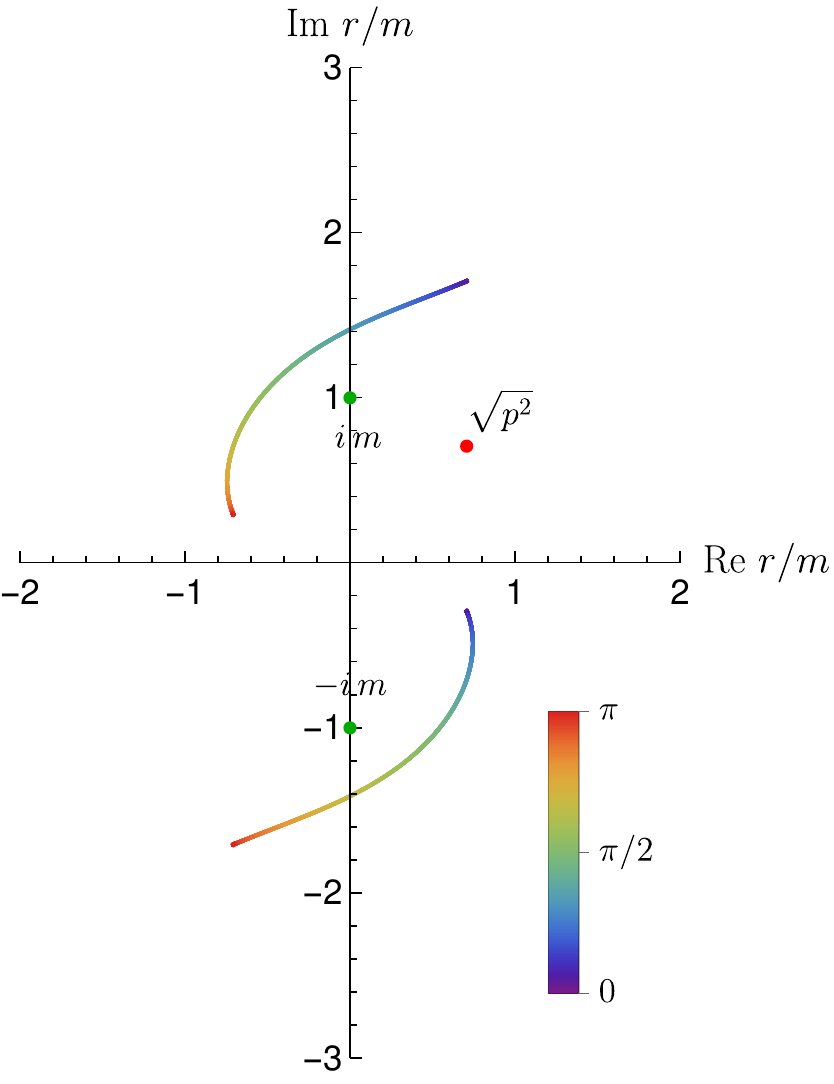}\hfill\includegraphics[width=0.48\textwidth]{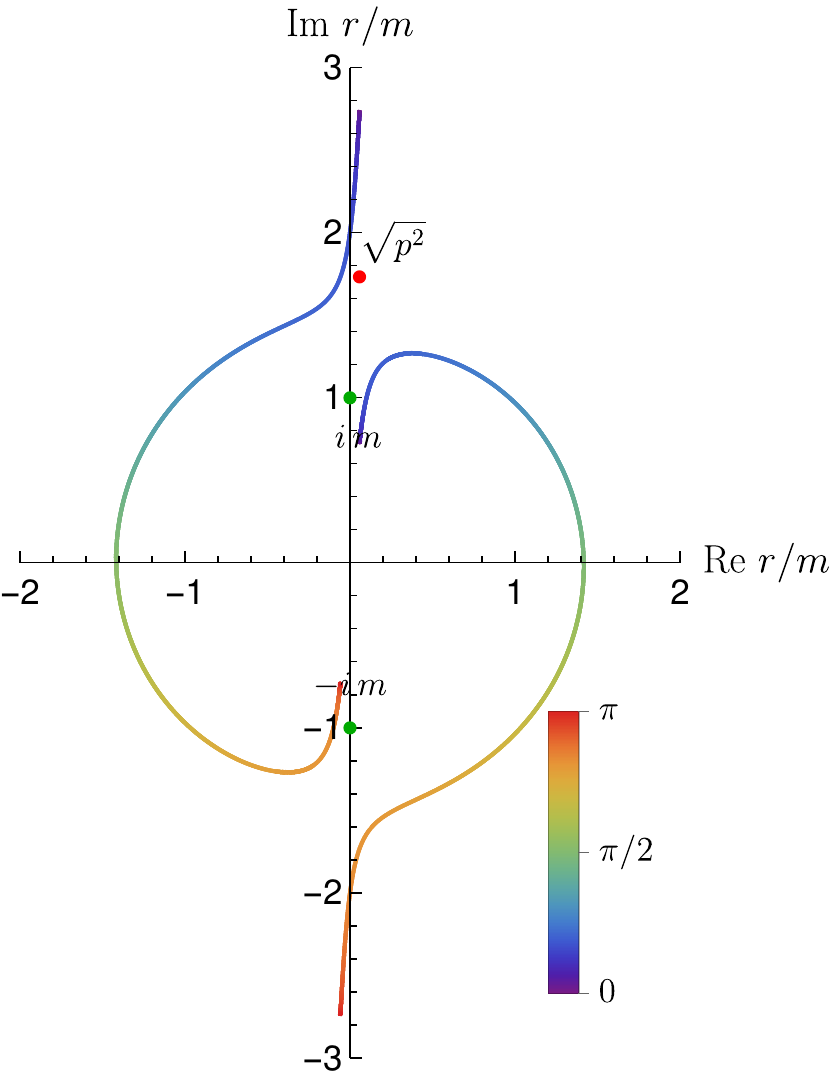}
\caption{Examples for the singularity structure $\gamma_{\pm}(z_1;p^2,m^2)$ of the propagator $r=\sqrt{q^2}$ integrand for $p^2=i\,m^2$ (left) and $p^2=(-3+0.2i)m^2$ (right).
The lines denote branch cuts stemming from the angular integral with the value of the angle $\theta_1$ indicated by the color.
$\gamma_{+}(z_1;p^2,m^2)$ is the line extending further to the top.
The green dots are the poles from the second propagator.
The red dots indicate the external $p^2$ and are only plotted for reference.}
\label{fig:sing_prop}
\end{figure*}

We will now turn to the evaluation of integrals using contour deformations.
Such deformations are necessary due to singularities in the integrand.
In analogy to the Landau analysis, we start with perturbative two- and three-point integrals.
Understanding this setup is key for nonperturbative calculations which are discussed afterwards.
In the perturbative case, only the poles of the propagators are relevant, but contour deformations can also handle poles from dressed $n$-point functions or branch cuts and are thus applicable to fully nonperturbative calculations as well.
The basic strategy is to deform the integration contour such as to avoid poles and branch cuts of the integrand.

The integration is most conveniently performed in hyperspherical coordinates.
The integration over angles $\theta_i$ can create cuts for the radial integration variable $r=\sqrt{q^2}$.
These cuts have to be distinguished from the physical cuts of correlation functions.
If these cuts in the integrand cross the positive real axis, the integration of the radial variable can no longer be performed directly along the real axis.
Instead, a contour deformation is necessary moving the integration path into the complex plane.
The position of the poles and branch cuts in the integrand can tell us about the analytic structure of the integral as will be detailed below.
We discuss how to determine them and how to choose the integration paths appropriately.
These paths can then also be employed in numerical calculations.
We will use the propagator, for which the corresponding techniques have already been studied to some extent, to introduce the basic idea and discuss some details of the method before we turn to the more complicated case of the vertex.

\subsection{Propagator}
\label{sec:CDM_prop}

\begin{figure*}
\centering
\includegraphics[width=0.48\textwidth]{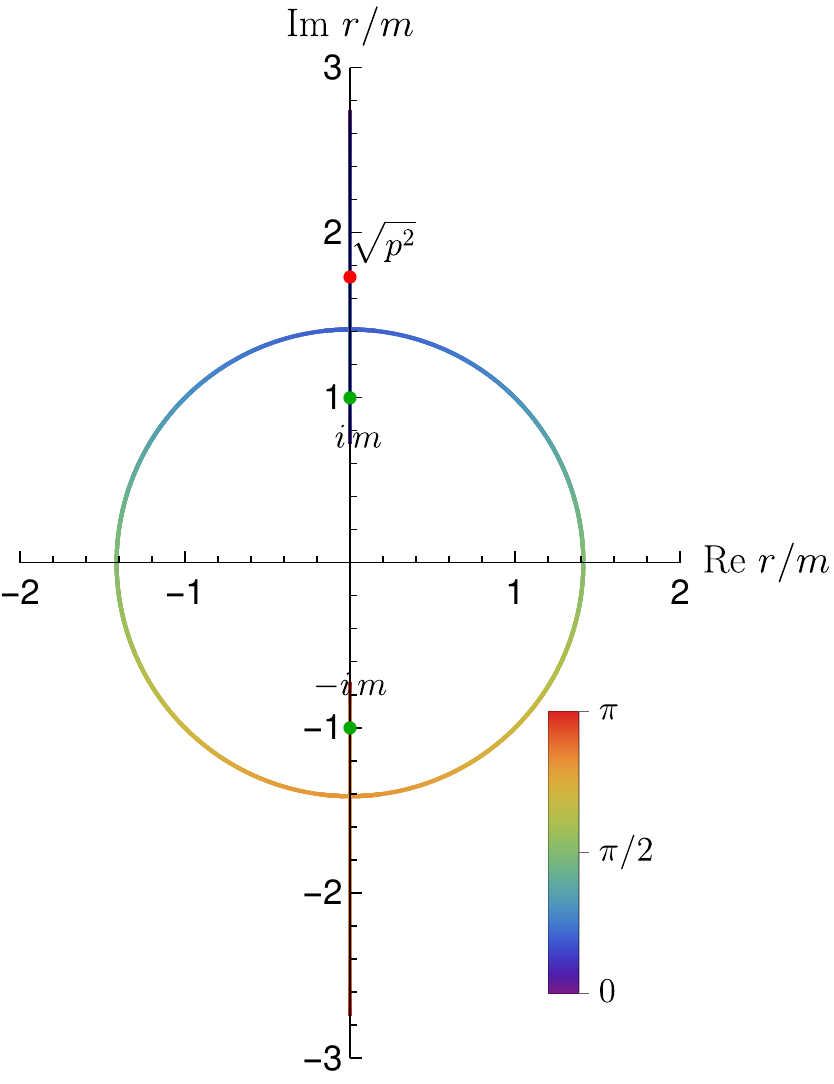}\hfill\includegraphics[width=0.48\textwidth]{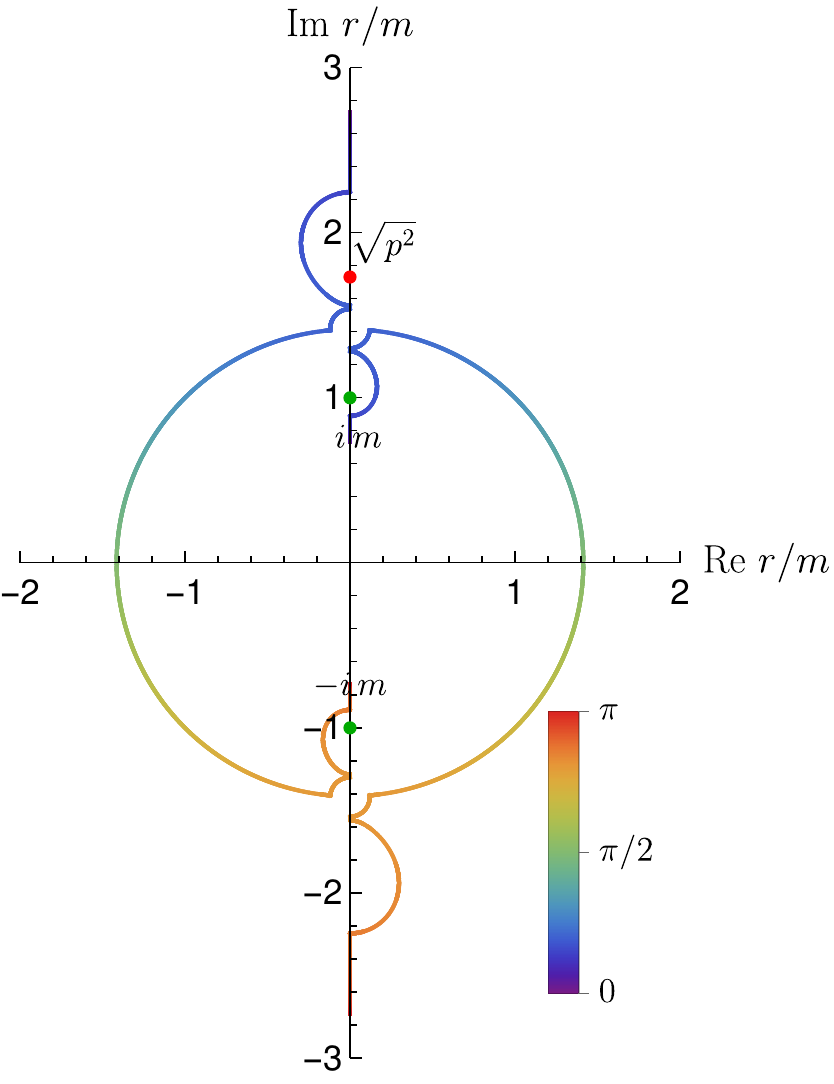}
\caption{Examples for the singularity structure of the propagator $r=\sqrt{q^2}$ integrand for $p^2=-3m^2$.
The left plot shows the cuts when the angle integral in $\theta_1$ is performed in a straight line from $0$ to $\pi$.
The right plot shows the cuts when the angle integral is performed such as to avoid the poles at $\pm i\,m$ and gaps are opened at $\pm\sqrt{-2}m$ using the path of \eref{eq:t1_complex}.
The two bulges at $\pm 2\,i\,m$ are a consequence of deforming $\theta_1$ around $\pm i\,m$.
}
\label{fig:sing_prop_real}
\end{figure*}

The one-loop two-point integral of \eref{eq:DSE_prop} was already discussed in Ref.~\cite{Windisch:2013dxa}, in particular how to determine the branch cuts.
Here, we will add some more details concerning deformations of the cuts necessary to avoid seemingly singular points in the integration.
We will also switch from the quadratic variable $q^2$ to $r=\sqrt{q^2}$.
This not only disentangles some ambiguities for the propagator analysis but is actually necessary for the three-point function as will be explained in Sec.~\ref{sec:vertex}.
We use a generic mass $m$ which could be the bare mass or a physical (renormalized) mass.
We set all dressings to one and consider the denominators of the propagators.
In general, for any one-loop diagram they are of the form $k_i^2+m^2$, where $k_i$ is a linear combination of the external and internal momenta.
The integrand is singular when any denominator is zero, hence we solve
\begin{align}
\label{eq:denom_cond}
 k_i^2+m^2=0
\end{align}
for $r$.
For the propagator, we can choose the $k_i$ as $q-p$ and $q$, see \eref{eq:DSE_prop}, where $p$ and $q$ are the external and internal momenta, respectively.
In the first case, the solution to \eref{eq:denom_cond} for a given value of $p$ depends on the angle $\theta_1$ between the momenta.
We can consider the solution as a parametric curve in $\theta_1\in[0,\pi]$ or equivalently in $z_1=\cos \theta_1\in[-1,1]$:
\begin{align}
\label{eq:denom_sol}
\gamma_{\pm}(z_1;p^2,m^2)=&\sqrt{p^2}\,z_1 \pm i\sqrt{m^2+p^2(1-z_1^2)} \nnnl
 =& \sqrt{p^2}\cos\theta_1 \pm i\sqrt{m^2+p^2\sin^2\theta_1}.
\end{align}
The parametrization for a cut will be central for the remainder of this analysis and will also appear for the three-point integral.
For reference and comparison with Ref.~\cite{Windisch:2013dxa} we also give the solution for $q^2$:
\begin{align}
\label{eq:denom_sol_squared}
\gamma^2_{\pm}(z_1;p^2,m^2)=&-m^2 -p^2(1-2z_1^2) \nnnl
 &\quad \pm 2\, i\sqrt{p^2\,z_1^2 \left(m^2+p^2 (1-z_1^2)\right) } \nnnl
 =& -m^2+p^2\cos(2\theta_1)\nnnl
 &\quad \pm 2\,i\sqrt{p^2\,\cos^2\theta_1(m^2+p^2\sin^2\theta_1)}.
\end{align}

Examples of these curves in the complex $r=\sqrt{q^2}$ plane are shown in \fref{fig:sing_prop}.
The singular points $r=\pm i\,m$ from the second denominator are also shown (green points).
When performing the $r$ integration from the origin to the UV cutoff, the branch cut must not be crossed which might require deforming the contour.
For the example when $p^2$ is purely imaginary, e.g., $p^2=i\,m^2$ as in the left plot of \fref{fig:sing_prop}, the integration can be done in the standard way along the real line.
In the second example on the right, however, the cuts cross the real line and the integration contour has to be deformed by moving it into the upper half-plane.
As discussed in Ref.~\cite{Windisch:2013dxa}, a branch point $p_B^2$ emerges for the external momentum if an end point of a curve touches a pole and the integration contour can not be deformed anymore.
We stress that the end points are important as the cut can be deformed by moving also the angle integration into the complex plane.
The end points, however, are fixed.
For values of $p^2$ beyond $p_B^2$, the possibility of going around the pole on either side leads to a discontinuous behavior and thus a cut in the external momentum starting at $p^2=p_B^2$.

\begin{figure}[tb]
 \includegraphics[width=0.49\textwidth]{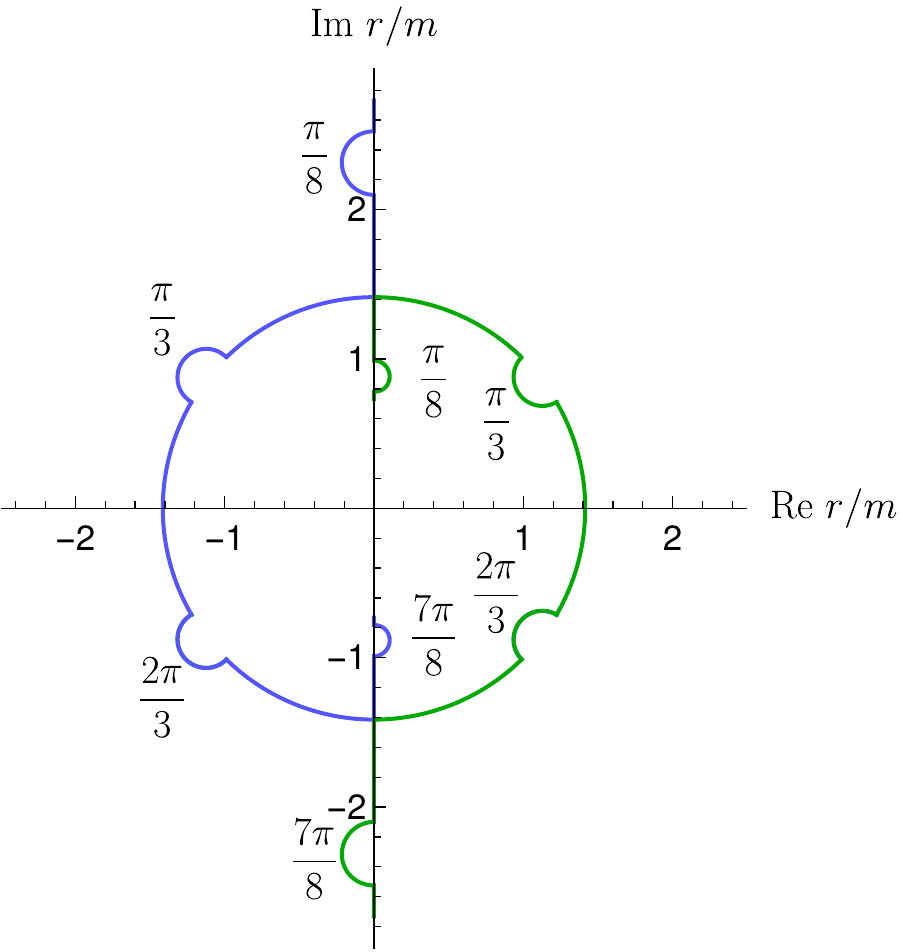}
 \caption{Bulges from deforming the angle integration in $\theta_1$ via $\theta_1^+$ as given in \eref{eq:t1_complex} for the indicated values of $\theta^*_1$.}
 \label{fig:bulges}
\end{figure}

As it will be useful for the analysis of the three-point integral, we have a more detailed look at the case when $p^2<0$ is real.
The cuts are then either lines along the imaginary axis if $-m^2<p^2<0$, as both terms in \eref{eq:denom_sol} are purely imaginary in this case, or they will be semicircles with attached lines on the imaginary axis.
An example for this is shown in \fref{fig:sing_prop_real}.
We have a problem now, because the inner parts of the cuts run over the poles at $\pm i\,m$ and we would like to make a contour integration \textit{between} the pole and the cut.
In addition, the two cuts touch at $\pm\sqrt{m^2+p^2}$.
The solution lies in the complexification of the angular integral.
As for the $r$ integral, we do not need to perform the integral in a straight line but can deviate from that path without changing the result.
To avoid the pole, we can use the following path which avoids a given value $\theta^*_1$ via a semicircle:
\begin{align}
 \theta_1\rightarrow \theta_{1}^\pm=
 \begin{cases}
    \theta^*_{1}+s\,e^{\pm i\frac{\theta_1-\theta^*_1-s}{s}\frac{\pi}{2}} &\quad  |\theta_1-\theta^*_{1}|<s \\
    \theta_1 &\quad  \text{otherwise}
 \end{cases}
 \label{eq:t1_complex}
\end{align}
$s$ is a radius which we can choose conveniently.
The sign of the phase determines on which side the semicircle passes $\theta^*_1$ and must be chosen appropriately.
The deformation leads to a bulge in the cut which moves with $\theta_1^{*}$.
Fig.~\ref{fig:bulges} illustrates this graphically for the positive sign.
As can be seen there, the created bulges always stay on one side of the cut.

We can use such deformations of the $\theta_1$ integration to avoid critical points in the $r$ plane.
For the propagator, we have four such critical values of $\theta_1$.
Two arise when the cut is at the pole and can be determined directly from equating the two denominators and setting $q^2=-m^2$ as
\begin{align}
 \theta^{*p\pm}_{1}=\arccos\left(\pm i\frac{\sqrt{p^2}}{2m}\right).
\end{align}
Two more are determined from the point where the four cuts touch.
There, the argument of the square root must vanish, which leads to
\begin{align}
 \theta^{*c\pm}_1=\arcsin\left(\pm i\frac{m}{\sqrt{p^2}}\right).
\end{align}
We distinguish the four cases by the letters $p$ and $c$ for 'pole' and 'crossing', respectively, as well as $\pm$ in the superscript.
A path with such deformations is illustrated in \fref{fig:sing_prop_real} for $p^2=-3m^2$.
According to the rules for the orientation of the created bulges described above, the signs of the phases of the deformations have the signature $1,1,-1,-1$ when $\theta_1$ goes from $0$ to $\pi$.
The first and fourth bulges are for circumventing the poles at $\theta^{*p\pm}_1$, while the second and third are for opening gaps at $\theta^{*c\pm}_1$.
Formerly, the deformations in the lower half-plane are not necessary when integrating via the gap in the upper half-plane but included here for completeness.
The deformations are only necessary for one cut, but naturally they introduce bulges for the other cut as well.

When $p^2$ moves to lower values, the end points inside the semicircle move towards $\pm i\,m$.
They arrive there for 
\begin{align}\label{eq:pb2_prop}
p_B^2=-4m^2.
\end{align}
Now there is no way left to perform the contour deformation as required, and a branch point $p_B^2$ appears in $p^2$.
It can be determined directly from
\begin{align}
\gamma_{-}(1;p^2,m^2) &= i\,m,\\
\gamma_{+}(1;p^2,m^2) &= -i\,m.
\end{align}
Depending on the imaginary part of $\sqrt{p^2}$, the upper or lower pole obstructs the contour deformation.
It is worth pointing out that in the formulation with $q$ there is no solution $p^2=0$ which arises in the formulation with $q^2$.
However, in that case a dedicated check confirms that for $p^2=0$ no problem for the integration arises and the solution can be discarded \cite{Windisch:2013dxa}.

For the numerical integration an explicit integration contour in the $r$ plane needs to be chosen.
A straightforward choice is along a ray from the origin through the value of the external $\sqrt{p^2}$ up to some fixed end point and from there to the UV cutoff, for details, see App.~A of \cite{Fischer:2020xnb}.
With this particular choice of the contour, the method is also known as the ray method, but any other contour not crossing any cuts is allowed as well.

It is illuminating to have a closer look at the case of vanishing mass.
The condition for the branch point, \eref{eq:pb2_prop}, yields a branch point at the origin.
Nevertheless, cuts appear in the form of two semicircles with starting points at $r=\sqrt{p^2}$.
So it might seem that no integration from the origin to the UV is possible.
However, as discussed in Ref.~\cite{Fischer:2020xnb}, the cut can be passed by $\sqrt{p^2}$ if the integrand or the integral measure counteract the singularity there, viz., the circle is open at exactly this point.
This is realized, for example, for QCD in three and four dimensions.
A counterexample is QCD in two dimensions, where the contour cannot be deformed in perturbative calculations and perturbation theory is hence ill-defined.
Nonperturbatively, the gluon propagator is sufficiently suppressed to overcome this \cite{Maas:2007uv,Dudal:2008xd,Huber:2012zj,Cucchieri:2016jwg}, and the theory is well-defined.

We close the discussion of the propagator with the consideration of additional nonanalyticities in the propagator.
Up to now we only considered a pole.
However, although the analysis was done perturbatively, it applies equally to any mass irrespective of its origin.
Of importance is only the fact that the propagator has a pole, which in a nonperturbative calculation will be at the dynamically determined value of the mass.
We will explicitly demonstrate this in Sec.~\ref{sec:results} for a scalar propagator where the bare mass is shifted to a lower value by the interaction.
The analysis also applies to complex values of masses.
The phase just leads to a rotation of the related cut but otherwise the structure is identical.
The case of complex conjugate poles in this context was investigated in Refs.~\cite{Baulieu:2009ha,Dudal:2010wn,Windisch:2012zd,Eichmann:2021vnj}.

\subsection{Vertex}
\label{sec:vertex}

\begin{figure*}
\includegraphics[width=0.48\textwidth]{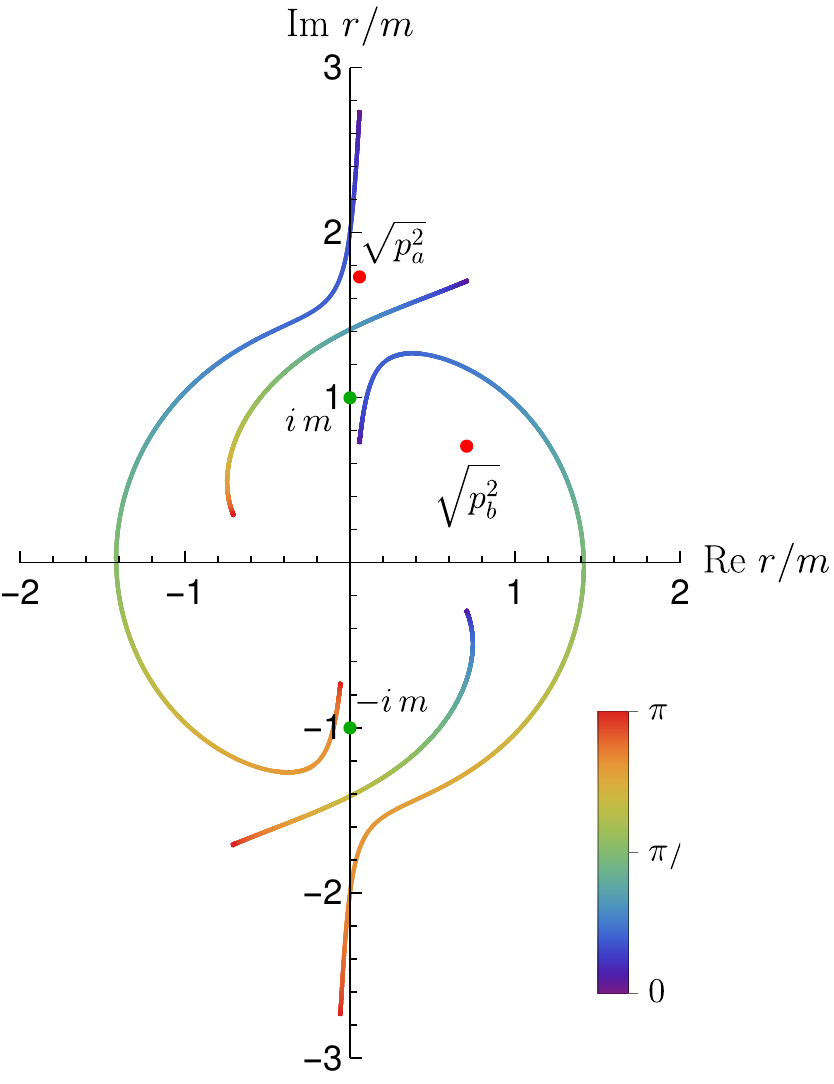}
\hfill
\includegraphics[width=0.48\textwidth]{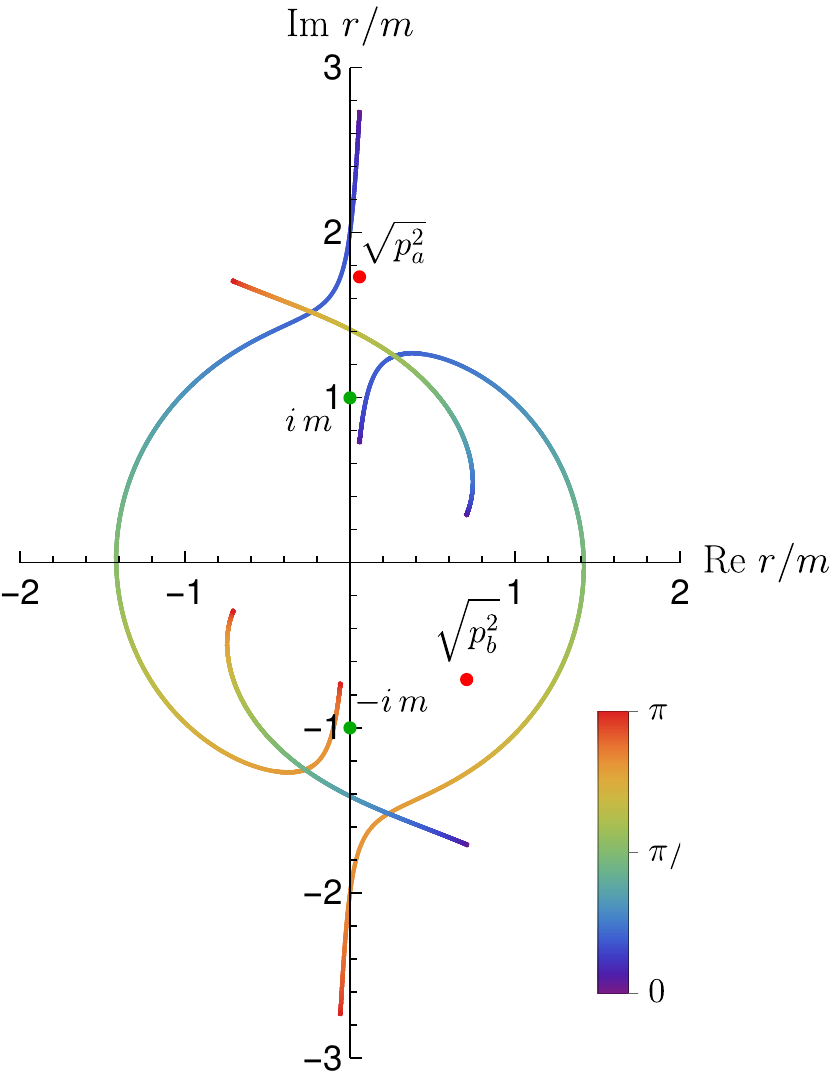}
\caption{
Examples for the singularity structure of the triangle $r=\sqrt{q^2}$ integrand.
The lines denote branch cuts stemming from the angle integral with the value of the angle $\theta_1$ indicated by the color.
The green dots are the pole from the third propagator, the red dots indicate the external $p_a^2$ and $p_b^2$.
The external angle is set to $\theta=\pi$.
}
\label{fig:sing_vert}
\end{figure*}

We now proceed to the triangle diagram, see \fref{fig:LC_routing}, with external momenta $p_a$, $p_b$ and $p_c$.
For the internal momenta $k_i$, $i=1,2,3$, we choose
\begin{subequations}\label{eq:vert_kinematics}
\begin{align}
k_1&=k_a=q-p_a,\\
k_2&=k_b=q+p_b,\\
k_3&=k_c=q.
\end{align}
\end{subequations}
The angle between the two external momenta $p_a$ and $p_b$ is $\theta$, and the internal momentum $q$ contains two relevant angles $\theta_1$ and $\theta_2$:
\begin{align}
p_a&=|p_a| \left( \begin{array}{c} 0 \\ 0 \\ 0 \\ 1 \end{array} \right), \quad p_b=|p_b|  \left( \begin{array}{c} 0 \\ 0 \\ \sin\theta \\ \cos\theta \end{array} \right),\nnnl
\quad q&=r \left( \begin{array}{c} \sin\theta_1 \sin\theta_2 \sin \phi \\ \sin\theta_1\, \sin \theta_2 \cos\phi\\ \sin\theta_1 \cos\theta_2 \\ \cos \theta_1 \end{array} \right).
\end{align}
We have taken the number of dimensions to be four.
The angle $\phi$ can be integrated out trivially as no quantity depends on it.
We thus set it to zero here.
The analysis can be straightforwardly extended to three or more than four dimensions.
For two dimensions, only one angle, $\theta_1$, appears, but as will become clear below, $\theta_2$ plays no crucial role and the same arguments apply.
Note that in \eref{eq:DSE_vert} the momenta squared were used as arguments of the vertex dressings.
In the following we will also use two momenta squared ($p_a^2$, $p_b^2$) and the angle between them ($\theta$) which is related to the third momentum square by $\cos\theta=(p_c^2-p_a^2-p_b^2)/\sqrt{p_a^2}\sqrt{p_b^2}$.
These two sets of variables are equivalent and we choose the one most suitable for the task at hand.

One propagator creates poles at $\pm i\,m$.
The others produce branch cuts for
\begin{subequations}
\label{eq:vertex_denoms}
\begin{align}
k_a^2+m^2&=q^2+p_a^2 + m^2 - 2\sqrt{q^2 p_a^2} \cos{\theta_1}, \\
k_b^2+m^2&=q^2+p_b^2 + m^2\nnnl
  &+2\sqrt{q^2 p_b^2} \left(\cos{\theta}\,\cos{\theta_1}+\sin{\theta}\,\sin{\theta_1}\, \cos{\theta_2} \right).
\end{align}
\end{subequations}
We call the resulting cuts $\gamma_{a\pm}(z_1;p_a^2,m^2)$ and $\gamma_{b\pm}(\tilde z;p_b^2,m^2)$ where
\begin{align}\label{eq:thetatilde}
\tilde z=\cos \tilde \theta=\cos{\theta}\,\cos{\theta_1}+\sin{\theta}\,\sin{\theta_1}\, \cos{\theta_2}.
\end{align}
Again, we use the cosines of the angles where convenient.
As required, $\tilde z\in[-1,1]$.
The branch cuts have the same form as in \eref{eq:denom_sol} but with their own values for the angles and external momenta, viz.,
\begin{subequations}
\label{eq:gamma_ab}
\begin{align}
\gamma_{a\pm}(z_1;p_a^2,m^2)&=\gamma_{\pm}(z_1;p_a^2,m^2),\\
\gamma_{b\pm}(\tilde z;p_b^2,m^2)&=\gamma_{\pm}(-\tilde z;p_b^2,m^2).
\end{align}
\end{subequations}
The minus before $\tilde z$ stems from the different sign of the external momentum in $k_b$.
The two cuts lead to a complicated structure of the $r$ integrand in the complex plane as illustrated in \fref{fig:sing_vert}.

Before we continue, it is useful to clarify the different types of cuts that appear and their interrelation to the integration contours.
The considered triangle integral has the following form reduced to the relevant parts:
\begin{align}\label{eq:int_q2_f}
 \int dr\, r^{d-1}f(q^2,p_a^2,p_b^2,p_c^2)
\end{align}
with
\begin{align}
 f(q^2,p_a^2,p_b^2,p_c^2)=&\int d\Omega \, \frac{1}{q^2+m^2}\frac{1}{(q-p_a)^2+m^2}\nnnl
 &\times\frac{1}{(q+p_b)^2+m^2}
\end{align}
where $d\Omega$ represents the angular part of the $d$-dimensional integration.
The radial integration is in $r=\sqrt{q^2}$.
The angular integrations lead to cuts of the function $f$ in the \textit{internal} variable $r$ due to the poles of the second and third propagators.
The first propagator, on the other hand, creates poles.
The cuts and poles shown in the figures correspond to these cuts and poles.
The standard integration for the angles is from $0$ to $\pi$, but it can be deformed leading to a different form of the cuts.
For the integration of the radial variable in \eref{eq:int_q2_f}, we need to find an integration contour that avoids the cuts and the pole.
This determines the integration contour for $r$.
Finally, if such contour deformations are not possible, nonanalyticities in the \textit{external} momenta emerge in the form of a threshold surface (corresponding to a simple branch point for the two-point integral).

We will now discuss this in detail for a kinematically simpler case before we come to the general one.

\subsubsection{Restricted kinematics}
\label{sec:restrKin}

The situation is simplified by restricting the kinematics to $p_b^2=p_a^2=p^2$.
The second cut (from $k_b$) is then always on top of the first (from $k_a$) and the third momentum squared is $p_c^2=2p^2(1+\cos\theta)$.

\begin{figure*}
\begin{minipage}{0.49\textwidth}
\includegraphics[width=0.98\textwidth]{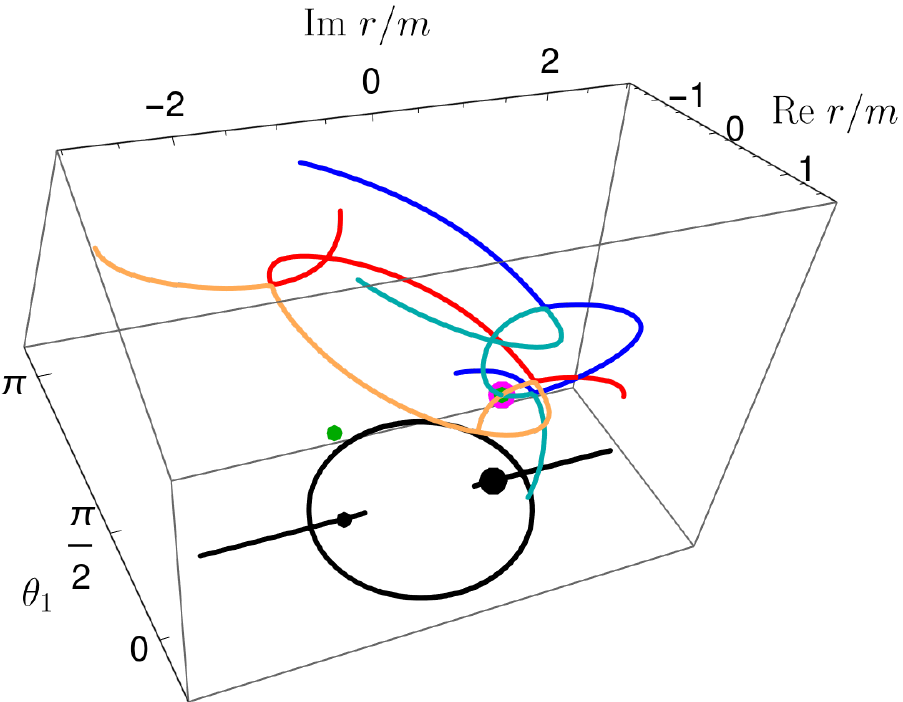}
\end{minipage}
\hfill
\begin{minipage}{0.41\textwidth}
\includegraphics[width=0.98\textwidth]{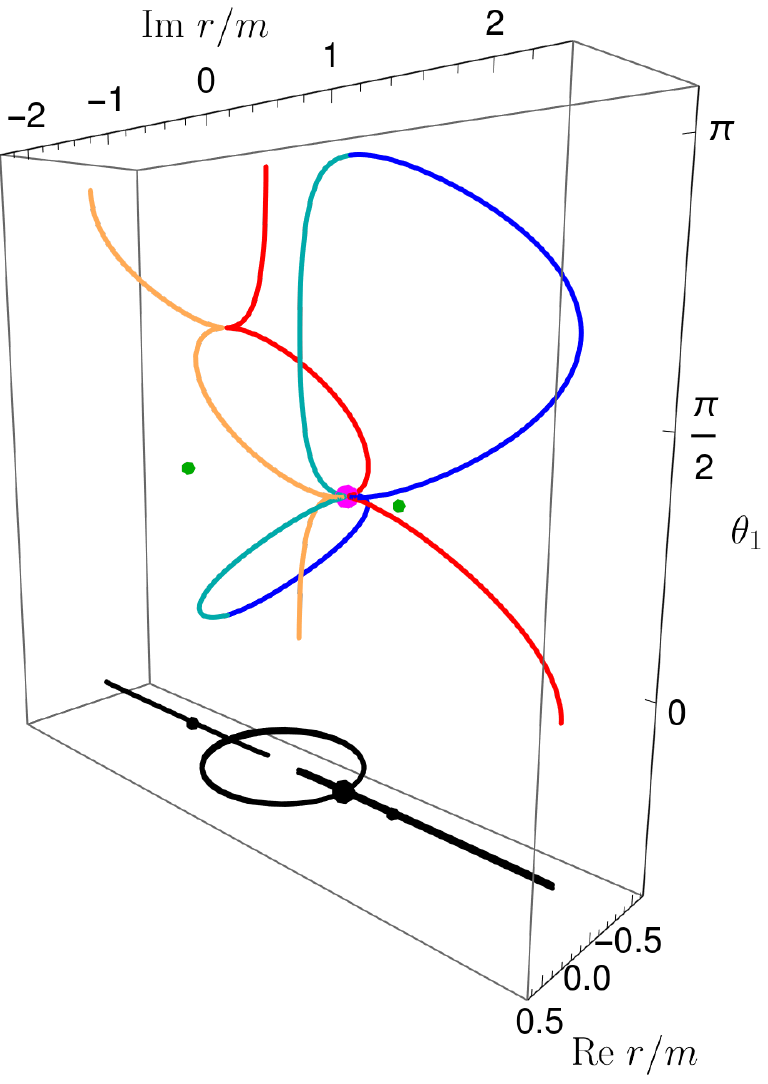}
\end{minipage}
\caption{
Examples for the singularity structure of the triangle $r=\sqrt{q^2}$ integrand.
The lines denote branch cuts stemming from the $\theta_1$ angle integral.
The red line is $\gamma_{a+}$, the orange line $\gamma_{a-}$, the blue line $\gamma_{b+}$, and the cyan line $\gamma_{b-}$.
The green dots are the poles from a propagator, the magenta ones indicate where the relevant crossings of cuts/poles are.
Left: $p^2=-3m^2$, $\theta=2\pi/3$, $\theta_2=\pi$, two cuts cross at $i\,m$ so the magenta dot is at the same point as a green one.
Right: $p^2=-4m^2/3$, $\theta=\pi/3$, $\theta_2=\pi$, four cuts touch at $-m^2/3$.
The black lines are projections of the cuts into one plane.
}
\label{fig:sing_vert_3d}
\end{figure*}

We start with a single specific point for which the situation is not only trivial but also physically clear.
When $\theta=\pi$, the external momenta $p_a$ and $p_b$ are anti-aligned.
The two denominators agree then and effectively we have a two-point integral.
Hence, the branch point for $p^2$ is at $p_B^2=-4m^2$.\footnote{When the present case is generalized beyond perturbation theory, one has to ensure that no kinematic divergences occur in special cases.
Then, as discussed in Ref.~\cite{Alkofer:2008dt}, the limit $\theta\rightarrow 0$ and the integration cannot be interchanged.}

For $\theta<\pi$, one needs to determine the points in $r$ which affect the contour deformation.
We call them singular points.
In the case of the two-point integral this happens when the end point of the cut lies at a pole of the other propagator, viz., the singular points are $\pm i\,m$.
Here, however, we can also encounter a nonanalyticity in the integrand when there is a set of $\theta_1$ and $\theta_2$ for which both cuts are at the pole or cuts coincide otherwise in a certain way.
For the two-point integral, the end points of the angle integral are relevant, because they cannot be changed 
whereas the integration contour between them can be deformed.
For the three-point integral, there are two angles and the end points of the innermost angle integral become important, viz., $\theta_2=0$ or $\pi$, while the $\theta_1$ integral requires more care.
For values of $\theta_2$ between $0$ and $\pi$, the related cut $\gamma_{b\pm}$ is restricted to a subset of the other one, $\gamma_{a\pm}$.
Only $\gamma_{b\pm}$ depends on the external angle $\theta$.
Varying $\theta$ moves the starting and end points and can also restrict $\gamma_{b\pm}$ to a subset of $\gamma_{a\pm}$.

In \fref{fig:sing_vert_3d} the cut structure in the complex $r$ plane is illustrated for $\theta_2=\pi$.
There are four cuts (two from each $\gamma$) which are partially on top of each other.
To disentangle them, the value of $\theta_1$ is shown along the third axis.
The decisive observation is that a single line can be deformed by taking the angle $\theta_1$ into the complex plane, but if there are several lines, this will affect the others as well.
Hence, it is possible that no suitable deformation of the angle integration contour exists.
When this happens, we have found a threshold in the external momenta.
Clearly, such a situation can only arise if two cuts come close to each other for a certain value of the angle.
Thus, we will first determine when the two cuts agree for general $\theta_1$.

To this end, we compare the solutions for $r$, see \eref{eq:gamma_ab}.
They agree if the following equation is fulfilled for $p_b^2=p_a^2=p^2$ and $\theta_2=0$ or $\pi$ (denoted by the two signs):\footnote{It should be noted that $\gamma_{a+}$ can only agree with $\gamma_{b+}$ (and similar for $\gamma_{a-}$ and $\gamma_{b-}$) except when all four cuts come together for the same value of $\theta_1$. However, the two solutions lead to the same result which is why we ignore this subtlety here.}
\begin{align}\label{eq:3p_twoCut_cond}
 -\cos \theta_1 = \cos \theta\cos\theta_1\pm\sin\theta\sin\theta_1 \, .
\end{align}
Dividing by $\cos\theta_1$ (excluding $\theta_1=\pi/2$ for now) we can rewrite this as
\begin{align}
 -1-\cos\theta&=-2\cos^2\left(\frac{\theta}{2}\right)=\pm2\cos\left(\frac{\theta}{2}\right)\sin\left(\frac{\theta}{2}\right)\tan\theta_1 \, ,\nnnl
 -\cot\left(\frac{\theta}{2}\right)&=\pm\tan\theta_1=\tan(\pm \theta_1).
\end{align}
Using $\cot(x)=-\tan(x-\pi/2)$, we find the solution
\begin{align}\label{eq:sol_t1_1}
 \theta_{1,c\pm}=\pm\frac{\theta}{2}+\frac{\pi}{2}.
\end{align}
Alternatively, one can show this directly from \eref{eq:3p_twoCut_cond} by using an elementary angle addition theorem.

We distinguish now two cases for the agreement of the two cuts.
One possibility is that they meet at the position of one of the poles of the third denominator, $r=\pm i\,m$.
To determine the branch point in the external momentum, we set the first branch cut equal to $\pm i\,m$.
This leads to
\begin{align}
 \sqrt{p_{B,1}^2}=\mp 2\,i\,m\cos\theta_1,
\end{align}
where the upper/lower sign is the solution for $\gamma_{a+}/\gamma_{a-}$.
Note that this is in agreement with the two-point case before setting $\theta_1$ to zero.
The difference here is that $\theta_1$ is determined by the agreement of the two cuts.
Thus, we plug in the solution $\theta_{1,c\pm}$, \eref{eq:sol_t1_1}.
Here, one needs to choose $\theta_2$ as $0$ or $\pi$ based on which pole one wants to avoid.
However, the final result for $p_B^2$ is independent of that choice:
\begin{align}\label{eq:sol_p1}
 p_{B,1}^2&=-4m^2\cos^2\left(\pm\frac{\theta}{2}+\frac{\pi}{2}\right) = -4m^2 \sin^2\frac{\theta}{2}.
\end{align}
This is one possible branch point, arising from the end points in the $\theta_2$ integration touching a pole at $\pm i\,m$ for the given $\theta_1$.

\begin{figure}
\includegraphics[width=0.48\textwidth]{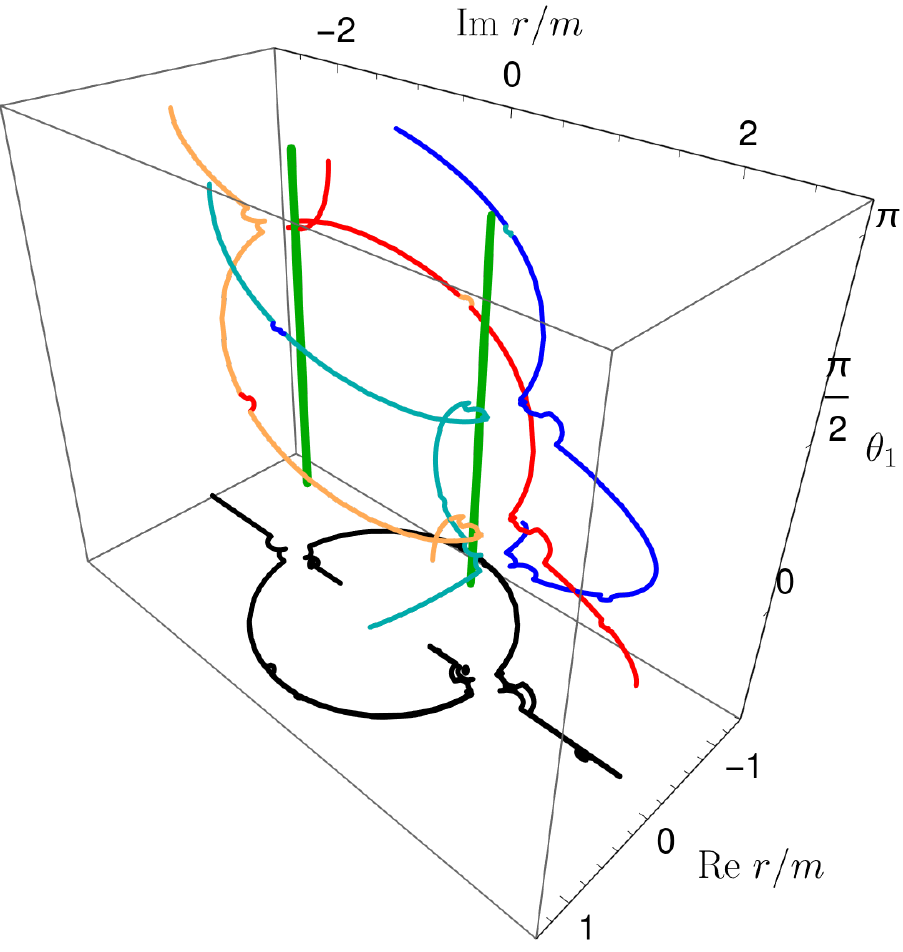}
\caption{
Cut structure for the triangle $r=\sqrt{q^2}$ integrand at $p_a^2=p_b^2=-2.5m^2$, $\theta=2\pi/3$, and $\theta_2=\pi$ when $\theta_1$ is deformed such as to avoid the singular points.
The threshold for this value of $\theta$ would start at $p^2=-3m^2$.
The avoidance of singular points can clearly be seen in the projection (black).
Colors are as in \fref{fig:sing_vert_3d}.
The poles at $\pm i\,m$ are denoted by the green lines.
The jumps in the colors for some deformations are due to the chosen parametrization of the cuts.
} 
\label{fig:vert_avoidPole}
\end{figure}

We illustrate now why the situation with the cuts from two propagators agreeing at the pole of the third propagator leads to a threshold.
We start with a point $p^2$ before the threshold so that the crossing of the cuts happens before $\pm\,i\,m$.
It is then possible to deform each cut such that it bypasses the pole in the upper half-plane, e.g., by using semicircles as given by \eref{eq:t1_complex}.
The pole is passed for the following values of $\theta_1$ for the first and second cut:
\begin{align}
 \theta^{*a\pm}_{1,\text{pole}}&=\arccos\left(\pm i\frac{\sqrt{p^2}}{2m}\right),\\
 \theta^{*b\pm}_{1,\text{pole}}&=\arccos\left(\pm i\frac{\sqrt{p^2}}{2m}\right)\pm\theta,
\end{align}
respectively.
As it turns out, we need to choose opposite signs for the phases of the semicircles for the two cuts.
Of course, every deformation in one cut also introduces a deformation in the other.
Thus, as the crossing of the cuts comes closer to $\pm i\,m$, the two deformations start to interfere until they are no longer possible if eventually the crossing is at $\pm i\,m$.
The situation with $p^2$ slightly before the threshold is illustrated in \fref{fig:vert_avoidPole}.
For completeness, also the necessary deformations to open a gap where all cuts meet, which happens for
\begin{align}\label{eq:theta1_gap}
 \theta^{*a\pm}_{1,\text{gap}}&=\arcsin\left(\pm i\frac{m}{\sqrt{p^2}}\right),\\
 \theta^{*b\pm}_{1,\text{gap}}&=\arcsin\left(\pm i\frac{m}{\sqrt{p^2}}\right)\pm\theta \, ,
\end{align}
are included.

The second relevant possibility is realized differently, and the poles at $\pm i\,m$ are not involved because a singular point is reached for $r$ closer to the origin.
As illustrated in \fref{fig:vert_avoidPole} from the previous example, it is also necessary to deform the contour at the point where all four cuts meet on the imaginary axis.
Again, this is possible when different signs for the phases in the deformations, \eref{eq:t1_complex}, can be chosen, but it is prevented when the cuts meet for the same values of $\theta_1$.
Then it is impossible to open a gap through which the $r$ integration can be performed and a branch point emerges.

\begin{figure*}[tb]
 \includegraphics[width=0.49\textwidth]{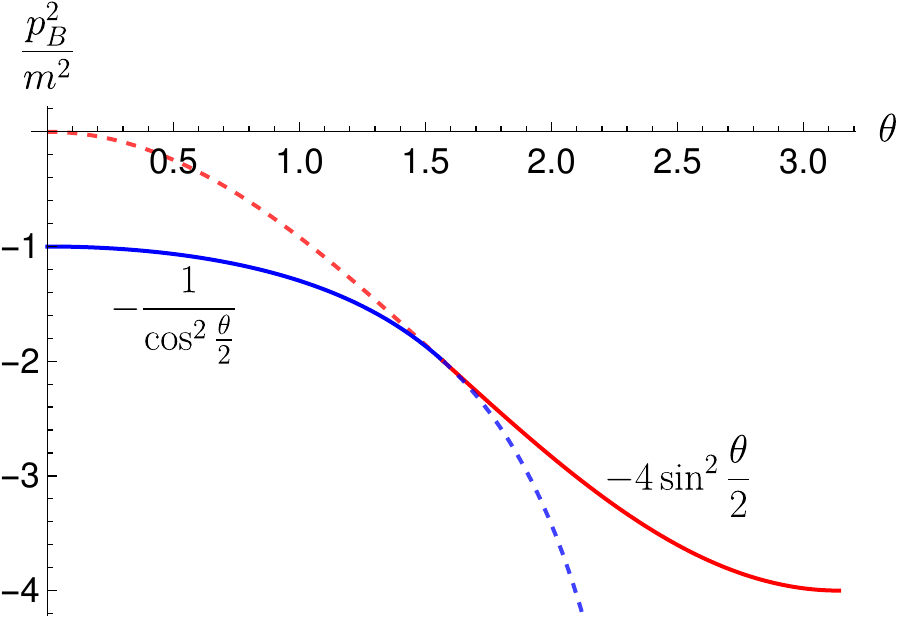}\hfill
 \includegraphics[width=0.49\textwidth]{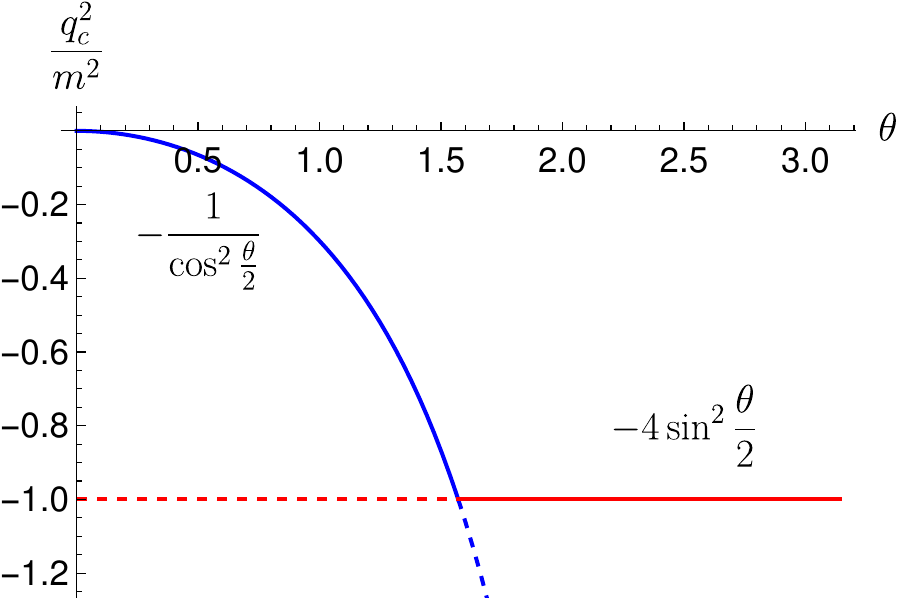}
 \caption{Left: The positions of the two potential branch points from Eqs.~(\ref{eq:sol_p1}) and (\ref{eq:sol_p2}).
 Right: The critical points as functions of $\theta$.
 The dashed lines correspond to the irrelevant cases and the continuous ones to the physical solutions.}
 \label{fig:twoSols}
\end{figure*}

To find this point, we take the solutions for the branch cuts from the first denominator, $\gamma_{a\pm}$, and determine when the last term, which distinguishes the two solutions, is zero.
From setting the argument of the square root to zero, see \eref{eq:denom_sol}, we obtain the condition
\begin{align}
 p_{B,2}^2 &= -\frac{m^2}{1-z_1^2}=-\frac{m^2}{\sin^2\theta_1}.
\end{align}
(NB: Eq.~(\ref{eq:theta1_gap}) was actually determined from this condition).
Plugging in the solutions \eref{eq:sol_t1_1} for $\theta_1$, this leads to
\begin{align}\label{eq:sol_p2}
 p_{B,2}^2 &=-\frac{m^2}{\sin^2\left(\pm\frac{\theta}{2}+\frac{\pi}{2}\right)}=-\frac{m^2}{\cos^2\frac{\theta}{2}}.
\end{align}

There is an alternative approach to obtain this result.
From a naive analysis that takes into account only the two propagators with the momenta $k_a$ and $k_c$, which actually corresponds to the analysis of a contracted diagram, we know that there is a branch point for $p_a^2=-4m^2$.
However, we discarded it because of the branch point at $p_a^2>-4m^2$.
The former branch point also exists for $p_c^2$.
Thus, using $p_c^2=4\,p_a^2\,\cos^2(\theta/2)$, we can rewrite this to a condition for $p_a^2$: $p_a^2=-m^2/\cos^2(\theta/2)$.
This is \eref{eq:sol_p2} from before.

We now have identified two possible branch points in $p^2$: $p_{B,1}^2$ and $p_{B,2}^2$.
It remains to be checked under what conditions which one is relevant.
In \fref{fig:twoSols} we show the two solutions and the corresponding singular points as a function of $q^2$ which prevent a proper integration, viz., the points where cuts or poles intersect.
For the first solution this happens at $q_{c,1}^2=-m^2$, for the second at
\begin{align}\label{eq:qc2}
 q_{c,2}^2=\gamma_{a\pm}^2(\theta_{1,c};p_{B,2}^2,m^2)=-m^2 \tan^2\frac{\theta}{2},
\end{align}
where we plugged in the solutions for $\theta_1$ and $p^2$, Eqs.~(\ref{eq:sol_t1_1}) and (\ref{eq:sol_p2}), respectively, into the expression for the cut, \eref{eq:denom_sol}.

At $\theta=\pi/2$ both solutions agree.
For $\theta>\pi/2$, it is clear from \fref{fig:twoSols} that the relevant solution is $p_{B,1}^2$, as the pole in $q^2$ at $-m^2$ is the first critical point hit.
For $\theta<\pi/2$, on the other hand, the first singular point hit comes from the second solution which is, consequently, relevant in this case.

\begin{figure*}[tb]
 \includegraphics[width=0.48\textwidth]{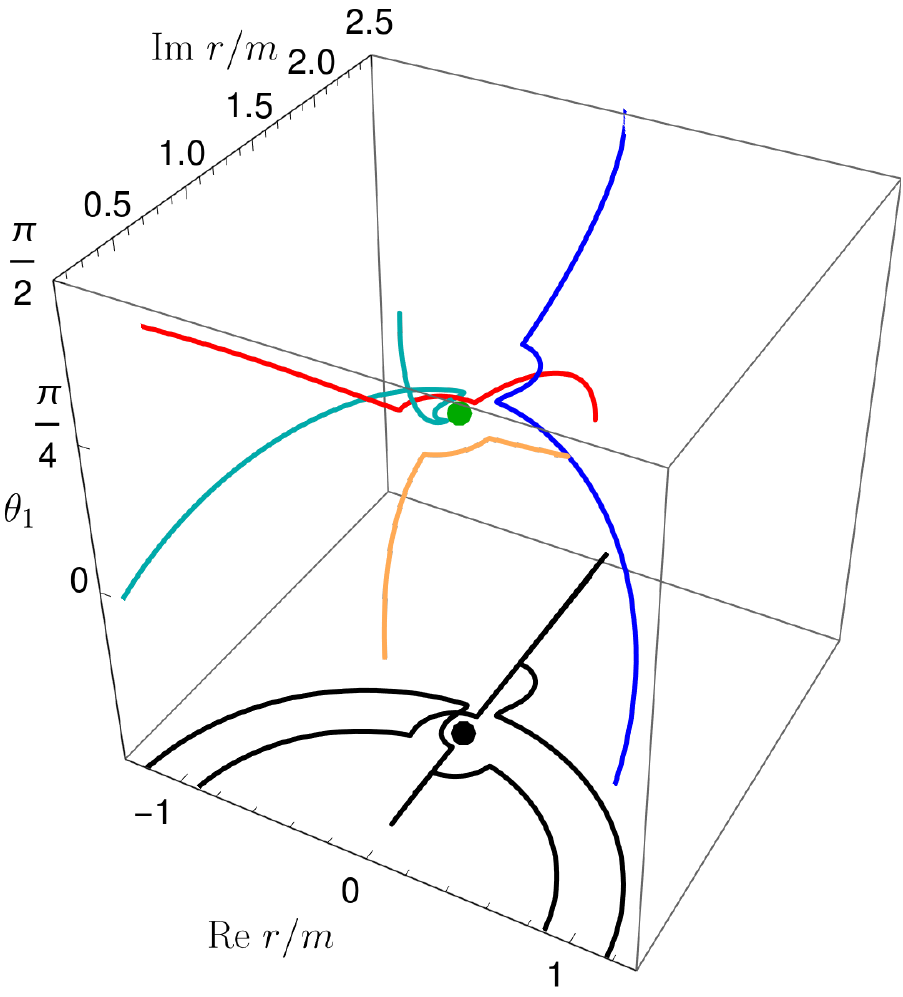}
 \hfill
 \includegraphics[width=0.48\textwidth]{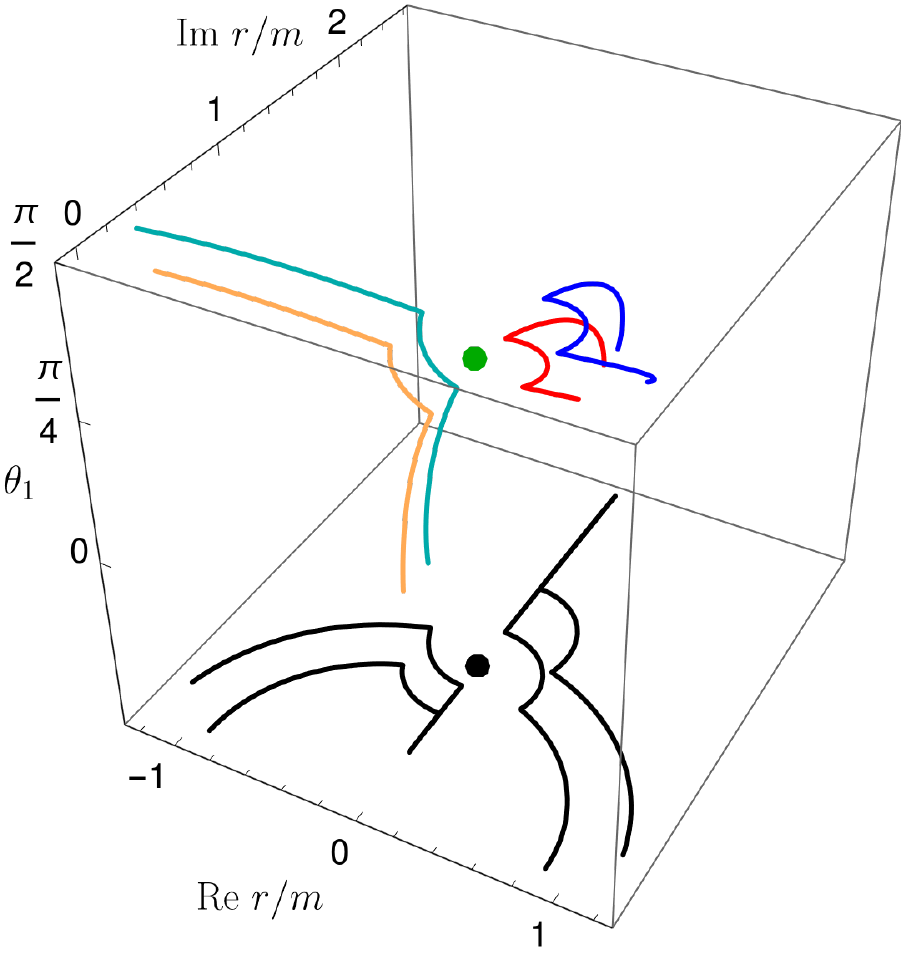}
 \caption{The cuts for $p_a^2=-1.8m^2$, $p_b^2=-2.4m^2$ and $\theta_1 \in [0,\pi/2]$.
 In the left/right plot, $\theta_+/\theta_-$ is used.
 The contours are deformed around the point where the cuts touch.
 This opens a path for $\theta_-$ but not for $\theta_+$ as can be seen in the projections (black).
 Colors are as in \fref{fig:sing_vert_3d}.
 }
 \label{fig:cp_cm}
\end{figure*}

\begin{figure*}
\includegraphics[width=0.48\textwidth]{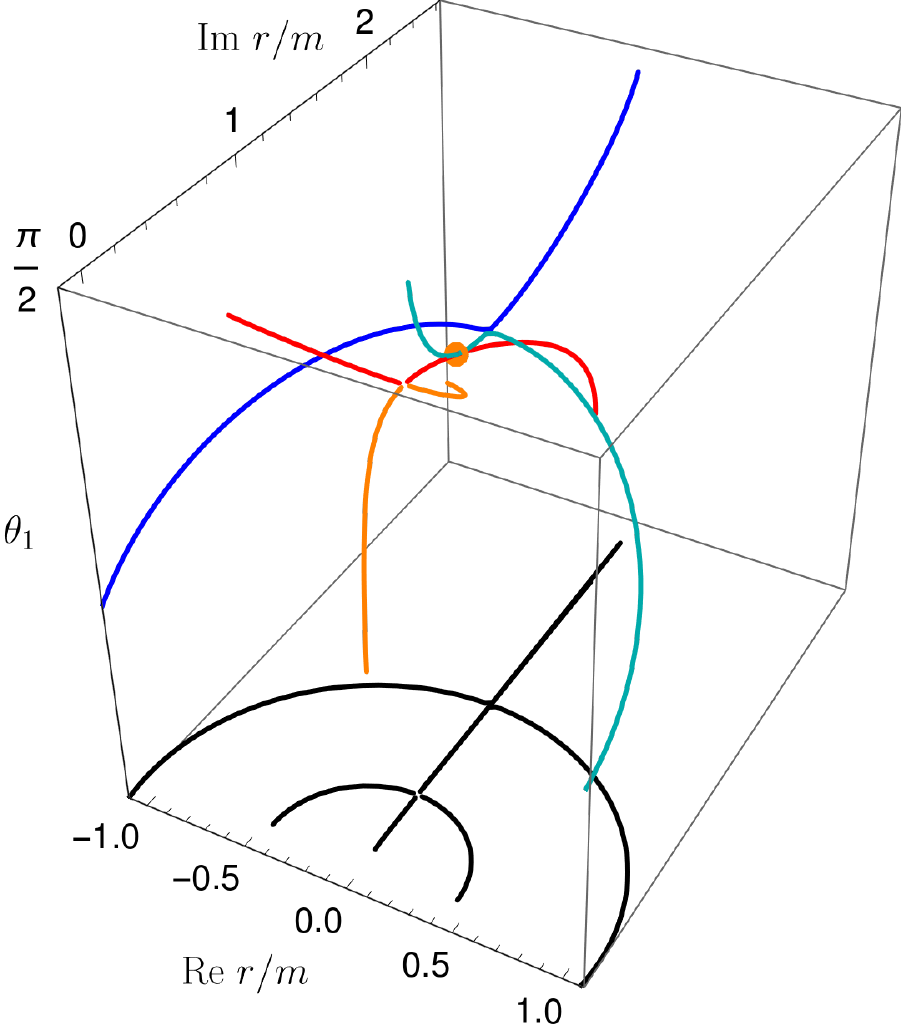}
\hfill
\includegraphics[width=0.48\textwidth]{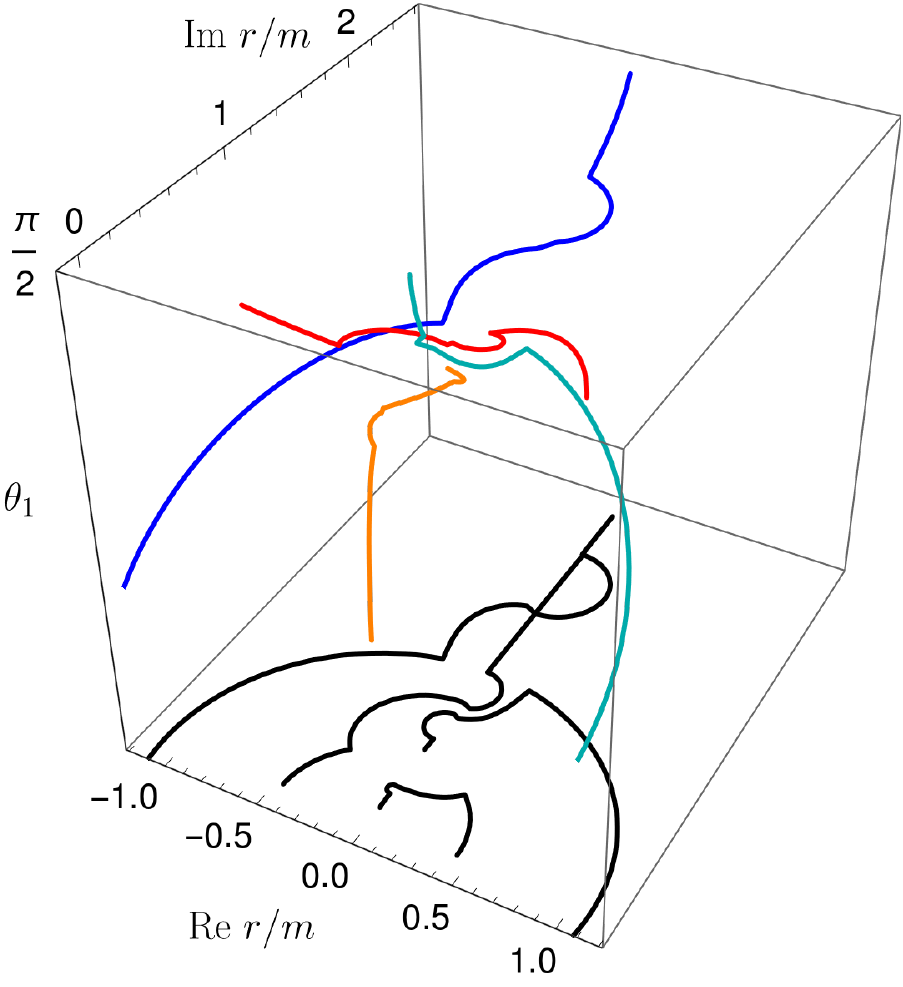}
\caption{Cuts of the triangle diagram for $p_a^2=-1.2m^2$, $p_b^2=-2.2m^2$ and $\theta_1 \in [0,\pi/2]$.
In the left plot, $p_c^2=-4m^2$, and the cuts touch at $\sqrt{(p_a^2+p_b^2)/2+m^2}$.
In the right plot, the value for $\theta$ is slightly shifted compared to the left plot and the cuts cross at two points.
However, a contour deformation can be found that allows to lead the integration out of the two circles as can be seen by the projected cuts in black.
Colors are as in \fref{fig:sing_vert_3d}.
}
\label{fig:vert_abm2}
\end{figure*}

We want to add that all of these considerations were checked graphically by plotting the branch cuts in the $r$ plane for and around the respective solutions to confirm that the contour deformation is indeed prohibited by these configurations.
In this context it should also be mentioned why we chose to work with $r=\sqrt{q^2}$ instead of $q^2$.
The reason is that ambiguities can arise when $q^2$ is used because one value of $q^2$ corresponds to two values of $r$.
Hence, it is possible that a contour deformation is not possible in $q^2$ but it is in $r$.
This can be seen in the parametrizations of the cuts in $r$ and $q^2$, 
\eref{eq:denom_sol} and \eref{eq:denom_sol_squared}, respectively.
The latter is oblivious of the sign of $\cos\theta_1$.
Solving the condition for agreement between the two cuts requires then to solve also for the opposite sign and leads to the additional solutions $\theta_1=\frac{\theta}{2}$ and $\pi-\frac{\theta}{2}$.
It can indeed happen that for certain kinematic configurations no contour deformation in $q^2$ is possible for these values but in $r$ it is.
Hence this is an artifact, and we use $r$ throughout.
For the propagator selfenergy this problem did not appear and working in $q^2$ is thus possible.

To summarize, for the chosen kinematics there is a branch point at
\begin{align}\label{eq:pB2_restKin}
p_B^2=\left \lbrace \begin{array}{c c} -4m^2 \sin\left(\frac{\theta}{2}\right)^2 & \quad \frac{\pi}{2}\leq \theta \leq \pi  \\ \frac{-m^2}{\cos\left(\frac{\theta}{2}\right)^2} & \quad  0\leq \theta \leq \frac{\pi}{2} \end{array} \right. .
\end{align}
This agrees with the Landau analysis discussed in App.~\ref{sec:Landau_cond_3p}.
The existence of different solutions below and above $\theta=\pi/2$ in both approaches can be directly related.
For $\theta>\pi/2$, the branch points come from the interplay of all three denominators.
On the other hand, the solution for $\theta<\pi/2$ arises from a contracted diagram in the Landau analysis.
This is also evident in the contour analysis, as this branch point is created from only two propagators.

\subsubsection{General kinematics}
\label{sec:genKin}

Finally, we consider general kinematics.
We start with the case where all three propagators are involved in creating the threshold.
As for the restricted kinematics discussed above, this corresponds to the case of the triangle diagram without any contractions for the Landau analysis.
In our choice of routing, there is one propagator that creates poles at $\pm i\, m$.
We thus have to determine when the other two cuts cross that point.
This can be derived, for example, by setting Eqs.~(\ref{eq:gamma_ab}) to $\pm i\,m$.
This leads to the conditions
\begin{subequations}
\label{eq:papb}
\begin{align}
\label{eq:pa}
 \sqrt{p_a^2}=&2i\,m \cos\theta_1,\\
 \sqrt{p_b^2}=&-2i\,m (\cos\theta \cos\theta_1+\sin \theta \sin\theta_1),
\end{align}
\end{subequations}
where we have chosen the specific end point $\theta_2=0$ in the second case and used the pole at $i\,m$.
$\theta_1$ is now fixed from
\begin{align}
 2i\,m=\frac{\sqrt{p_a^2}}{\cos\theta_1}=-\frac{\sqrt{p_b^2}}{\cos\theta \cos\theta_1+\sin \theta \sin\theta_1}
\end{align}
as
\begin{align}
 \tan\theta_{1,s}=\frac{-\frac{\sqrt{p_b^2}}{\sqrt{p_a^2}}-\cos\theta}{\sin\theta}.
\end{align}
Plugging this into \eref{eq:pa} yields
\begin{align}
 p_a^2=-\frac{4\,p_a^2\,m^2\sin^2\theta}{p_a^2+p_b^2+2\sqrt{p_a^2}\sqrt{p_b^2}\cos\theta},
\end{align}
where we used $\cos\arctan x=1/\sqrt{1+x^2}$.
Finally, we express the angle $\theta$ by the external momenta, $\cos\theta=(p_c^2-p_a^2-p_b^2)/(2\,\sqrt{p_a^2\,p_b^2})$, leading to
\begin{align}\label{eq:genKinCond}
 p_a^2=&-4\,m^2\,p_a^2\frac{4p_a^2p_b^2-(p_c^2-p_a^2-p_b^2)^2}{4\,p_a^2\,p_b^2}\frac{1}{p_c^2},\nnnl
 p_a^2\, p_b^2\, p_c^2=&m^2(p_a^4+p_b^4+p_c^4-2(p_a^2\,p_b^2+p_a^2\,p_c^2+p_b^2\,p_c^2)).
\end{align}
This is identical to the Landau condition \eref{eq:LC_triangle_app}, derived in App.~\ref{sec:Landau_cond_3p}.

We know from the Landau analysis in App.~\ref{sec:Landau_cond_3p} that this solution only applies for certain values of the external momenta.
In summary, only one solution of this quadratic equation corresponds to a threshold and the condition $p_a^2+p_b^2<-4m^2$ needs to be fulfilled.
We will first explain why only one solution is relevant.
To this end, we solve the condition for $p_c^2$:
\begin{align}
 p_{c\pm}^2=&\frac{1}{2m^2} \Big( 2(p_a^2+p_b^2)m^2 + p_a^2 \,p_b^2\nnnl
 &\pm \sqrt{p_a^2}\sqrt{4m^2+p_a^2} \sqrt{p_b^2}\sqrt{4m^2+p_b^2} \Big).
\end{align}
For given $p_a^2$ and $p_b^2$, one can calculate the angle between $p_a$ and $p_b$ from this which we call $\theta_\pm$.
This angle influences the cut of only one propagator.
As it turns out, there is a decisive difference for the position where the cuts from $\theta_+$ and $\theta_-$ touch, see \fref{fig:cp_cm} for an example.
For $\theta_-$, the contour can be deformed around this singular point and thus no threshold emerges.
For $\theta_+$, on the other hand, one cut has the opposite direction and similar to the discussion for the restricted kinematics, this leads to the opposite orientations of the semicircles.
The solution leading to a threshold is thus $p_{c+}^2$ in agreement with the analysis in App.~\ref{sec:Landau_cond_3p}.

Now we turn to the question of how the condition $p_a^2+p_b^2<-4m^2$ emerges.
Again, the reason is that for certain configurations there are singular points in the $r$ plane which are closer to the origin than $\pm i\,m$.
We first motivate the existence of further thresholds by considering only two propagators.
For one we choose the form $q^2+m^2$ and for the other $(q-p_i)^2+m^2$, where $i\in \{a,b,c\}$.
This leads to the following three thresholds:
\begin{align}
 p_a^2=&-4m^2,\\
 p_b^2=&-4m^2,\\
 p_c^2=&-4m^2.
\end{align}
Consequently, there is a threshold for all three external momenta squared at $-4m^2$.
However, changing the routing during an actual numerical calculation would be tedious.
Furthermore, we do not know if these thresholds really come from singular points closer to the origin than $\pm i\,m$.
Thus we continue now with the original kinematics, \eref{eq:vert_kinematics}, and determine the singular point.

In the following we assume $p_b^2<p_a^2$ without loss of generality.
First, we distinguish three cases depending on where the poles at $\pm i\,m$ are in relation to the cuts.
Since they can be inside or outside of the circles, there are three possible combinations.
For $p_b^2<p_a^2<-2m^2$, the poles are inside both circles and $p_a^2+p_b^2<-4m^2$.
From the results obtained above we know that we can deform the contour appropriately in this case.
For $-2m^2<p_b^2<p_a^2$, the pole is outside of the circles.
However, the cuts touch before creating a singular point.
Also for $p_b^2<-2m^2<p_a^2$ the cuts touch before the pole which lies inside the circle related to $p_a$ and outside of the one related to $p_b$.
Testing $p_c^2=-4m^2$, as determined from the considerations above, we find that the cuts touch at one point but do not cross, see \fref{fig:vert_abm2} for an example.
Shifting $\theta$ determined from $c=-4m^2$, the cuts either do not cross anymore or cross at two points.
The crucial point is that we need to deform the contour such that a gap opens between the two cuts.
This is not possible when the cuts only touch, but when the cuts cross in two places, it can be realized, because we can use several deformations with opposite orientations.
Since the effect of deforming $\theta$ is different for the two cuts, viz., the distance to the original path differs, a gap can be opened by choosing the parameters appropriately, see \fref{fig:vert_abm2} for an example.

The position of the touching point is $\sqrt{(p_a^2+p_b^2)/2+m^2}$.
The poles at $\pm i\,m$ also create thresholds.
Thus, they create the highest threshold if $p_a^2+p_b^2<-4m^2$.
Otherwise, the two cuts touching before the poles create it.
This explains the result of the Landau analysis in terms of singular points in the $r$ integration.
Finally, we can set $p_b^2=p_a^2=p^2$ to compare with the results from \ref{sec:restrKin}.
The singular point is then at $\sqrt{p^2+m^2}$.
Plugging in the value for $p^2$ at the branch point from \eref{eq:sol_p2}, we recover \eref{eq:qc2}.

\subsection{Generalization to the nonperturbative case}
\label{sec:nonperturbative}

The analyses of the propagator and the vertex were using perturbative expressions.
For nonperturbative calculations, we need to consider a few generalizations.
However, it should be emphasized that in the nonperturbative case only one-loop diagrams appear for the case considered here, see \fref{fig:DSEs}.

We start by discussing the propagator.
In the perturbative case, it has a pole corresponding to the bare mass $m$.
Nonperturbatively, this mass is shifted, but besides a different numeric value of the mass, the analysis of the equation and the procedure to deform the contour remain the same.
Note that it is not even necessary to know the value of the mass extremely precisely, as the deformed contour is typically chosen not to pass the pole very closely.
For example, the extraction of the mass in the example of Sec.~\ref{sec:results} works reasonably well.

\begin{figure}[tb]
 \includegraphics[width=0.48\textwidth]{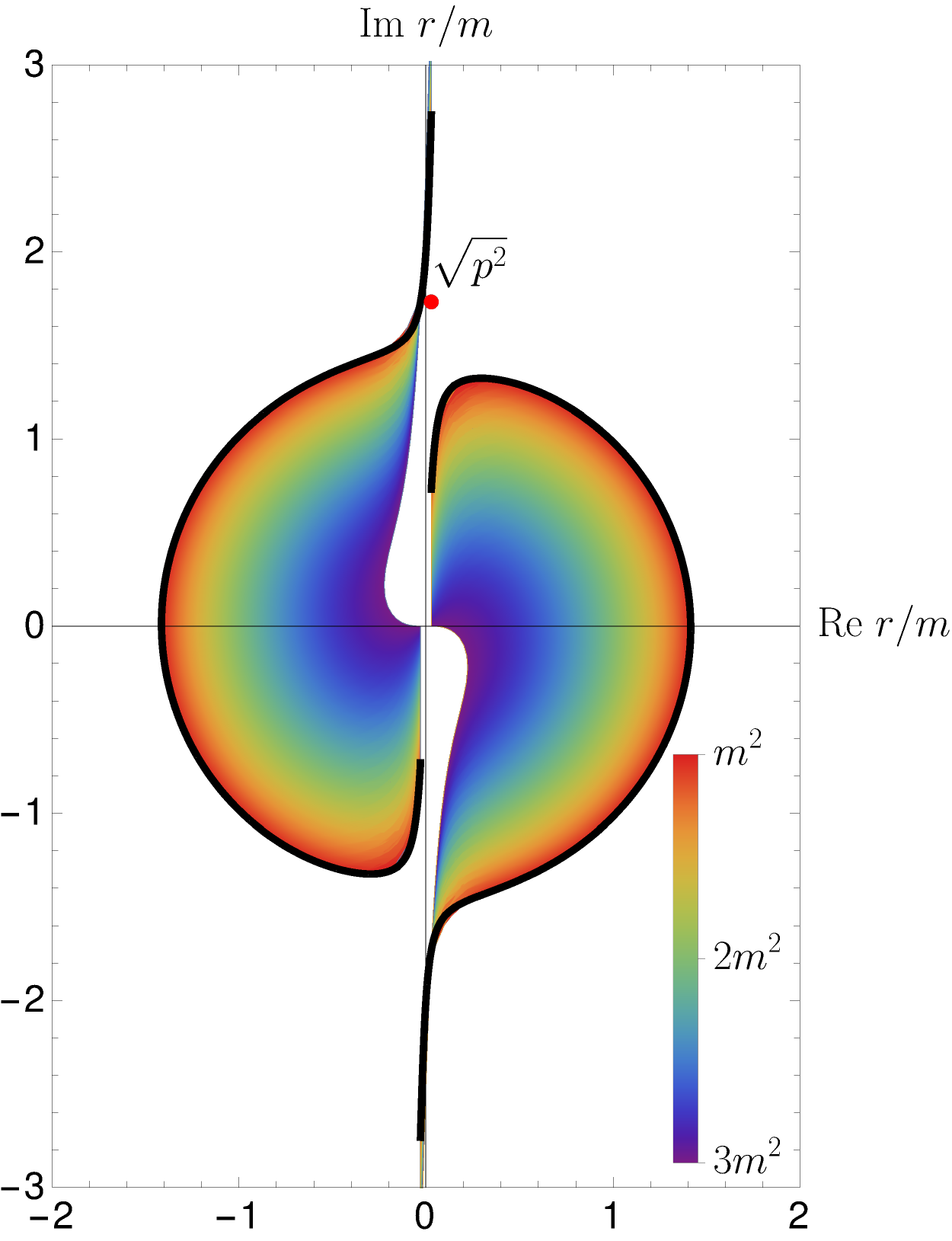}
 \caption{Forbidden region in the complex $r=\sqrt{q^2}$ plane for different values of the mass parameter $m_c^2$, which is indicated by color, and the external $p^2=(-3+0.1\,i)m^2$.}
 \label{fig:cutInProp}
\end{figure}

Beyond poles, we also need to consider branch cuts of the propagators which also create singularities in the $r$ integration that have to be avoided.
If such a cut exists, it can be treated in a similar way as poles by considering it as a continuum of points $\pm i\, m_c$ starting at a threshold value and going to $\pm i\,\infty$.
For each point of the propagator cuts, the angle integration results in cuts in the plane of the $r$ integration as given by \eref{eq:denom_sol} but with $m$ replaced by $m_c$.
Having a continuum of such points, this gives an area in the $r$ plane forbidden for the integration.
Such a region is illustrated in \fref{fig:cutInProp} where it is plotted for $m_c^2\in [m^2,3m^2]$.
Although it seems that the freedom for the integration is very restricted, a simple integration path passing $\sqrt{p^2}$ works.
Also curves for higher values of $m_c^2$ do not interfere with such an integration contour.
Formally, the area forbidden by a cut in the propagator is given by
\begin{align}
\label{eq:area_from_cut}
\gamma^\text{cut}_{\pm}(z_1;p^2,m_c^2)=&\sqrt{p^2}\,z_1 \pm i\sqrt{m_c^2+p^2(1-z_1^2)} \nonumber \\
 =& \sqrt{p^2}\cos\theta_1 \pm i\sqrt{m_c^2+p^2\sin^2\theta_1}
\end{align}
where $m_c$ varies from the start of the cut until infinity and $\theta_1\in[0,\pi]$.

\begin{figure*}
\includegraphics[width=0.48\textwidth]{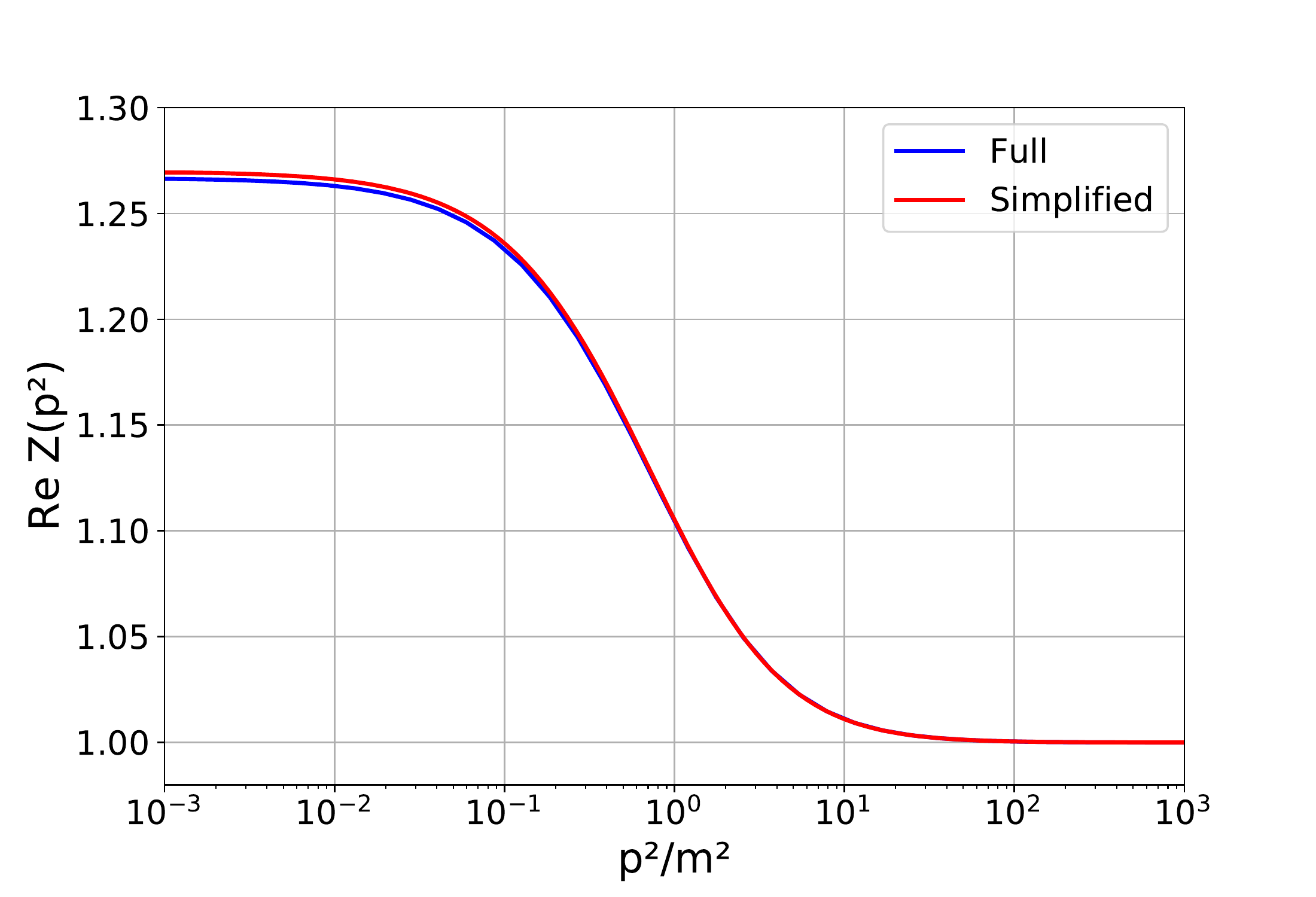}
\hfill
\includegraphics[width=0.48\textwidth]{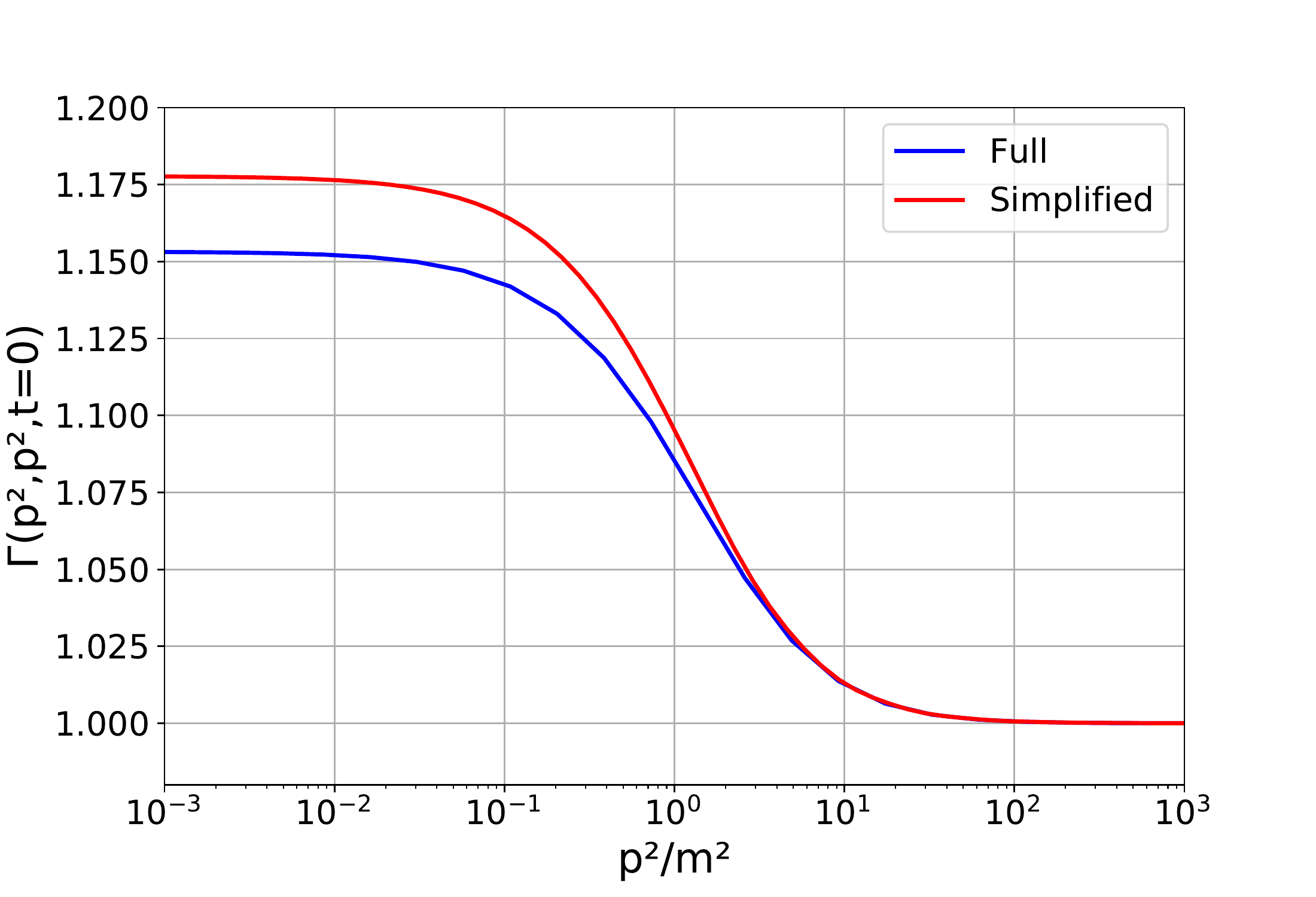}
\caption{The propagator dressing (left) and the vertex dressing with $\theta=\pi/2$ (right) for Euclidean momenta.
The blue curves were obtained with full kinematic dependence and the red ones with the approximation $p_a^2=p_b^2=p^2$.
}
\label{fig:prop_vert_comp}
\end{figure*}

It remains to discuss nonperturbative vertices.
Again, we identify singularities in the dressing functions.
In an iterative solution, one can start with the perturbative expressions and then include the created branch cuts in the next step.
Let us consider the dressed vertex in the propagator equation, see \eref{eq:DSE_prop} and \fref{fig:DSEs}, as the simplest case.
The three incoming momenta are $-p$, $-q+p$ and $q$.
For general kinematics, it is advantageous to employ convenient variables of the dressing.
We choose variables based on the permutation group $S_3$ \cite{Eichmann:2014xya}, which are also used in applications in QCD, e.g., \cite{Eichmann:2014xya,Williams:2015cvx,Huber:2020keu}.
We have then two angles and one scale variable $S_0=(p_a^2+p_b^2+p_c^2)/6$ in which the singularities are expected.
In the propagator equation, $S_0=(p^2+q^2+p\cdot q)/3$.
Note that this looks in structure similar to the arguments of the propagators with the exception of a missing factor 2 in front of $p\cdot q$ and the overall factor $1/3$.
We thus have to solve $S_0=-m_s^2$ where $-m_s$ is the singularity in the dressing.
The sign was chosen in analogy to poles in the propagator.
This yields
\begin{align}
\label{eq:sol_S0}
\gamma^{S_0}_{\pm}(z_1;p^2,m_s^2)=&\frac{1}{2}\sqrt{p^2}\,z_1 \pm i\sqrt{\frac{1}{3}m_s^2+p^2\left(1-\frac{1}{4}z_1^2\right)}\,.
\end{align}
The resulting forbidden region does not introduce a new singularity in the integral.
In general, one can convince oneself graphically that the solution of an equation of the form $p^2+q^2+n\,p\cdot q=-s\,m^2$ depends on the variable $n$ and $s$ as follows:
$s$ determines the form of the curve and $n$ varies its length.
Since we have studied the case $s=1$ and $n=2$ in detail, we can directly infer that changing $n$ and $s$ does not introduce any new relevant obstructions.
In addition, we also tested the effect of introducing a phase, although our results only show cuts for real $m_s^2$,
Even then we did not find additional relevant obstructions and conclude that the nonperturbative propagator can be calculated as determined in the purely perturbative case.
A convenient integration contour is from the origin via $\sqrt{p^2}$ beyond all the cuts and then in an arc to the UV cutoff.

Finally, the same procedure needs to be applied to the triangle diagram.
The $S_0$ variable for two vertices falls into the same class as the propagators with momenta $q-p_a$ and $q+p_b$.
The third one is slightly more complicated, but again a graphical analysis shows no new obstructions occur as long as the vertex singularity is further away from the origin as the one from the propagators.
For the numeric calculation of Sec.~\ref{sec:results}, where we restricted ourselves to the kinematics of Sec.~\ref{sec:CDM_prop}, this is automatically the case as can be seen from \eref{eq:pB2_restKin}.
Of course, $m$ is now the nonperturbative mass.
The numeric results confirm that conclusion.

\section{Numerical solution of the coupled system}
\label{sec:results}

\begin{figure*}
\includegraphics[width=1\textwidth]{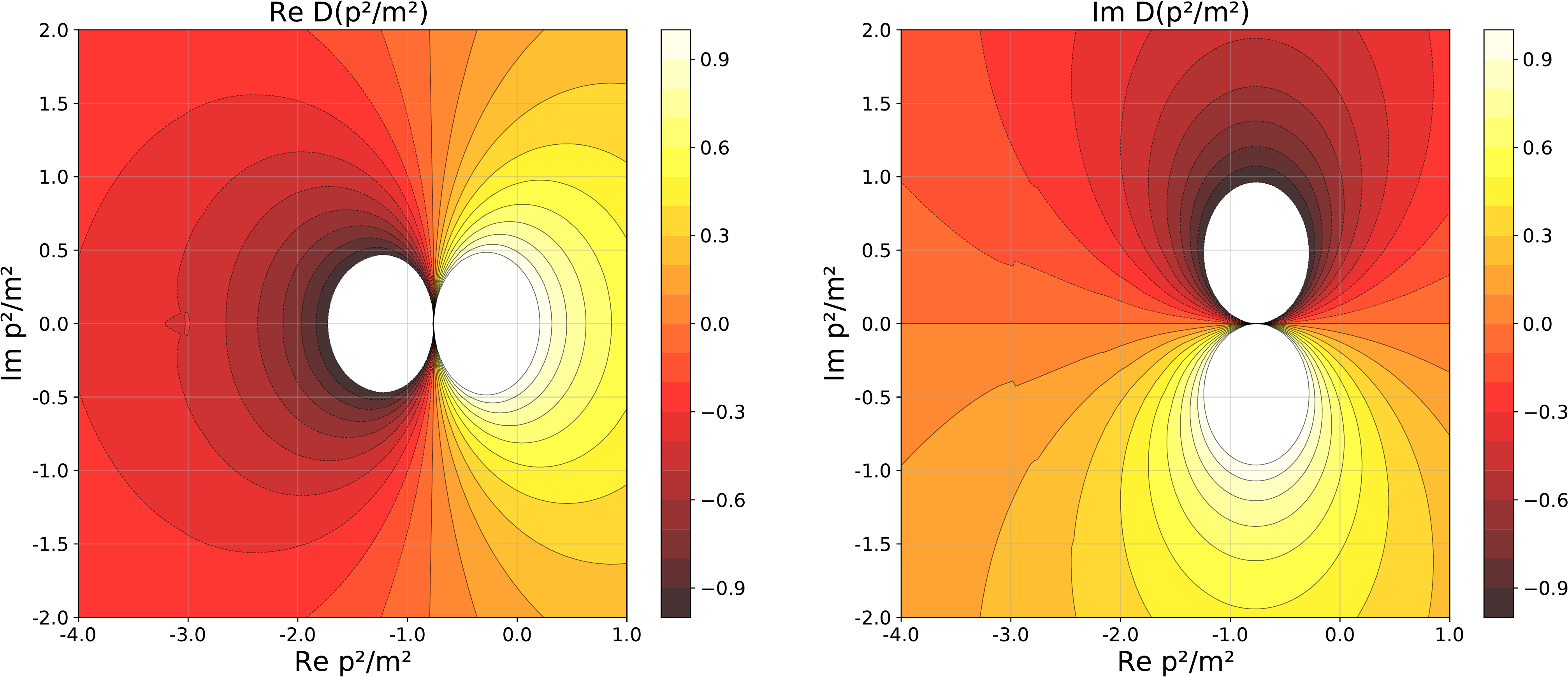}\\
\includegraphics[width=1\textwidth]{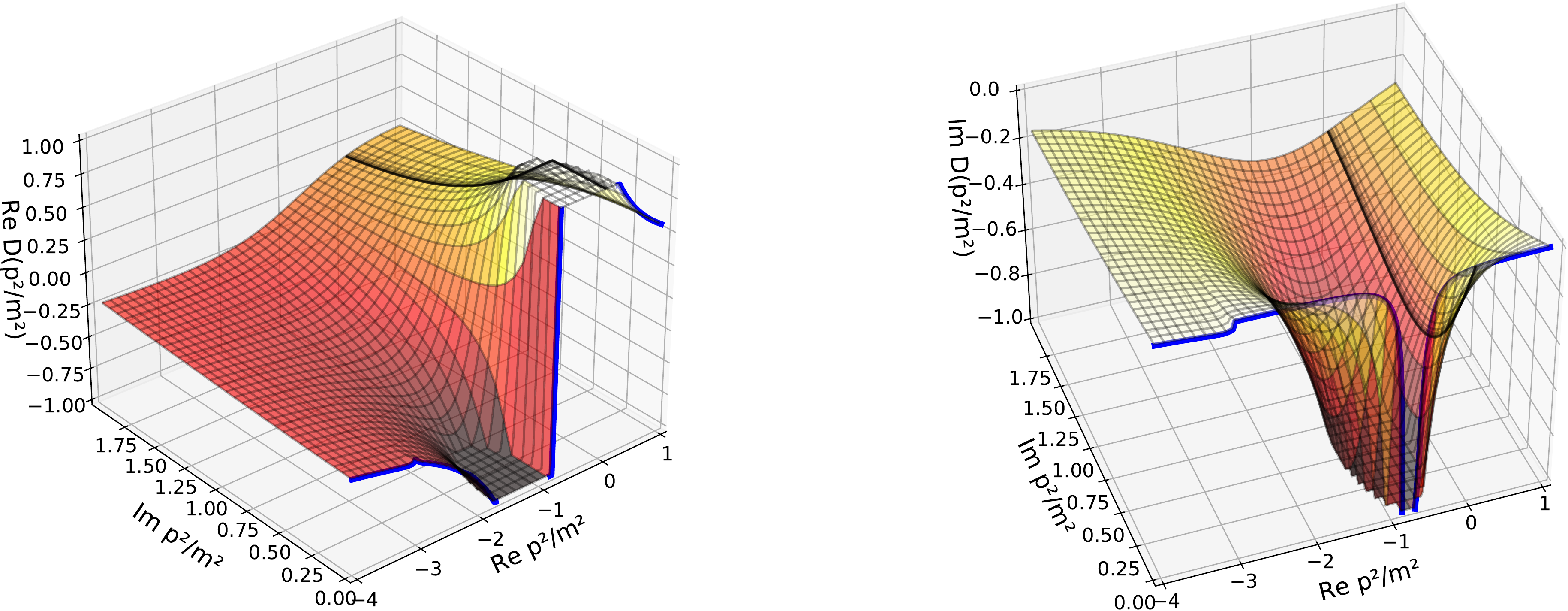}
\caption{Contour (top) and 3d (bottom) plots for the real (left) and imaginary (right) parts of the propagator $D(p^2)$ for $g=3$.
One can clearly see the pole at approximately $p^2=-0.75m^2=-m_r^2$ and a branch cut starting approximately at $-3m^2=-4m_r^2$.
}
\label{fig:prop_g3}
\end{figure*}

\begin{figure}[tb]
 \includegraphics[width=0.48\textwidth]{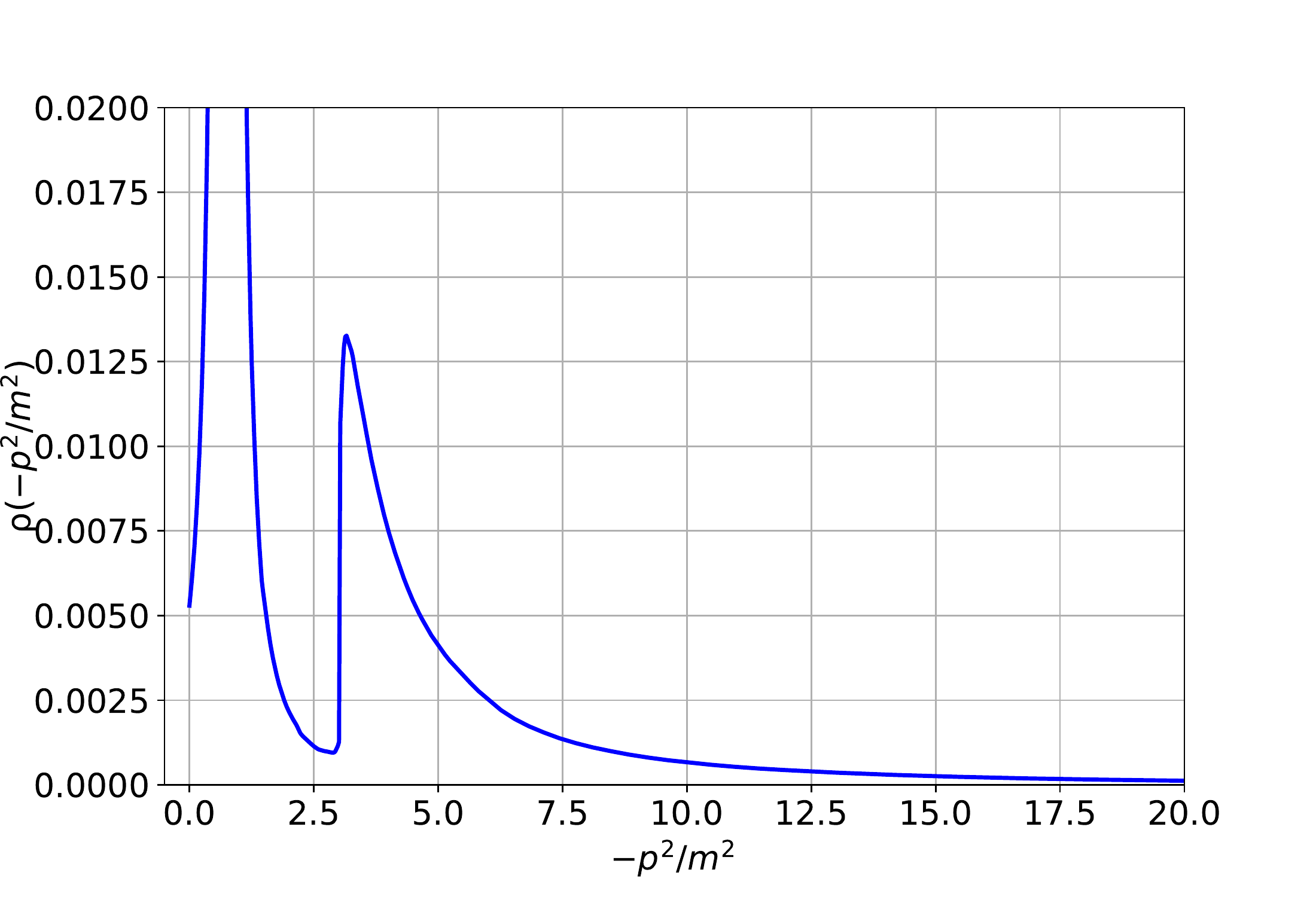} 
 \caption{Spectral density, \eref{eq:spectralDensity}, of the scalar propagator for $g=3$.
 Since we extract the spectral density at a finite distance from the real axis, the pole contribution has a finite width.}
 \label{fig:spectralDensity}
\end{figure}

\begin{figure}
\begin{center}
\includegraphics[width=.48\textwidth]{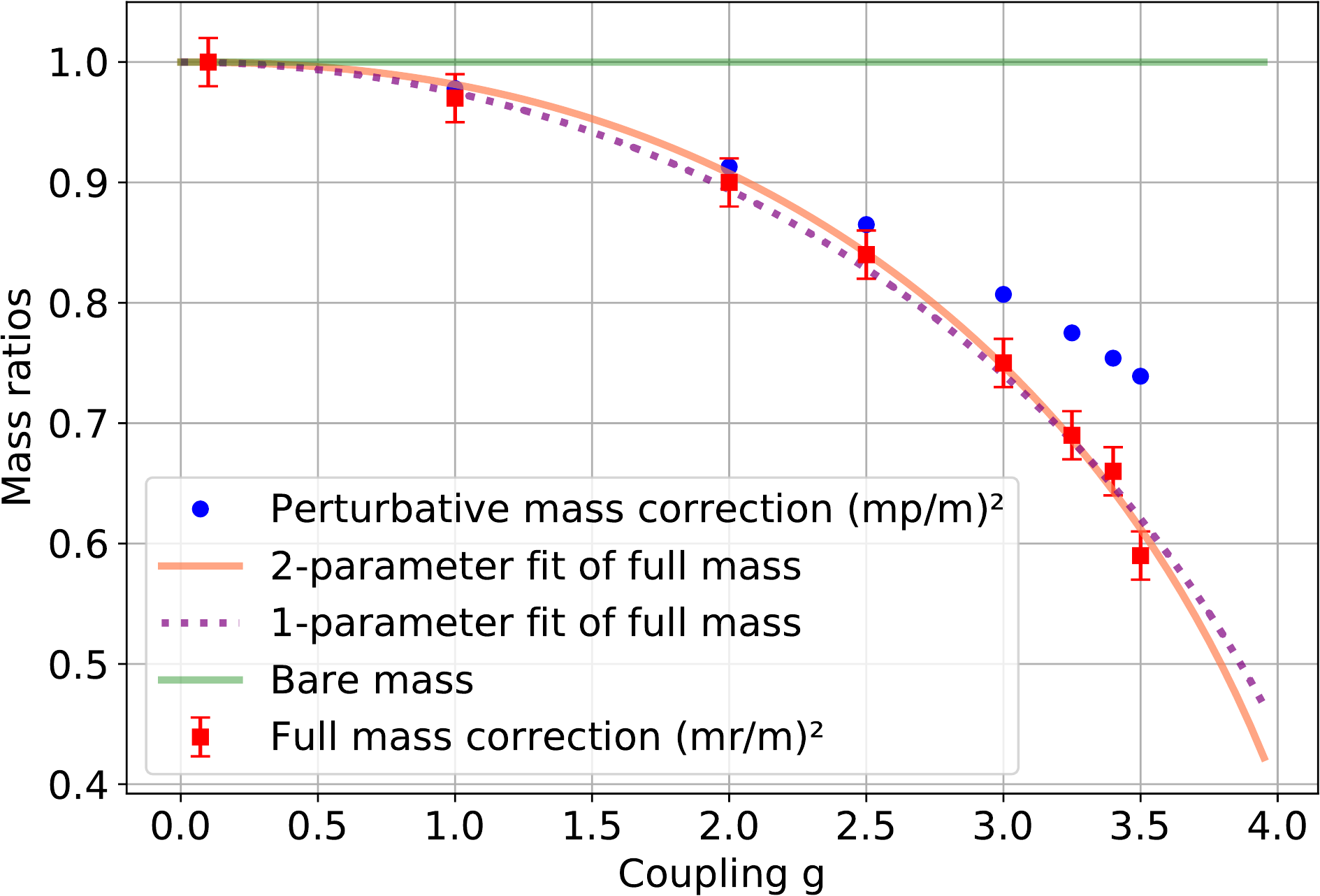}
\end{center}
\caption{Ratios of the perturbative and nonperturbative masses over the bare mass as a function of the coupling $g$.
The numerical fits \eqref{eq:Fit} with parameters \eqref{eq:Fit2} and \eqref{eq:Fit1} respectively, are also shown.}
\label{fig:masses}
\end{figure}

To test the analytic findings of the preceding section, we implement the coupled system of propagator and vertex DSEs numerically for the simplified kinematics $p_a^2=p_b^2=p^2$, see Sec.~\ref{sec:restrKin} for details.
The system is solved by a fixed point iteration starting from $Z(p^2)=1=\overline\Gamma(p_a,p_b,p_c)$ using standard methods, see, e.g., \cite{Huber:2011xc}.
As the system is very stable, we can calculate directly on a momentum grid in the complex plane and no more complicated methods like rays \cite{Fischer:2020xnb} are required.
We work in three dimensions where all integrals are finite.
We will give all dimensionful quantities in units of mass which we set to one if not stated explicitly.

To get an idea about the effect of the kinematic approximation, we compare this setup with the full setup for Euclidean momenta.
In this case, the integration contour can be along the positive real axis.
The results are shown in \fref{fig:prop_vert_comp}.
As can be seen, the effect of the approximation on the propagator is tiny.
For the vertex, the dressing shows a quantitative change for small momenta, which, however, is also small.
Nevertheless, it should be noted that small differences in the spacelike region do not forbid qualitative differences on the timelike side, see Refs.~\cite{Miramontes:2021xgn,Alkofer:2022hln} for an example.

For the calculations in the complex plane, we continue with the restricted kinematics only.
For momenta with nonvanishing imaginary part, the contour deformation can be directly implemented as the cuts have broad openings as illustrated in \fref{fig:sing_prop}.
Only for real $p^2<-m^2$, more care is required due to the cut structure as shown in \fref{fig:sing_prop_real}.
Instead of deforming the angle integration as described in Sec.~\ref{sec:CDM_prop}, we simply avoid the negative real axis and always keep a small imaginary part.
This turned out to be sufficient.

As expected, we observed that the perturbative pole of the propagator moves to a smaller value, viz., $m_r<m$ where $m_r$ is the effective mass.
We also find a branch cut on the timelike semiaxis.
The discontinuity of the branch cut is maximal at the branch point.
The positions of the pole and the branch point depend on the coupling.
Results for the propagator dressing in the complex plane are shown in \fref{fig:prop_g3} for a chosen value of the coupling.

The spectral density, given by
\begin{align}\label{eq:spectralDensity}
 \rho(s)=&-\frac{D(-s+i\,\epsilon)-D(-s-i\epsilon)}{2\pi\,i}\nnnl
 =&-\frac{\text{disc}D(-s)}{2\pi\,i}=-\frac{1}{\pi}\text{Im}D(-s).
\end{align}
is shown in \fref{fig:spectralDensity}.
Since we extract it at a finite distance $\epsilon$ to the real axis, the pole at $m_r^2$ is not a delta peak but has a finite width.
The threshold at $4m_r^2$ is clearly visible.

\begin{table}[tb]
\begin{tabular}{|c|c|c|c|c|c|}
\hline 
$g$ & pert. mass $m_p^2$ & eff. mass $m_r^2$ & $\Delta m_r^2$  & $p_{B,\text{prop}}^2$ & $\Delta p_{B,\text{prop}}^2$ \\ 
\hline 
0.1 & 1.0 & 1.0 & 0.02  & -4.00 & 0.02\\ 
\hline 
1 & 0.978 & 0.97 & 0.02 & -3.90 & 0.02\\ 
\hline 
2 & 0.913 & 0.90  & 0.02 & -3.63 & 0.02\\ 
\hline 
2.5 & 0.865 & 0.84 & 0.02 & -3.41 & 0.02\\ 
\hline 
\bf{3} & \bf{0.807} & \bf{0.75} & \bf{0.02} & \bf{-3.09}  & \bf{0.02}\\ 
\hline 
3.25 & 0.775 & 0.69 & 0.02 & -2.81 & 0.02\\ 
\hline 
3.4 & 0.754 & 0.66 & 0.02 & -2.72 & 0.02\\ 
\hline 
3.5 & 0.739 & 0.59 & 0.02 & -2.41 & 0.02\\ 
\hline 
\end{tabular} 
\caption{Perturbative (one-loop) and nonperturbative masses, $m_p$ and $m_r$, respectively, as well as the position of the branch points $p_{B,\text{prop}}^2$ for different couplings $g$.
The line in bold corresponds to the solution shown in the figures.
All masses are given in multiples of the bare mass $m$.
}
\label{tab:masses}
\end{table}

We varied the coupling constant $g$ from the perturbative to the nonperturbative regime.
The positions of the obtained poles and branch points in the propagator are given in \tref{tab:masses}.
One can see that the branch points fulfill the relation $p_B=-4 m_r^2$ with the errors from determining the position of the pole.
The grid width is the main cause for the errors $\Delta m_r^2$ and $\Delta p_{B,\text{prop}}^2$.
We compare the effective masses $m_r$ with the one-loop perturbative mass $m_p$ in \tref{tab:masses}.
$m_p$ is determined by setting the denominator of the propagator to zero using the one-loop selfenergy from \eref{eq:prop_pert}.
The dependence of the effective mass on the coupling is also shown in \fref{fig:masses}.
We can see that the mass shift deviates significantly from the perturbative prediction for $g>3$ due to nonperturbative effects.
We fit the behavior of the effective mass by
\begin{equation}
f(g)=\sqrt{1-a\,g^b}.
\label{eq:Fit}
\end{equation}
For the parameters $a$ and $b$ we obtain:
\begin{equation}
\label{eq:Fit2}
a=0.0368, \qquad b=2.261.
\end{equation}
From the fit we can estimate the critical coupling, where the mass reaches zero, as $g^*\approx 4.30$.
Alternatively, we fix $b=2$ and perform a one-parameter fit which leads to
\begin{equation}
\label{eq:Fit1}
a=0.0502.
\end{equation}
From this fit, we obtain $g^*\approx 4.46$.

\begin{figure}[tb]
 \includegraphics[width=0.48\textwidth]{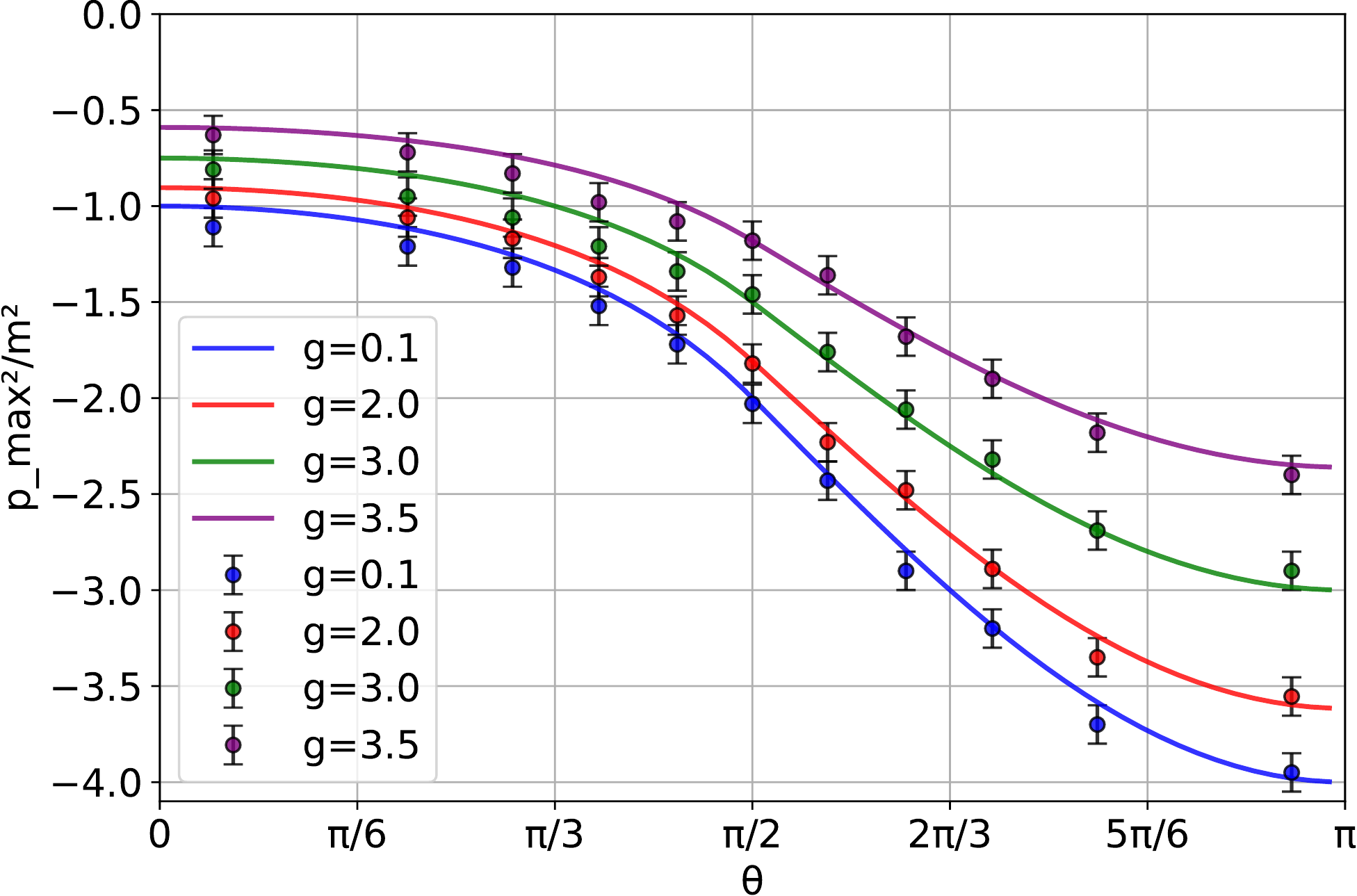}
 \caption{Branch points determined from calculations (dots) and as predicted by the Landau and contour deformation analyses (lines) for different values of the coupling $g$.}
 \label{fig:branchPointsVertex}
\end{figure}

\begin{figure*}[tb]
 \includegraphics[width=0.98\textwidth]{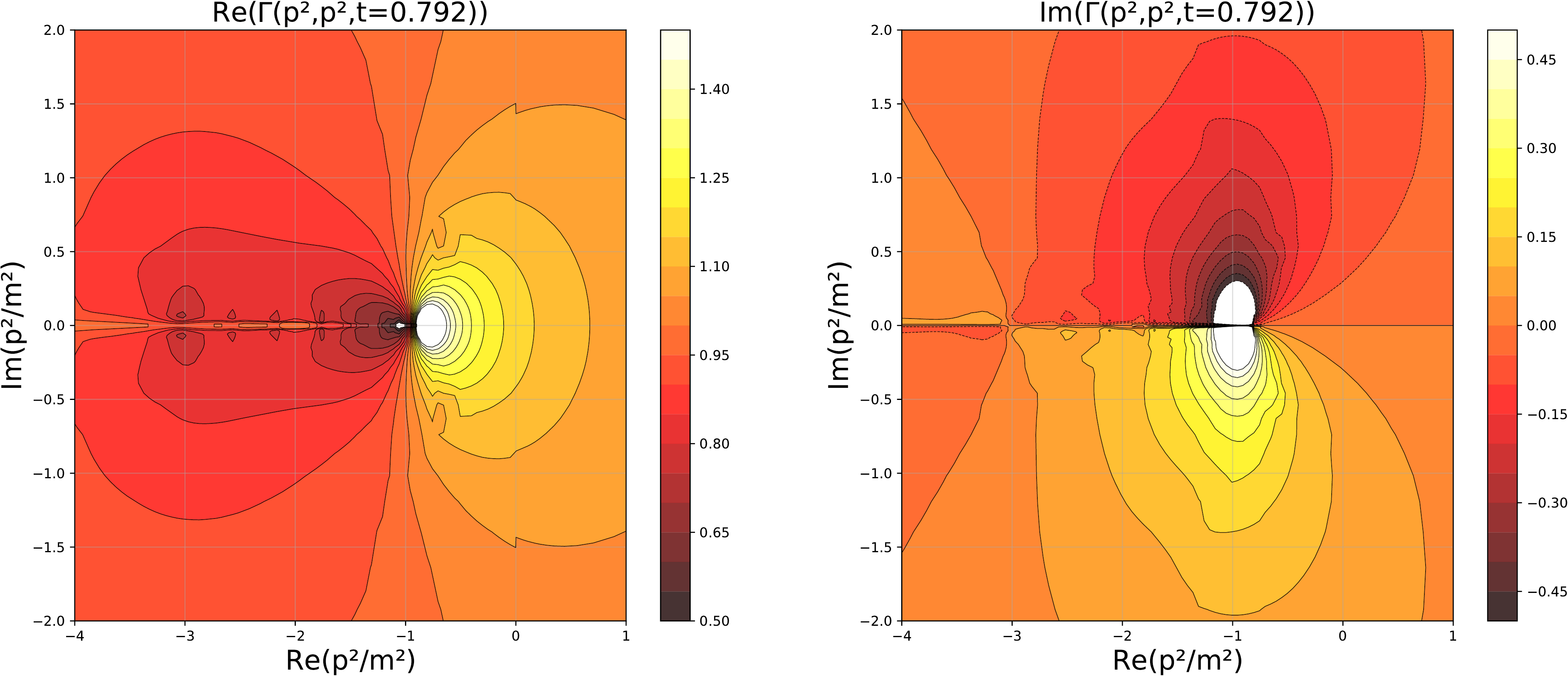}\\
 \includegraphics[width=0.98\textwidth]{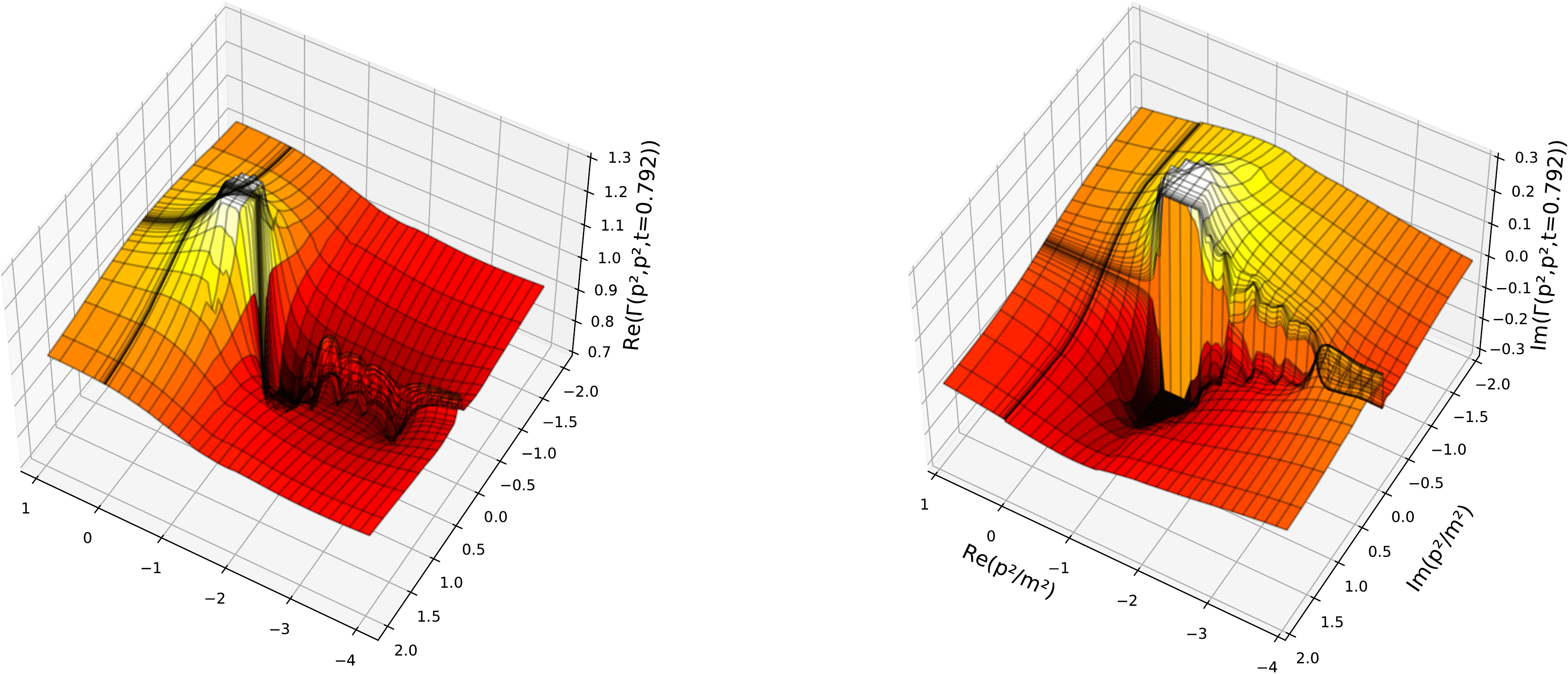}
 \caption{
 Contour (top) and 3d (bottom) plots of the vertex dressing function for $g$ and $\theta=0.66$.
 The fluctuations arise from integration close along the cut.
 }
 \label{fig:result_vert}
\end{figure*}

The calculation of the vertex is numerically more demanding due to the more complicated cut structure of the radial integrand.
For the propagator, it was possible to integrate quite close to the cuts and still get reasonable accuracy.
For the vertex, however, calculations are more sensitive to the neighborhood of cuts.
We tested how close one can calculate reliably and extrapolated beyond that.
Typically, imaginary parts of the order of $0.05$ could be reached.
The extracted branch point positions are shown in \fref{fig:branchPointsVertex} for different couplings.
The error is larger than for the propagator, owing to the fact that we cannot get as close to the real axis as for the propagator.
In \fref{fig:result_vert} we show an exemplary result.
Close to the cut, fluctuations are clearly visible.
However, the start of the cut can be seen by the increase, which we checked to be not caused by a pole.
It should also be noted that, within the employed kinematic approximation, the vertex dressing arguments can be chosen such that the numerically unstable region does not influence the region away from the cut and the error does thus not propagate.
Hence, this calculation is sufficient for illustrational purposes.
In a future calculation this will be overcome.

\section{Summary and conclusions}
\label{sec:summary}

In this work we explored the contour deformation method as a tool for analyzing the analytic structure of one-loop integrals of two- and three-point functions.
An advantage of this method is that it can also be applied numerically providing access to the evaluation of these integrals for arbitrary complex momentum variables.
As a new aspect, we treated the deformation of the angle integration contour to explain the analytic structure of the total integral in detail.
Such deformations are necessary to route the cuts around the poles in the integrand.
For the triangle integral of three-point functions their analysis is crucial as one needs to determine if such deformations are possible or not.
In the latter case, a threshold surface emerges in the external momenta.
This happens when two required deformations are in conflict as any deformation of the angle integral affects all cuts.
We also explained that using $q^2$ as radial integration variable can lead to ambiguities and thus $r=\sqrt{q^2}$ should be used instead.

With the CDM we could reproduce the solution for the threshold surface as known from the Landau analysis for general kinematics.
We found a direct correspondence between singularities coming from contracted diagrams in the Landau analysis and their emergence from two instead of three propagators in the CDM.
We illustrated the numerical applicability of the method by solving the coupled propagator and vertex equations of motion of $\phi^3$ theory for simplified kinematics in a three-loop truncation of the 3PI effective action.

The analytical part of the analysis was first performed based on the perturbative form of the propagators, because it elucidates the general mechanism which can be transferred to the nonperturbative setting as well.
We explained the generalization to the nonperturbative setup, for which shifted poles and cuts in the dressings need to be taken into account, in Sec.~\ref{sec:nonperturbative}.
This was explicitly illustrated numerically for the scalar propagator in Sec.~\ref{sec:results}.

While we explained the general mechanism for the emergence of thresholds, we also saw that the calculational complexity increases for three-point functions.
It will be challenging to implement this fully generally for theories like QCD.
Any simplifications will be helpful.
A good starting point might be the three-gluon vertex which can be described remarkably well by only one single kinematic variable \cite{Eichmann:2014xya,Blum:2014gna,Cyrol:2016tym,Huber:2020keu,Eichmann:2021zuv,Pinto-Gomez:2022brg}.
Automatization of contour deformations would also be helpful, for instance, using machine learning for identifying the cuts \cite{Windisch:2021mem} and finding appropriate contours \cite{Windisch:2019byg}.

\section*{Acknowledgments}

We thank Christian S. Fischer and Gernot Eichmann for a critical reading of the manuscript.
This work was supported by the DFG (German Research Foundation) grant No. FI 970/11-2 and by the BMBF under contract No. 05P21RGFP3.

\appendix

\section{Landau condition for the three-point function}
\label{sec:Landau_cond_3p}

For the triangle diagram, depicted in \fref{fig:LC_routing}, we get the following for the matrix $Q$:
\begin{align}
Q=\left( \begin{array}{cc c} k_1^2 & k_1 \cdot k_2 & k_1 \cdot k_3 \\ k_1 \cdot k_2  & k_2^2 & k_2 \cdot k_3 \\ k_1 \cdot k_3 & k_2 \cdot k_3 & k_3^2 \end{array} \right).
\end{align}
Assuming that the $\alpha_i$ do not vanish, we set
\begin{align}
k_1^2=k_2^2=k_3^2=-m^2.
\end{align}

With the choice of momenta given in \eref{eq:vert_kinematics} and using momentum conservation at each vertex of the triangle, we write the mixed terms as
\begin{align}
k_1 \cdot k_3 &=-m^2-\frac{p_a^2}{2},\\
k_2 \cdot k_3 &=-m^2-\frac{p_b^2}{2},\\
k_1 \cdot k_2 &=-m^2-\frac{p_c^2}{2}.
\end{align}

\begin{figure}[tb]
\centering
\includegraphics[width=0.5\textwidth]{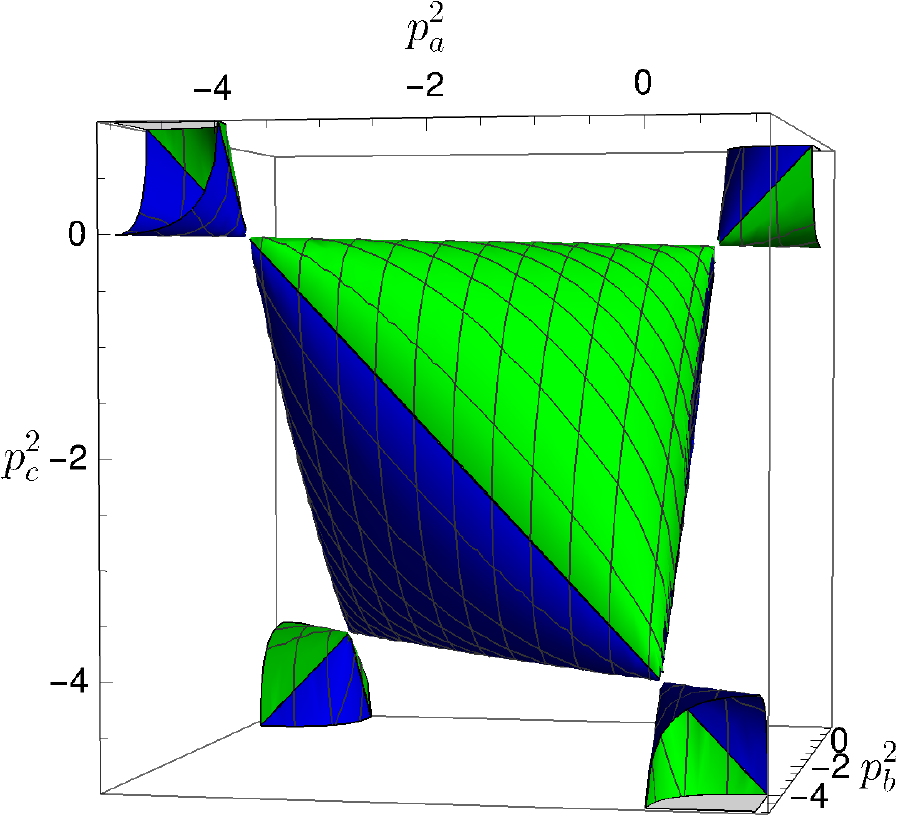}
\caption{Surface described by \eref{eq:LC_triangle_app} in $(p_a^2,p_b^2,p_c^2)$-space.
Blue corresponds to $c_+$ and green to $c_-$.
}
\label{fig:surface_triangle}
\end{figure}

We now set the determinant of $Q$ to zero:
\begin{align}\label{eq:LC_triangle_app}
p_a^2\, p_b^2\, p_c^2 = m^2\left(p_a^4 + p_b^4 +p_c^4 -2(p_a^2\,p_b^2+p_a^2\,p_c^2+p_b^2\,p_c^2)\right).
\end{align}
This equation describes a two-dimensional surface in three-dimensional $(a,b,c)$-space, see \fref{fig:surface_triangle}.
It can also be written as
\begin{align}
 p_a^2\,p_b^2\,p_c^2=m^2 \lambda(p_a^2,p_b^2,p_c^2).
\end{align}
where $\lambda(a,b,c)$ is the K\"all\`en function \cite{Kallen:1964epp},
\begin{align}\label{eq:Kaellen}
 \lambda(p_a^2,p_b^2,p_c^2):=p_a^4+p_b^4+p_c^4-2\,p_a^2\,p_b^2-2\,p_b^2\,p_c^2-2\,p_a^2\,p_c^2.
\end{align}
It is related to the angle $\theta$ via
\begin{align}
 \sin^2\theta=\frac{-\lambda(p_a^2,p_b^2,p_c^2)}{4p_a^2p_b^2}.
\end{align}

\begin{figure}[tb]
 \includegraphics[width=0.48\textwidth]{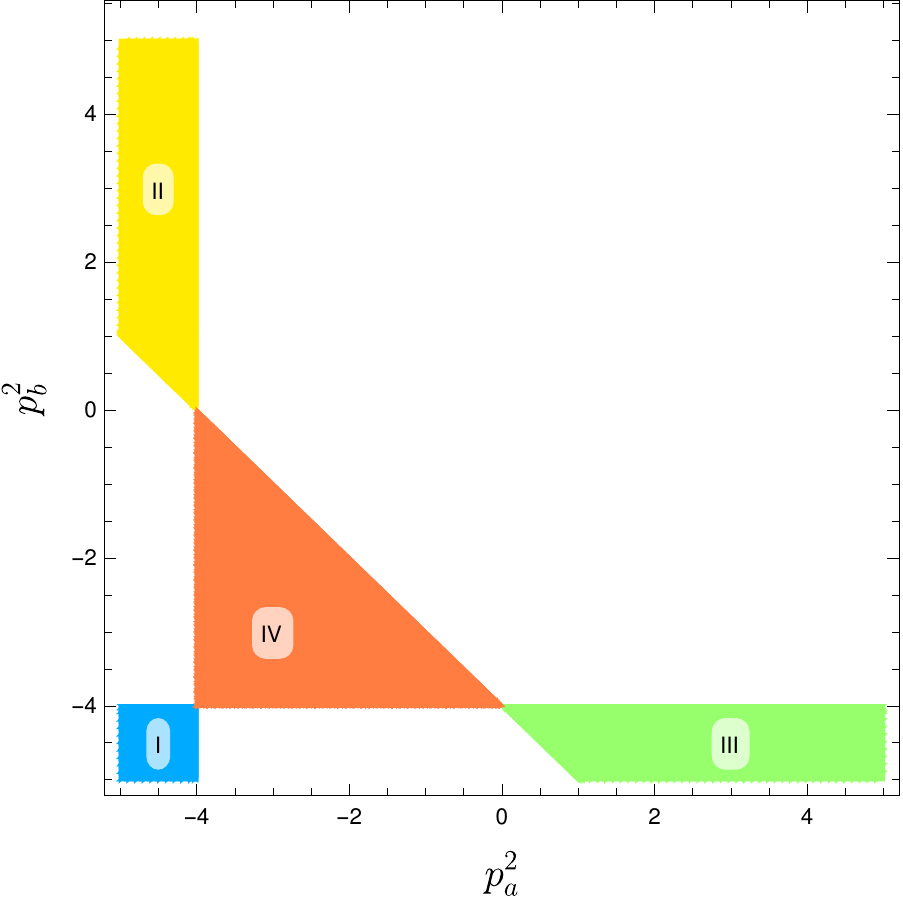}
 \caption{Thresholds for $p_a^2$ and $p_b^2$.
 The allowed values for $p_c^2$ follow from \eref{eq:c_sol}.}
 \label{fig:triangle_regions} 
\end{figure}

We now have to find solutions of \eref{eq:LC_triangle_app} for which $0\leq\alpha_i\leq1$ and $\alpha_1+\alpha_2+\alpha_3=1$. 
The surface described by \eref{eq:LC_triangle_app} consists of five disjunct regions, see \fref{fig:surface_triangle}.
We can distinguish them as follows:
\begin{itemize}
\item Region 1 is in the first octant for which $p_a^2>0$, $p_b^2>0$ and $p_c^2>0$.
\item For three regions, two momentum squares are smaller than $-4m^2$ and one is positive, e.g., $p_a^2<-4m^2$, $p_b^2<-4m^2$ and $p_c^2>0$.
\item For the fifth region (the central one), for all momentum squares $-4m^2<p_a^2,p_b^2,p_c^2<0$ holds. 
\end{itemize}

Solving \eref{eq:LC_triangle_app} for $p_c^2$ yields
\begin{align}\label{eq:c_sol}
p_{c\pm}^2=&\frac{1}{2} \Big( 2m^2(p_a^2+p_b^2) + p_a^2 \,p_b^2 \nnnl
 &\pm \sqrt{p_a^2(4m^2+p_a^2)} \sqrt{p_b^2(4m^2+p_b^2)} \Big).
\end{align}
 
We plug the solutions for $p_c^2$ into $Q$ and solve $Q\cdot \vec{\alpha}=0$ together with $\alpha_1+\alpha_2+\alpha_3=1$.
This leads to the following expressions for $\vec\alpha$:
{\small 
\begin{align}
 \alpha_{1-}&=\frac{2m^2(4m^2+p_a^2)}{\sqrt{p_a^2\,p_b^2\,(4m^2+p_a^2)(4m^2+p_b^2)}-p_b^2(4m^2+p_a^2)},\\
 \alpha_{2-}&=\frac{2m^2\sqrt{(4m^2+p_a^2)(4m^2+p_b^2)}}{4m^2\sqrt{p_a^2\,p_b^2}+\sqrt{p_a^6\,p_b^2}-p_a^2\sqrt{(4m^2+p_a^2)(4m^2+p_b^2)}},\\
 \alpha_{1+}&=-\frac{2m^2(4m^2+p_a^2)}{\sqrt{p_a^2\,p_b^2\,(4m^2+p_a^2)(4m^2+p_b^2)}+p_b^2(4m^2+p_a^2)},\\
 \alpha_{2+}&=-\frac{2m^2\sqrt{(4m^2+p_a^2)(4m^2+p_b^2)}}{4m^2\sqrt{p_a^2\,p_b^2}+\sqrt{p_a^6\,p_b^2}+p_a^2\sqrt{(4m^2+p_a^2)(4m^2+p_b^2)}},\\
 \alpha_{3\pm}&=1-\alpha_{1\pm}-\alpha_{2\pm}.
\end{align}
}

We require $0\leq\alpha_i\leq1$.
Plotting these constraints, we find that the following four regions allow solutions:
\begin{itemize}
 \item Region I: $p_a^2<-4m^2$, $p_b^2<-4m^2$,
 \item Region II: $p_a^2<-4m^2$, $-p_a^2-4m^2<p_b^2$,
 \item Region III: $p_b^2<-4m^2$, $-p_b^2-4m^2<p_a^2$,
 \item Region IV: $-4m^2<p_a^2$, $-4m^2<p_b^2<-4m^2-p_a^2$.
\end{itemize}
They are illustrated in \fref{fig:triangle_regions}.
Regions I-III belong to $p_{c-}^2$.
As will be seen shortly, they are not relevant.
Region IV belongs to $p_{c+}^2$.

\begin{figure}
\includegraphics[width=.49\textwidth]{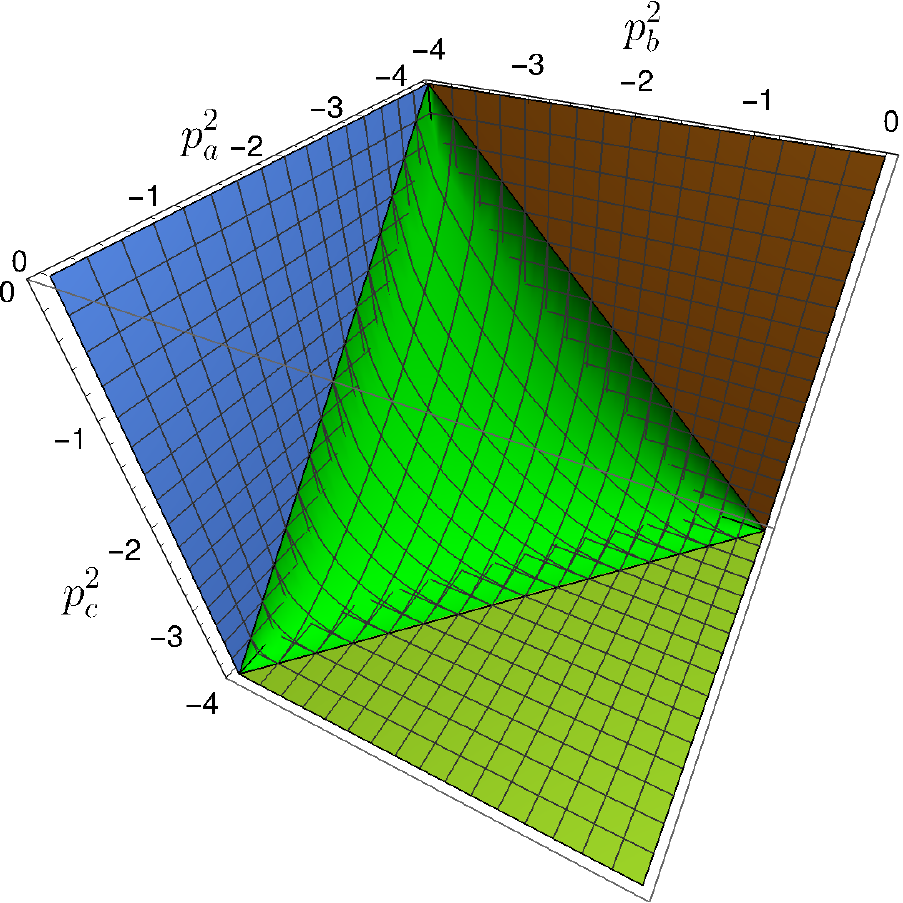}
\caption{Full solution for thresholds of the triangle diagram including contracted diagrams. }
\label{fig:threshold_3p}
\end{figure}

We now consider the contracted diagram, viz. the swordfish diagram, see \fref{fig:LC_routing}.
The resulting matrix $Q$ is the same matrix as for the propagator with $p$ one of the external momenta $p_a$, $p_b$ or $p_c$.
The thresholds are then $p_a^2,p_b^2,p_c^2=-4m^2$.
Thus regions I-III from above are excluded because they are beyond that.

The final threshold surface consists thus of the walls at $-4m^2$ plus the surface of region IV from the $p_{c+}^2$ solution of the triangle diagram:
\begin{align}
\left. \begin{array}{l}
p_c^2=\frac{2m^2(p_a^2+p_b^2) + p_a^2 p_b^2 + \sqrt{p_a^2(4m^2+p_a^2)} \sqrt{p_b^2(4m^2+p_b^2)}}{2m^2} \\  \qquad \text{for} \quad -4m^2\leq p_a^2,p_b^2 \leq 0, \quad \text{and} \quad p_a^2+p_b^2\leq-4m^2  \\ ~\\
p_a^2=p_b^2=p_c^2=-4m^2 \qquad \text{else} .
\label{eq:Vert_LC_full}
\end{array} \right. 
\end{align}
The last line is not a point but a shorthand notation representing the walls as illustrated in \fref{fig:threshold_3p}.
The detailed expression can be found in \eref{eq:Landau_sol_3p_loc}.

Finally, we extract the Landau conditions for the momentum configuration used in the numerical calculations.
To this end, we set $p_b^2=p_a^2=p^2$ and $p_c^2=2p^2(1+\cos\theta)$.
The Landau condition \eref{eq:LC_triangle_app} then becomes
\begin{align}
\label{eq:LC_Vert_Sym}
p^4 (1+\cos\theta)\left(p^2+2m^2(1-\cos \theta)\right)=0.
\end{align}
The solutions $p^2=0$ and $\theta=\pi$ are again on an unphysical sheet, while the remaining solution is given by
\begin{align}
p^2_{B,1}=-2m^2(1-\cos \theta).
\end{align}
The solution for $\vec{\alpha}$ is then
\begin{align}
\alpha_1=\frac{-\cos \theta}{1-\cos \theta}, \qquad \alpha_2=\alpha_3=\frac{1}{2(1-\cos \theta)}.
\end{align}
Taking into account the condition $0\leq\alpha_i\leq1$, we see that this singularity is physical for $\theta\geq \pi/2$.
For $\theta<\pi/2$, the contracted diagrams are relevant which we consider next.

We can distinguish two cases of contractions, see \fref{fig:LC_routing}: 
First, the momentum square at the top is $p^2$ or $2p^2(1+\cos\theta)$.
This case leads to the same matrix as for the propagators,  \eref{eq:LC_Q_prop}, and hence exhibits the same threshold $p^2=-4m^2$.
The second case has the threshold at $2p^2(1+\cos\theta)=-4m^2$ which can be reformulated as
\begin{align}
 p^2_{B,2}=-\frac{2m^2}{1+\cos\theta}.
\end{align}
This is valid for $0\leq\theta\leq\pi$.
Alternatively, one can also derive this via the matrix $Q$ again.

For the two thresholds $p^2_{B,1}$ and $p^2_{B,2}$ the following inequality holds for $\pi/2\leq\theta\leq \pi$:
\begin{align}
p_{B,2}^2=\frac{-2m^2}{1+\cos\theta} \leq -2m^2(1-\cos\theta)=p_{B,1}^2.
\end{align}
Consequently, the relevant singularity is given by $p_{B,1}^2$ for $\theta\geq\pi/2$.

To summarize, we find the following behavior for the thresholds of the restricted momentum configuration:
\begin{align}
p^2_{B}=\left \lbrace \begin{array}{c c} -2m^2(1-\cos\theta) &\quad \pi/2\leq \theta \leq \pi  \\ \frac{-2m^2}{1+\cos\theta} & \quad 0\leq \theta \leq \frac{\pi}{2} \end{array} \right. .
\end{align}

This can be rewritten to
\begin{align}
p^2_{B}=\left \lbrace \begin{array}{c c} -4m^2 \sin\left(\frac{\theta}{2}\right)^2 & \quad \frac{\pi}{2}\leq \theta \leq \pi  \\ \frac{-m^2}{\cos\left(\frac{\theta}{2}\right)^2} & \quad  0\leq \theta \leq \frac{\pi}{2} \end{array} \right. .
\end{align}

\section{Triangle integral with three different masses}

The analysis of Sec.~\ref{sec:genKin} can be generalized to three different masses as follows.
We call the three masses $m_1$, $m_2$ and $m_3$ belonging to the three propagators with momenta given in \eref{eq:vertex_denoms}.
Without loss of generality we take $m_1\geq m_2\geq m_3$.
We then put the first two cuts at the poles $\pm i \,m_3$.
In contrast to the equal mass case, the masses do not cancel and we obtain
\begin{subequations}
\begin{align}
\label{eq:3masses_denom1}
 0&=-m_3^2+p_a^2-2\,i\,m_3\sqrt{p_a^2}\cos\theta_1+m_1^2,\\
 0&=-m_3^2+p_b^2+2\,i\,m_3\sqrt{p_b^2}\cos\tilde\theta+m_2^2.
\end{align}
\end{subequations}
Bringing both expressions into the form $2\,i\,m_3=\ldots$, we can equate them and solve for $\theta_1$.
This leads to
\begin{align}
  \tan\theta_{1,s}=\frac{-\frac{B}{A}-\cos\theta}{\sin\theta},
\end{align}
where
\begin{subequations}
\begin{align}
 A&=(m_2^2-m_3^2+p_b^2)/\sqrt{p_a^2},\nnnl
 B&=(m_1^2-m_3^2+p_a^2)/\sqrt{p_b^2}.
\end{align}
\end{subequations}
From this we can calculate
\begin{align}
 \cos^2\theta_1=\frac{A^2\sin^2\theta}{A^2+B^2+2\,A\,B\cos\theta}.
\end{align}
Plugging it into \eref{eq:3masses_denom1}, we obtain
\begin{align}
 &m_1^2-m_3^2+p_a^2=\frac{i\,m_3 \sqrt{- \lambda(p_a^2,p_b^2,p_c^2)}}{\sqrt{p_b^2}}\nnnl
 &\quad\times \frac{A}{(A^2+B^2+2\,A\,B (p_c^2-p_a^2-p_b^2)/(2\sqrt{p_a^2}\sqrt{p_b^2}))^{\frac{1}{2}}}.
\end{align}
The K\"all\`en function $\lambda$ is given in \eref{eq:Kaellen}.
The thresholds corresponding to the contracted diagrams are inferred from $-(m_i+m_j)^2$ \cite{Cutkosky:1960sp,Dudal:2010wn,Windisch:2013dxa}, where $m_i$ and $m_j$ are the masses involved in its creation.
Following the kinematics depicted in \fref{fig:LC_routing}, the thresholds are
\begin{align}
 p_a^2=&-(m_1+m_3)^2,\\
 p_b^2=&-(m_2+m_3)^2,\\
 p_c^2=&-(m_1+m_2)^2.
\end{align}

\bibliographystyle{utphys_mod}
\bibliography{literature_anStruct_phi3}

\providecommand{\href}[2]{#2}\begingroup\raggedright\begin{thebibliography}{100}

\bibitem{Cloet:2013jya}
I.~C. Cloet and C.~D. Roberts, ``{Explanation and Prediction of Observables
  using Continuum Strong QCD}'',
  \href{http://dx.doi.org/10.1016/j.ppnp.2014.02.001}{{\em Prog. Part. Nucl.
  Phys.} {\bfseries 77} (2014) 1--69},
\href{http://arxiv.org/abs/1310.2651}{{\ttfamily arXiv:1310.2651 [nucl-th]}}.

\bibitem{Eichmann:2016yit}
G.~Eichmann, H.~Sanchis-Alepuz, R.~Williams, R.~Alkofer, and C.~S. Fischer,
  ``{Baryons as relativistic three-quark bound states}'',
  \href{http://dx.doi.org/10.1016/j.ppnp.2016.07.001}{{\em Prog. Part. Nucl.
  Phys.} {\bfseries 91} (2016) 1--100},
\href{http://arxiv.org/abs/1606.09602}{{\ttfamily arXiv:1606.09602 [hep-ph]}}.

\bibitem{Eichmann:2020oqt}
G.~Eichmann, C.~S. Fischer, W.~Heupel, N.~Santowsky, and P.~C. Wallbott,
  ``{Four-Quark States from Functional Methods}'',
  \href{http://dx.doi.org/10.1007/s00601-020-01571-3}{{\em Few Body Syst.}
  {\bfseries 61} no.~4, (2020) 38},
  \href{http://arxiv.org/abs/2008.10240}{{\ttfamily arXiv:2008.10240
  [hep-ph]}}.

\bibitem{Williams:2015cvx}
R.~Williams, C.~S. Fischer, and W.~Heupel, ``{Light mesons in QCD and
  unquenching effects from the 3PI effective action}'',
  \href{http://dx.doi.org/10.1103/PhysRevD.93.034026}{{\em Phys. Rev.}
  {\bfseries D93} no.~3, (2016) 034026},
\href{http://arxiv.org/abs/1512.00455}{{\ttfamily arXiv:1512.00455 [hep-ph]}}.

\bibitem{Cyrol:2016tym}
A.~K. Cyrol, L.~Fister, M.~Mitter, J.~M. Pawlowski, and N.~Strodthoff,
  ``{Landau gauge Yang-Mills correlation functions}'',
  \href{http://dx.doi.org/10.1103/PhysRevD.94.054005}{{\em Phys. Rev.}
  {\bfseries D94} no.~5, (2016) 054005},
\href{http://arxiv.org/abs/1605.01856}{{\ttfamily arXiv:1605.01856 [hep-ph]}}.

\bibitem{Cyrol:2017ewj}
A.~K. Cyrol, M.~Mitter, J.~M. Pawlowski, and N.~Strodthoff, ``{Nonperturbative
  quark, gluon, and meson correlators of unquenched QCD}'',
  \href{http://dx.doi.org/10.1103/PhysRevD.97.054006}{{\em Phys. Rev.}
  {\bfseries D97} no.~5, (2018) 054006},
\href{http://arxiv.org/abs/1706.06326}{{\ttfamily arXiv:1706.06326 [hep-ph]}}.

\bibitem{Huber:2018ned}
M.~Q. Huber, ``{Nonperturbative properties of Yang\textendash{}Mills
  theories}'', \href{http://dx.doi.org/10.1016/j.physrep.2020.04.004}{{\em
  Phys. Rept.} {\bfseries 879} (2020) 1--92},
  \href{http://arxiv.org/abs/1808.05227}{{\ttfamily arXiv:1808.05227
  [hep-ph]}}.

\bibitem{Huber:2020keu}
M.~Q. Huber, ``{Correlation functions of Landau gauge Yang-Mills theory}'',
  \href{http://dx.doi.org/10.1103/PhysRevD.101.114009}{{\em Phys. Rev. D}
  {\bfseries 101} no.~11, (2020) 114009},
  \href{http://arxiv.org/abs/2003.13703}{{\ttfamily arXiv:2003.13703
  [hep-ph]}}.

\bibitem{Gao:2021wun}
F.~Gao, J.~Papavassiliou, and J.~M. Pawlowski, ``{Fully coupled functional
  equations for the quark sector of QCD}'',
  \href{http://dx.doi.org/10.1103/PhysRevD.103.094013}{{\em Phys. Rev. D}
  {\bfseries 103} no.~9, (2021) 094013},
  \href{http://arxiv.org/abs/2102.13053}{{\ttfamily arXiv:2102.13053
  [hep-ph]}}.

\bibitem{Pawlowski:2022oyq}
J.~M. Pawlowski, C.~S. Schneider, and N.~Wink, ``{On Gauge Consistency In
  Gauge-Fixed Yang-Mills Theory}'',
  \href{http://arxiv.org/abs/2202.11123}{{\ttfamily arXiv:2202.11123
  [hep-th]}}.

\bibitem{Huber:2020ngt}
M.~Q. Huber, C.~S. Fischer, and H.~Sanchis-Alepuz, ``{Spectrum of scalar and
  pseudoscalar glueballs from functional methods}'',
  \href{http://dx.doi.org/10.1140/epjc/s10052-020-08649-6}{{\em Eur. Phys. J.
  C} {\bfseries 80} no.~11, (2020) 1077},
  \href{http://arxiv.org/abs/2004.00415}{{\ttfamily arXiv:2004.00415
  [hep-ph]}}.

\bibitem{Huber:2021yfy}
M.~Q. Huber, C.~S. Fischer, and H.~Sanchis-Alepuz, ``{Higher spin glueballs
  from functional methods}'',
  \href{http://dx.doi.org/10.1140/epjc/s10052-021-09864-5}{{\em Eur. Phys. J.
  C} {\bfseries 81} no.~12, (2021) 1083},
  \href{http://arxiv.org/abs/2110.09180}{{\ttfamily arXiv:2110.09180
  [hep-ph]}}.

\bibitem{Maris:1995ns}
P.~Maris, ``{Confinement and complex singularities in QED in
  three-dimensions}'', \href{http://dx.doi.org/10.1103/PhysRevD.52.6087}{{\em
  Phys. Rev.} {\bfseries D52} (1995) 6087--6097},
\href{http://arxiv.org/abs/hep-ph/9508323}{{\ttfamily arXiv:hep-ph/9508323
  [hep-ph]}}.

\bibitem{Alkofer:2003jj}
R.~Alkofer, W.~Detmold, C.~S. Fischer, and P.~Maris, ``{Analytic properties of
  the Landau gauge gluon and quark propagators}'',
  \href{http://dx.doi.org/10.1103/PhysRevD.70.014014}{{\em Phys. Rev.}
  {\bfseries D70} (2004) 014014},
\href{http://arxiv.org/abs/hep-ph/0309077}{{\ttfamily arXiv:hep-ph/0309077}}.

\bibitem{Eichmann:2007nn}
G.~Eichmann, A.~Krassnigg, M.~Schwinzerl, and R.~Alkofer, ``{A Covariant view
  on the nucleons' quark core}'',
  \href{http://dx.doi.org/10.1016/j.aop.2008.02.007}{{\em Annals Phys.}
  {\bfseries 323} (2008) 2505--2553},
  \href{http://arxiv.org/abs/0712.2666}{{\ttfamily arXiv:0712.2666 [hep-ph]}}.

\bibitem{Windisch:2012zd}
A.~Windisch, R.~Alkofer, G.~Haase, and M.~Liebmann, ``{Examining the Analytic
  Structure of Green's Functions: Massive Parallel Complex Integration using
  GPUs}'', \href{http://dx.doi.org/10.1016/j.cpc.2012.09.003}{{\em Comput.
  Phys. Commun.} {\bfseries 184} (2013) 109--116},
\href{http://arxiv.org/abs/1205.0752}{{\ttfamily arXiv:1205.0752 [hep-ph]}}.

\bibitem{Windisch:2012sz}
A.~Windisch, M.~Q. Huber, and R.~Alkofer, ``{On the analytic structure of
  scalar glueball operators at the Born level}'',
  \href{http://dx.doi.org/10.1103/PhysRevD.87.065005}{{\em Phys. Rev.}
  {\bfseries D87} no.~6, (2013) 065005},
\href{http://arxiv.org/abs/1212.2175}{{\ttfamily arXiv:1212.2175 [hep-ph]}}.

\bibitem{Strauss:2012dg}
S.~Strauss, C.~S. Fischer, and C.~Kellermann, ``{Analytic structure of the
  Landau gauge gluon propagator}'',
  \href{http://dx.doi.org/10.1103/PhysRevLett.109.252001}{{\em Phys.Rev.Lett.}
  {\bfseries 109} (2012) 252001},
\href{http://arxiv.org/abs/1208.6239}{{\ttfamily arXiv:1208.6239 [hep-ph]}}.

\bibitem{Windisch:2013dxa}
A.~Windisch, M.~Q. Huber, and R.~Alkofer, ``{How to determine the branch points
  of correlation functions in Euclidean space}'',
  \href{http://dx.doi.org/10.5506/APhysPolBSupp.6.887}{{\em Acta Phys. Polon.
  Supp.} {\bfseries 6} no.~3, (2013) 887--892},
\href{http://arxiv.org/abs/1304.3642}{{\ttfamily arXiv:1304.3642 [hep-ph]}}.

\bibitem{Eichmann:2019dts}
G.~Eichmann, P.~Duarte, M.~Peña, and A.~Stadler, ``{Scattering amplitudes and
  contour deformations}'',
  \href{http://dx.doi.org/10.1103/PhysRevD.100.094001}{{\em Phys. Rev. D}
  {\bfseries 100} no.~9, (2019) 094001},
  \href{http://arxiv.org/abs/1907.05402}{{\ttfamily arXiv:1907.05402
  [hep-ph]}}.

\bibitem{Fischer:2020xnb}
C.~S. Fischer and M.~Q. Huber, ``{Landau gauge Yang-Mills propagators in the
  complex momentum plane}'',
  \href{http://dx.doi.org/10.1103/PhysRevD.102.094005}{{\em Phys. Rev. D}
  {\bfseries 102} no.~9, (2020) 094005},
  \href{http://arxiv.org/abs/2007.11505}{{\ttfamily arXiv:2007.11505
  [hep-ph]}}.

\bibitem{Fischer:2008sp}
C.~S. Fischer, D.~Nickel, and R.~Williams, ``{On Gribov's supercriticality
  picture of quark confinement}'',
  \href{http://dx.doi.org/10.1140/epjc/s10052-008-0821-1}{{\em Eur. Phys. J.}
  {\bfseries C60} (2009) 47--61},
\href{http://arxiv.org/abs/0807.3486}{{\ttfamily arXiv:0807.3486 [hep-ph]}}.

\bibitem{GimenoSegovia:2008sx}
M.~Gimeno-Segovia and F.~J. Llanes-Estrada, ``{From Euclidean to Minkowski
  space with the Cauchy-Riemann equations}'',
  \href{http://dx.doi.org/10.1140/epjc/s10052-008-0676-5}{{\em Eur. Phys. J.}
  {\bfseries C56} (2008) 557--569},
\href{http://arxiv.org/abs/0805.4145}{{\ttfamily arXiv:0805.4145 [hep-th]}}.

\bibitem{Biernat:2018khd}
E.~P. Biernat, F.~Gross, M.~T. Pe\~na, A.~Stadler, and S.~Leit\~ao, ``Quark
  mass function from a one-gluon-exchange-type interaction in minkowski
  space'', \href{http://dx.doi.org/10.1103/PhysRevD.98.114033}{{\em Phys. Rev.
  D} {\bfseries 98} no.~11, (2018) 114033},
  \href{http://arxiv.org/abs/1811.01003}{{\ttfamily arXiv:1811.01003
  [hep-ph]}}.

\bibitem{Fischer:2005en}
C.~S. Fischer, P.~Watson, and W.~Cassing, ``{Probing unquenching effects in the
  gluon polarisation in light mesons}'',
  \href{http://dx.doi.org/10.1103/PhysRevD.72.094025}{{\em Phys. Rev.}
  {\bfseries D72} (2005) 094025},
\href{http://arxiv.org/abs/hep-ph/0509213}{{\ttfamily arXiv:hep-ph/0509213
  [hep-ph]}}.

\bibitem{Krassnigg:2009gd}
A.~Krassnigg, ``{Excited mesons in a Bethe-Salpeter approach}'', {\em PoS}
  {\bfseries CONFINEMENT8} (2008) 075,
\href{http://arxiv.org/abs/0812.3073}{{\ttfamily arXiv:0812.3073 [nucl-th]}}.

\bibitem{Nakanishi:1963zz}
N.~Nakanishi, ``Partial-wave bethe-salpeter equation'',
  \href{http://dx.doi.org/10.1103/PhysRev.130.1230}{{\em Phys. Rev.} {\bfseries
  130} (1963) 1230--1235}.

\bibitem{Nakanishi:1969ph}
N.~Nakanishi, ``A general survey of the theory of the bethe-salpeter
  equation'', \href{http://dx.doi.org/10.1143/PTPS.43.1}{{\em Prog. Theor.
  Phys. Suppl.} {\bfseries 43} (1969) 1--81}.

\bibitem{Nakanish:i1971gtf}
N.~Nakanishi, {\em Graph Theory and Feynman Integrals}.
\newblock Gordon and Breach, 1971.

\bibitem{Sauli:2001mb}
V.~Sauli and J.~Adam, ``{Solving the Schwinger-Dyson equation for a scalar
  propagator in Minkowski space}'',
  \href{http://dx.doi.org/10.1016/S0375-9474(01)00884-3}{{\em Nucl. Phys. A}
  {\bfseries 689} (2001) 467--470},
  \href{http://arxiv.org/abs/hep-ph/0110298}{{\ttfamily arXiv:hep-ph/0110298}}.

\bibitem{Sauli:2002tk}
V.~Sauli, ``{Minkowski solution of Dyson-Schwinger equations in momentum
  subtraction scheme}'',
  \href{http://dx.doi.org/10.1088/1126-6708/2003/02/001}{{\em JHEP} {\bfseries
  02} (2003) 001}, \href{http://arxiv.org/abs/hep-ph/0209046}{{\ttfamily
  arXiv:hep-ph/0209046}}.

\bibitem{Sauli:2006ba}
V.~Sauli, J.~Adam, Jr., and P.~Bicudo, ``Dynamical chiral symmetry breaking
  with integral minkowski representations'',
  \href{http://dx.doi.org/10.1103/PhysRevD.75.087701}{{\em Phys. Rev. D}
  {\bfseries 75} (2007) 087701},
  \href{http://arxiv.org/abs/hep-ph/0607196}{{\ttfamily arXiv:hep-ph/0607196}}.

\bibitem{Jia:2017niz}
S.~Jia and M.~R. Pennington, ``{Exact Solutions to the Fermion Propagator
  Schwinger-Dyson Equation in Minkowski space with on-shell Renormalization for
  Quenched QED}'', \href{http://dx.doi.org/10.1103/PhysRevD.96.036021}{{\em
  Phys. Rev. D} {\bfseries 96} no.~3, (2017) 036021},
  \href{http://arxiv.org/abs/1705.04523}{{\ttfamily arXiv:1705.04523
  [nucl-th]}}.

\bibitem{Solis:2019fzm}
E.~L. Solis, C.~S.~R. Costa, V.~V. Luiz, and G.~Krein, ``Quark propagator in
  minkowski space'', \href{http://dx.doi.org/10.1007/s00601-019-1517-9}{{\em
  Few Body Syst.} {\bfseries 60} no.~3, (2019) 49},
  \href{http://arxiv.org/abs/1905.08710}{{\ttfamily arXiv:1905.08710
  [hep-ph]}}.

\bibitem{Frederico:2019noo}
T.~Frederico, D.~C. Duarte, W.~de~Paula, E.~Ydrefors, S.~Jia, and P.~Maris,
  ``{Towards Minkowski space solutions of Dyson-Schwinger Equations through
  un-Wick rotation}'', \href{http://arxiv.org/abs/1905.00703}{{\ttfamily
  arXiv:1905.00703 [hep-ph]}}.

\bibitem{Horak:2020eng}
J.~Horak, J.~M. Pawlowski, and N.~Wink, ``{Spectral functions in the
  $\phi^4$-theory from the spectral DSE}'',
  \href{http://dx.doi.org/10.1103/PhysRevD.102.125016}{{\em Phys. Rev. D}
  {\bfseries 102} (2020) 125016},
  \href{http://arxiv.org/abs/2006.09778}{{\ttfamily arXiv:2006.09778
  [hep-th]}}.

\bibitem{Mezrag:2020iuo}
C.~Mezrag and G.~Salm\`e, ``Fermion and photon gap-equations in minkowski space
  within the nakanishi integral representation method'',
  \href{http://dx.doi.org/10.1140/epjc/s10052-020-08806-x}{{\em Eur. Phys. J.
  C} {\bfseries 81} no.~1, (2021) 34},
  \href{http://arxiv.org/abs/2006.15947}{{\ttfamily arXiv:2006.15947
  [hep-ph]}}.

\bibitem{Horak:2021pfr}
J.~Horak, J.~Papavassiliou, J.~M. Pawlowski, and N.~Wink, ``{Ghost spectral
  function from the spectral Dyson-Schwinger equation}'',
  \href{http://dx.doi.org/10.1103/PhysRevD.104.074017}{{\em Phys. Rev. D}
  {\bfseries 104} (2021) }, \href{http://arxiv.org/abs/2103.16175}{{\ttfamily
  arXiv:2103.16175 [hep-th]}}.

\bibitem{Horak:2022myj}
J.~Horak, J.~M. Pawlowski, and N.~Wink, ``{On the complex structure of
  Yang-Mills theory}'', \href{http://arxiv.org/abs/2202.09333}{{\ttfamily
  arXiv:2202.09333 [hep-th]}}.

\bibitem{Duarte:2022yur}
D.~C. Duarte, T.~Frederico, W.~de~Paula, and E.~Ydrefors, ``{Dynamical mass
  generation in Minkowski space at QCD scale}'',
  \href{http://dx.doi.org/10.1103/PhysRevD.105.114055}{{\em Phys. Rev. D}
  {\bfseries 105} no.~11, (2022) 114055},
  \href{http://arxiv.org/abs/2204.08091}{{\ttfamily arXiv:2204.08091
  [hep-ph]}}.

\bibitem{Horak:2022aza}
J.~Horak, J.~M. Pawlowski, and N.~Wink, ``{On the quark spectral function in
  QCD}'', \href{http://arxiv.org/abs/2210.07597}{{\ttfamily arXiv:2210.07597
  [hep-ph]}}.

\bibitem{Cucchieri:2011ig}
A.~Cucchieri, D.~Dudal, T.~Mendes, and N.~Vandersickel, ``{Modeling the Gluon
  Propagator in Landau Gauge: Lattice Estimates of Pole Masses and
  Dimension-Two Condensates}'',
  \href{http://dx.doi.org/10.1103/PhysRevD.85.094513}{{\em Phys. Rev. D}
  {\bfseries 85} (2012) 094513},
  \href{http://arxiv.org/abs/1111.2327}{{\ttfamily arXiv:1111.2327 [hep-lat]}}.

\bibitem{Dudal:2013yva}
D.~Dudal, O.~Oliveira, and P.~J. Silva, ``{Källén-Lehmann spectroscopy for
  (un)physical degrees of freedom}'',
  \href{http://dx.doi.org/10.1103/PhysRevD.89.014010}{{\em Phys. Rev.}
  {\bfseries D89} no.~1, (2014) 014010},
\href{http://arxiv.org/abs/1310.4069}{{\ttfamily arXiv:1310.4069 [hep-lat]}}.

\bibitem{Cucchieri:2016jwg}
A.~Cucchieri, D.~Dudal, T.~Mendes, and N.~Vandersickel, ``{Modeling the
  Landau-gauge ghost propagator in 2, 3, and 4 spacetime dimensions}'',
  \href{http://dx.doi.org/10.1103/PhysRevD.93.094513}{{\em Phys. Rev. D}
  {\bfseries 93} no.~9, (2016) 094513},
  \href{http://arxiv.org/abs/1602.01646}{{\ttfamily arXiv:1602.01646
  [hep-lat]}}.

\bibitem{Siringo:2016jrc}
F.~Siringo, ``{Analytic structure of QCD propagators in Minkowski space}'',
  \href{http://dx.doi.org/10.1103/PhysRevD.94.114036}{{\em Phys. Rev. D}
  {\bfseries 94} no.~11, (2016) 114036},
  \href{http://arxiv.org/abs/1605.07357}{{\ttfamily arXiv:1605.07357
  [hep-ph]}}.

\bibitem{Cyrol:2018xeq}
A.~K. Cyrol, J.~M. Pawlowski, A.~Rothkopf, and N.~Wink, ``{Reconstructing the
  gluon}'', \href{http://dx.doi.org/10.21468/SciPostPhys.5.6.065}{{\em SciPost
  Phys.} {\bfseries 5} no.~6, (2018) 065},
  \href{http://arxiv.org/abs/1804.00945}{{\ttfamily arXiv:1804.00945
  [hep-ph]}}.

\bibitem{Tripolt:2018xeo}
R.-A. Tripolt, P.~Gubler, M.~Ulybyshev, and L.~Von~Smekal, ``{Numerical
  analytic continuation of Euclidean data}'',
  \href{http://dx.doi.org/10.1016/j.cpc.2018.11.012}{{\em Comput. Phys.
  Commun.} {\bfseries 237} (2019) 129--142},
\href{http://arxiv.org/abs/1801.10348}{{\ttfamily arXiv:1801.10348 [hep-ph]}}.

\bibitem{Dudal:2019gvn}
D.~Dudal, O.~Oliveira, M.~Roelfs, and P.~Silva, ``{Spectral representation of
  lattice gluon and ghost propagators at zero temperature}'',
  \href{http://dx.doi.org/10.1016/j.nuclphysb.2019.114912}{{\em Nucl. Phys. B}
  {\bfseries 952} (2020) 114912},
  \href{http://arxiv.org/abs/1901.05348}{{\ttfamily arXiv:1901.05348
  [hep-lat]}}.

\bibitem{Binosi:2019ecz}
D.~Binosi and R.-A. Tripolt, ``{Spectral functions of confined particles}'',
  \href{http://dx.doi.org/10.1016/j.physletb.2019.135171}{{\em Phys. Lett. B}
  {\bfseries 801} (2020) 135171},
  \href{http://arxiv.org/abs/1904.08172}{{\ttfamily arXiv:1904.08172
  [hep-ph]}}.

\bibitem{Li:2019hyv}
S.~W. Li, P.~Lowdon, O.~Oliveira, and P.~J. Silva, ``{The generalised infrared
  structure of the gluon propagator}'',
  \href{http://dx.doi.org/10.1016/j.physletb.2020.135329}{{\em Phys. Lett. B}
  {\bfseries 803} (2020) 135329},
  \href{http://arxiv.org/abs/1907.10073}{{\ttfamily arXiv:1907.10073
  [hep-th]}}.

\bibitem{Falcao:2020vyr}
A.~F. Falc\~ao, O.~Oliveira, and P.~J. Silva, ``{Analytic structure of the
  lattice Landau gauge gluon and ghost propagators}'',
  \href{http://dx.doi.org/10.1103/PhysRevD.102.114518}{{\em Phys. Rev. D}
  {\bfseries 102} no.~11, (2020) 114518},
  \href{http://arxiv.org/abs/2008.02614}{{\ttfamily arXiv:2008.02614
  [hep-lat]}}.

\bibitem{Horak:2021syv}
J.~Horak, J.~M. Pawlowski, J.~Rodr\'\i{}guez-Quintero, J.~Turnwald, J.~M.
  Urban, N.~Wink, and S.~Zafeiropoulos, ``{Reconstructing QCD spectral
  functions with Gaussian processes}'',
  \href{http://dx.doi.org/10.1103/PhysRevD.105.036014}{{\em Phys. Rev. D}
  {\bfseries 105} no.~3, (2022) 036014},
  \href{http://arxiv.org/abs/2107.13464}{{\ttfamily arXiv:2107.13464
  [hep-ph]}}.

\bibitem{Lechien:2022ieg}
T.~Lechien and D.~Dudal, ``{Neural network approach to reconstructing spectral
  functions and complex poles of confined particles}'',
  \href{http://arxiv.org/abs/2203.03293}{{\ttfamily arXiv:2203.03293
  [hep-lat]}}.

\bibitem{Falcao:2022gxt}
A.~F. Falc\~ao and O.~Oliveira, ``{The analytic structure of the Landau gauge
  quark propagator from Pad\'e analysis}'',
  \href{http://arxiv.org/abs/2209.14815}{{\ttfamily arXiv:2209.14815
  [hep-lat]}}.

\bibitem{Boito:2022rad}
D.~Boito, A.~Cucchieri, C.~Y. London, and T.~Mendes, ``{Probing the
  singularities of the Landau-gauge gluon and ghost propagators with rational
  approximants}'', \href{http://arxiv.org/abs/2210.10490}{{\ttfamily
  arXiv:2210.10490 [hep-lat]}}.

\bibitem{Horak:2023xfb}
J.~Horak, J.~M. Pawlowski, J.~Turnwald, J.~M. Urban, N.~Wink, and
  S.~Zafeiropoulos, ``{Non-perturbative strong coupling at timelike momenta}'',
  \href{http://arxiv.org/abs/2301.07785}{{\ttfamily arXiv:2301.07785
  [hep-ph]}}.

\bibitem{Lowdon:2017uqe}
P.~Lowdon, ``{Nonperturbative structure of the photon and gluon propagators}'',
  \href{http://dx.doi.org/10.1103/PhysRevD.96.065013}{{\em Phys. Rev. D}
  {\bfseries 96} no.~6, (2017) 065013},
  \href{http://arxiv.org/abs/1702.02954}{{\ttfamily arXiv:1702.02954
  [hep-th]}}.

\bibitem{Lowdon:2018mbn}
P.~Lowdon, ``{Dyson--Schwinger equation constraints on the gluon propagator in
  BRST quantised QCD}'',
  \href{http://dx.doi.org/10.1016/j.physletb.2018.10.023}{{\em Phys. Lett. B}
  {\bfseries 786} (2018) 399--402},
  \href{http://arxiv.org/abs/1801.09337}{{\ttfamily arXiv:1801.09337
  [hep-th]}}.

\bibitem{Hayashi:2018giz}
Y.~Hayashi and K.-I. Kondo, ``{Complex poles and spectral function of
  Yang-Mills theory}'',
  \href{http://dx.doi.org/10.1103/PhysRevD.99.074001}{{\em Phys. Rev. D}
  {\bfseries 99} no.~7, (2019) 074001},
  \href{http://arxiv.org/abs/1812.03116}{{\ttfamily arXiv:1812.03116
  [hep-th]}}.

\bibitem{Kondo:2019rpa}
K.-I. Kondo, M.~Watanabe, Y.~Hayashi, R.~Matsudo, and Y.~Suda, ``{Reflection
  positivity and complex analysis of the Yang-Mills theory from a viewpoint of
  gluon confinement}'',
  \href{http://dx.doi.org/10.1140/epjc/s10052-020-7632-4}{{\em Eur. Phys. J. C}
  {\bfseries 80} no.~2, (2020) 84},
  \href{http://arxiv.org/abs/1902.08894}{{\ttfamily arXiv:1902.08894
  [hep-th]}}.

\bibitem{Hayashi:2020few}
Y.~Hayashi and K.-I. Kondo, ``{Complex poles and spectral functions of Landau
  gauge QCD and QCD-like theories}'',
  \href{http://dx.doi.org/10.1103/PhysRevD.101.074044}{{\em Phys. Rev. D}
  {\bfseries 101} no.~7, (2020) 074044},
  \href{http://arxiv.org/abs/2001.05987}{{\ttfamily arXiv:2001.05987
  [hep-th]}}.

\bibitem{Hayashi:2021nnj}
Y.~Hayashi and K.-I. Kondo, ``{Reconstructing confined particles with complex
  singularities}'', \href{http://dx.doi.org/10.1103/PhysRevD.103.L111504}{{\em
  Phys. Rev. D} {\bfseries 103} no.~11, (2021) L111504},
  \href{http://arxiv.org/abs/2103.14322}{{\ttfamily arXiv:2103.14322
  [hep-th]}}.

\bibitem{Hayashi:2021jju}
Y.~Hayashi and K.-I. Kondo, ``{Reconstructing propagators of confined particles
  in the presence of complex singularities}'',
  \href{http://dx.doi.org/10.1103/PhysRevD.104.074024}{{\em Phys. Rev. D}
  {\bfseries 104} no.~7, (2021) 074024},
  \href{http://arxiv.org/abs/2105.07487}{{\ttfamily arXiv:2105.07487
  [hep-th]}}.

\bibitem{Carbonell:2010zw}
J.~Carbonell and V.~A. Karmanov, ``Solving bethe-salpeter equation for two
  fermions in minkowski space'',
  \href{http://dx.doi.org/10.1140/epja/i2010-11055-4}{{\em Eur. Phys. J. A}
  {\bfseries 46} (2010) 387--397},
  \href{http://arxiv.org/abs/1010.4640}{{\ttfamily arXiv:1010.4640 [hep-ph]}}.

\bibitem{Frederico:2013vga}
T.~Frederico, G.~Salme', and M.~Viviani, ``Quantitative studies of the
  homogeneous bethe-salpeter equation in minkowski space'',
  \href{http://dx.doi.org/10.1103/PhysRevD.89.016010}{{\em Phys. Rev. D}
  {\bfseries 89} (2014) 016010},
  \href{http://arxiv.org/abs/1312.0521}{{\ttfamily arXiv:1312.0521 [hep-ph]}}.

\bibitem{dePaula:2017ikc}
W.~de~Paula, T.~Frederico, G.~Salm\`e, M.~Viviani, and R.~Pimentel, ``Fermionic
  bound states in minkowski-space: Light-cone singularities and structure'',
  \href{http://dx.doi.org/10.1140/epjc/s10052-017-5351-2}{{\em Eur. Phys. J. C}
  {\bfseries 77} no.~11, (2017) 764},
  \href{http://arxiv.org/abs/1707.06946}{{\ttfamily arXiv:1707.06946
  [hep-ph]}}.

\bibitem{Ydrefors:2019jvu}
E.~Ydrefors, J.~H. Alvarenga~Nogueira, V.~A. Karmanov, and T.~Frederico,
  ``Solving the three-body bound-state bethe-salpeter equation in minkowski
  space'', \href{http://dx.doi.org/10.1016/j.physletb.2019.02.046}{{\em Phys.
  Lett. B} {\bfseries 791} (2019) 276--280},
  \href{http://arxiv.org/abs/1903.01741}{{\ttfamily arXiv:1903.01741
  [hep-ph]}}.

\bibitem{Sauli:2001we}
V.~Sauli and J.~Adam, Jr., ``{Study of relativistic bound states for scalar
  theories in the Bethe-Salpeter and Dyson-Schwinger formalism}'',
  \href{http://dx.doi.org/10.1103/PhysRevD.67.085007}{{\em Phys. Rev. D}
  {\bfseries 67} (2003) 085007},
  \href{http://arxiv.org/abs/hep-ph/0111433}{{\ttfamily arXiv:hep-ph/0111433}}.

\bibitem{Kusaka:1997xd}
K.~Kusaka, K.~M. Simpson, and A.~G. Williams, ``{Solving the Bethe-Salpeter
  equation for bound states of scalar theories in Minkowski space}'',
  \href{http://dx.doi.org/10.1103/PhysRevD.56.5071}{{\em Phys. Rev. D}
  {\bfseries 56} (1997) 5071--5085},
  \href{http://arxiv.org/abs/hep-ph/9705298}{{\ttfamily arXiv:hep-ph/9705298}}.

\bibitem{Karmanov:2005nv}
V.~A. Karmanov and J.~Carbonell, ``{Solving Bethe-Salpeter equation in
  Minkowski space}'', \href{http://dx.doi.org/10.1140/epja/i2005-10193-0}{{\em
  Eur. Phys. J. A} {\bfseries 27} (2006) 1--9},
  \href{http://arxiv.org/abs/hep-th/0505261}{{\ttfamily arXiv:hep-th/0505261}}.

\bibitem{Karmanov:2005yg}
V.~A. Karmanov and J.~Carbonell, ``{Bethe-Salpeter equation in Minkowski space
  with cross-ladder kernel}'',
  \href{http://dx.doi.org/10.1016/j.nuclphysbps.2006.08.068}{{\em Nucl. Phys. B
  Proc. Suppl.} {\bfseries 161} (2006) 123--129},
  \href{http://arxiv.org/abs/nucl-th/0510051}{{\ttfamily
  arXiv:nucl-th/0510051}}.

\bibitem{Frederico:2015ufa}
T.~Frederico, G.~Salm\`e, and M.~Viviani, ``{Solving the inhomogeneous
  Bethe\textendash{}Salpeter equation in Minkowski space: the zero-energy
  limit}'', \href{http://dx.doi.org/10.1140/epjc/s10052-015-3616-1}{{\em Eur.
  Phys. J. C} {\bfseries 75} no.~8, (2015) 398},
  \href{http://arxiv.org/abs/1504.01624}{{\ttfamily arXiv:1504.01624
  [hep-ph]}}.

\bibitem{dePaula:2016oct}
W.~de~Paula, T.~Frederico, G.~Salm\`e, and M.~Viviani, ``{Advances in solving
  the two-fermion homogeneous Bethe-Salpeter equation in Minkowski space}'',
  \href{http://dx.doi.org/10.1103/PhysRevD.94.071901}{{\em Phys. Rev. D}
  {\bfseries 94} no.~7, (2016) 071901},
  \href{http://arxiv.org/abs/1609.00868}{{\ttfamily arXiv:1609.00868
  [hep-th]}}.

\bibitem{AlvarengaNogueira:2019zcs}
J.~H. Alvarenga~Nogueira, D.~Colasante, V.~Gherardi, T.~Frederico, E.~Pace, and
  G.~Salm\`e, ``{Solving the Bethe-Salpeter Equation in Minkowski Space for a
  Fermion-Scalar system}'',
  \href{http://dx.doi.org/10.1103/PhysRevD.100.016021}{{\em Phys. Rev. D}
  {\bfseries 100} no.~1, (2019) 016021},
  \href{http://arxiv.org/abs/1907.03079}{{\ttfamily arXiv:1907.03079
  [hep-ph]}}.

\bibitem{Gutierrez:2016ixt}
C.~Gutierrez, V.~Gigante, T.~Frederico, G.~Salm\`e, M.~Viviani, and L.~Tomio,
  ``{Bethe\textendash{}Salpeter bound-state structure in Minkowski space}'',
  \href{http://dx.doi.org/10.1016/j.physletb.2016.05.066}{{\em Phys. Lett. B}
  {\bfseries 759} (2016) 131--137},
  \href{http://arxiv.org/abs/1605.08837}{{\ttfamily arXiv:1605.08837
  [hep-ph]}}.

\bibitem{Moita:2022lfu}
R.~M. Moita, J.~P. B.~C. de~Melo, T.~Frederico, and W.~de~Paula, ``{Pion
  inspired by QCD: Nakanishi and light-front integral representations}'',
  \href{http://dx.doi.org/10.1103/PhysRevD.106.016016}{{\em Phys. Rev. D}
  {\bfseries 106} no.~1, (2022) 016016},
  \href{http://arxiv.org/abs/2208.03845}{{\ttfamily arXiv:2208.03845
  [hep-ph]}}.

\bibitem{Weil:2017knt}
E.~Weil, G.~Eichmann, C.~S. Fischer, and R.~Williams, ``{Electromagnetic decays
  of the neutral pion}'',
  \href{http://dx.doi.org/10.1103/PhysRevD.96.014021}{{\em Phys. Rev. D}
  {\bfseries 96} no.~1, (2017) 014021},
  \href{http://arxiv.org/abs/1704.06046}{{\ttfamily arXiv:1704.06046
  [hep-ph]}}.

\bibitem{Williams:2018adr}
R.~Williams, ``{Vector mesons as dynamical resonances in the Bethe--Salpeter
  framework}'', \href{http://dx.doi.org/10.1016/j.physletb.2019.134943}{{\em
  Phys. Lett. B} {\bfseries 798} (2019) 134943},
  \href{http://arxiv.org/abs/1804.11161}{{\ttfamily arXiv:1804.11161
  [hep-ph]}}.

\bibitem{Miramontes:2019mco}
A.~S. Miramontes and H.~Sanchis-Alepuz, ``{On the effect of resonances in the
  quark-photon vertex}'',
  \href{http://dx.doi.org/10.1140/epja/i2019-12847-6}{{\em Eur. Phys. J. A}
  {\bfseries 55} no.~10, (2019) 170},
  \href{http://arxiv.org/abs/1906.06227}{{\ttfamily arXiv:1906.06227
  [hep-ph]}}.

\bibitem{Miramontes:2021xgn}
A.~S. Miramontes, H.~Sanchis~Alepuz, and R.~Alkofer, ``Elucidating the effect
  of intermediate resonances in the quark interaction kernel on the timelike
  electromagnetic pion form factor'',
  \href{http://dx.doi.org/10.1103/PhysRevD.103.116006}{{\em Phys. Rev. D}
  {\bfseries 103} no.~11, (2021) 116006},
  \href{http://arxiv.org/abs/2102.12541}{{\ttfamily arXiv:2102.12541
  [hep-ph]}}.

\bibitem{Alkofer:2022hln}
R.~Alkofer, A.~S. Miramontes, and H.~Sanchis-Alepuz, ``Elucidating the
  \ensuremath{\rho}-meson\textquoteright{}s role as intermediate resonance in
  the time-like electromagnetic pion form factor'',
  \href{http://dx.doi.org/10.1051/epjconf/202226201020}{{\em EPJ Web Conf.}
  {\bfseries 262} (2022) 01020},
  \href{http://arxiv.org/abs/2202.05056}{{\ttfamily arXiv:2202.05056
  [hep-ph]}}.

\bibitem{Eichmann:2021vnj}
G.~Eichmann, E.~Ferreira, and A.~Stadler, ``{Going to the light front with
  contour deformations}'',
  \href{http://dx.doi.org/10.1103/PhysRevD.105.034009}{{\em Phys. Rev. D}
  {\bfseries 105} no.~3, (2022) 034009},
  \href{http://arxiv.org/abs/2112.04858}{{\ttfamily arXiv:2112.04858
  [hep-ph]}}.

\bibitem{Leitao:2017esb}
S.~Leit\~ao, Y.~Li, P.~Maris, M.~T. Pe\~na, A.~Stadler, J.~P. Vary, and E.~P.
  Biernat, ``Comparison of two minkowski-space approaches to heavy quarkonia'',
  \href{http://dx.doi.org/10.1140/epjc/s10052-017-5248-0}{{\em Eur. Phys. J. C}
  {\bfseries 77} no.~10, (2017) 696},
  \href{http://arxiv.org/abs/1705.06178}{{\ttfamily arXiv:1705.06178
  [hep-ph]}}.

\bibitem{Leitao:2017mlx}
S.~Leit\~ao, A.~Stadler, M.~T. Pe\~na, and E.~P. Biernat, ``{Covariant
  spectator theory of quark-antiquark bound states: Mass spectra and vertex
  functions of heavy and heavy-light mesons}'',
  \href{http://dx.doi.org/10.1103/PhysRevD.96.074007}{{\em Phys. Rev. D}
  {\bfseries 96} no.~7, (2017) 074007},
  \href{http://arxiv.org/abs/1707.09303}{{\ttfamily arXiv:1707.09303
  [hep-ph]}}.

\bibitem{Landau:1959fi}
L.~D. Landau, ``{On analytic properties of vertex parts in quantum field
  theory}'', \href{http://dx.doi.org/10.1016/B978-0-08-010586-4.50103-6}{{\em
  Sov. Phys. JETP} {\bfseries 10} no.~1, (1959) 45--50}.

\bibitem{Collins:2020euz}
J.~Collins, ``{A new and complete proof of the Landau condition for pinch
  singularities of Feynman graphs and other integrals}'',
  \href{http://arxiv.org/abs/2007.04085}{{\ttfamily arXiv:2007.04085
  [hep-ph]}}.

\bibitem{Huber:2023uzd}
M.~Q. Huber, W.~J. Kern, and R.~Alkofer, ``{How to determine the branch points
  of correlation functions in Euclidean space II: Three-point functions}'',
  \href{http://dx.doi.org/10.3390/sym15020414}{{\em Symmetry} {\bfseries 15(2)}
  (2, 2023) 414}, \href{http://arxiv.org/abs/2302.01350}{{\ttfamily
  arXiv:2302.01350 [hep-ph]}}.

\bibitem{Maas:2007uv}
A.~Maas, ``{Two- and three-point Green's functions in two-dimensional
  Landau-gauge Yang-Mills theory}'',
  \href{http://dx.doi.org/10.1103/PhysRevD.75.116004}{{\em Phys. Rev.}
  {\bfseries D75} (2007) 116004},
\href{http://arxiv.org/abs/0704.0722}{{\ttfamily arXiv:0704.0722 [hep-lat]}}.

\bibitem{Maas:2011se}
A.~Maas, ``{Describing gauge bosons at zero and finite temperature}'',
  \href{http://dx.doi.org/10.1016/j.physrep.2012.11.002}{{\em Phys.Rept.}
  {\bfseries 524} (2013) 203--300},
\href{http://arxiv.org/abs/1106.3942}{{\ttfamily arXiv:1106.3942 [hep-ph]}}.

\bibitem{Maas:2015nva}
A.~Maas, ``{More on the properties of the first Gribov region in Landau
  gauge}'', \href{http://dx.doi.org/10.1103/PhysRevD.93.054504}{{\em Phys.
  Rev.} {\bfseries D93} no.~5, (2016) 054504},
\href{http://arxiv.org/abs/1510.08407}{{\ttfamily arXiv:1510.08407 [hep-lat]}}.

\bibitem{Huber:2007kc}
M.~Q. Huber, R.~Alkofer, C.~S. Fischer, and K.~Schwenzer, ``{The infrared
  behavior of Landau gauge Yang-Mills theory in d=2, 3 and 4 dimensions}'',
  \href{http://dx.doi.org/10.1016/j.physletb.2007.10.073}{{\em Phys. Lett.}
  {\bfseries B659} (2008) 434--440},
\href{http://arxiv.org/abs/0705.3809}{{\ttfamily arXiv:0705.3809 [hep-ph]}}.

\bibitem{Dudal:2008xd}
D.~Dudal, S.~P. Sorella, N.~Vandersickel, and H.~Verschelde, ``{The effects of
  Gribov copies in 2D gauge theories}'',
  \href{http://dx.doi.org/10.1016/j.physletb.2009.08.055}{{\em Phys. Lett.}
  {\bfseries B680} (2009) 377--383},
\href{http://arxiv.org/abs/0808.3379}{{\ttfamily arXiv:0808.3379 [hep-th]}}.

\bibitem{Dudal:2008rm}
D.~Dudal, J.~A. Gracey, S.~P. Sorella, N.~Vandersickel, and H.~Verschelde,
  ``{The Landau gauge gluon and ghost propagator in the refined
  Gribov-Zwanziger framework in 3 dimensions}'',
  \href{http://dx.doi.org/10.1103/PhysRevD.78.125012}{{\em Phys. Rev.}
  {\bfseries D78} (2008) 125012},
\href{http://arxiv.org/abs/0808.0893}{{\ttfamily arXiv:0808.0893 [hep-th]}}.

\bibitem{Huber:2012zj}
M.~Q. Huber, A.~Maas, and L.~von Smekal, ``{Two- and three-point functions in
  two-dimensional Landau-gauge Yang-Mills theory: Continuum results}'',
  \href{http://dx.doi.org/10.1007/JHEP11(2012)035}{{\em JHEP} {\bfseries 1211}
  (2012) 035},
\href{http://arxiv.org/abs/1207.0222}{{\ttfamily arXiv:1207.0222 [hep-th]}}.

\bibitem{Huber:2016tvc}
M.~Q. Huber, ``{Correlation functions of three-dimensional Yang-Mills theory
  from Dyson-Schwinger equations}'',
  \href{http://dx.doi.org/10.1103/PhysRevD.93.085033}{{\em Phys. Rev.}
  {\bfseries D93} no.~8, (2016) 085033},
\href{http://arxiv.org/abs/1602.02038}{{\ttfamily arXiv:1602.02038 [hep-th]}}.

\bibitem{Corell:2018yil}
L.~Corell, A.~K. Cyrol, M.~Mitter, J.~M. Pawlowski, and N.~Strodthoff,
  ``{Correlation functions of three-dimensional Yang-Mills theory from the
  FRG}'', \href{http://dx.doi.org/10.21468/SciPostPhys.5.6.066}{{\em SciPost
  Phys.} {\bfseries 5} no.~6, (2018) 066},
  \href{http://arxiv.org/abs/1803.10092}{{\ttfamily arXiv:1803.10092
  [hep-ph]}}.

\bibitem{Collins:2008re}
J.~Collins, {\em Renormalization: An Introduction to Renormalization, the
  Renormalization Group and the Operator-product Expansion}.
\newblock Cambridge University Press, 2008.

\bibitem{Borinsky:2021jdb}
M.~Borinsky, J.~A. Gracey, M.~V. Kompaniets, and O.~Schnetz, ``{Five-loop
  renormalization of \ensuremath{\phi^3} theory with applications to the
  Lee-Yang edge singularity and percolation theory}'',
  \href{http://dx.doi.org/10.1103/PhysRevD.103.116024}{{\em Phys. Rev. D}
  {\bfseries 103} no.~11, (2021) 116024},
  \href{http://arxiv.org/abs/2103.16224}{{\ttfamily arXiv:2103.16224
  [hep-th]}}.

\bibitem{Wick:1954eu}
G.~C. Wick, ``{Properties of Bethe-Salpeter Wave Functions}'',
  \href{http://dx.doi.org/10.1103/PhysRev.96.1124}{{\em Phys. Rev.} {\bfseries
  96} (1954) 1124--1134}.

\bibitem{Berges:2004pu}
J.~Berges, ``{n-PI effective action techniques for gauge theories}'',
  \href{http://dx.doi.org/10.1103/PhysRevD.70.105010}{{\em Phys. Rev.}
  {\bfseries D70} (2004) 105010},
\href{http://arxiv.org/abs/hep-ph/0401172}{{\ttfamily arXiv:hep-ph/0401172}}.

\bibitem{Carrington:2010qq}
M.~Carrington and Y.~Guo, ``{Techniques for n-Particle Irreducible Effective
  Theories}'', \href{http://dx.doi.org/10.1103/PhysRevD.83.016006}{{\em
  Phys.Rev.} {\bfseries D83} (2011) 016006},
  \href{http://arxiv.org/abs/1010.2978}{{\ttfamily arXiv:1010.2978 [hep-ph]}}.

\bibitem{Alkofer:2008nt}
R.~Alkofer, M.~Q. Huber, and K.~Schwenzer, ``{Algorithmic derivation of
  Dyson-Schwinger Equations}'',
  \href{http://dx.doi.org/10.1016/j.cpc.2008.12.009}{{\em Comput. Phys.
  Commun.} {\bfseries 180} (2009) 965--976},
\href{http://arxiv.org/abs/0808.2939}{{\ttfamily arXiv:0808.2939 [hep-th]}}.

\bibitem{Huber:2011qr}
M.~Q. Huber and J.~Braun, ``{Algorithmic derivation of functional
  renormalization group equations and Dyson-Schwinger equations}'',
  \href{http://dx.doi.org/10.1016/j.cpc.2012.01.014}{{\em Comput.Phys.Commun.}
  {\bfseries 183} (2012) 1290--1320},
\href{http://arxiv.org/abs/1102.5307}{{\ttfamily arXiv:1102.5307 [hep-th]}}.

\bibitem{Huber:2019dkb}
M.~Q. Huber, A.~K. Cyrol, and J.~M. Pawlowski, ``{DoFun 3.0: Functional
  equations in Mathematica}'',
  \href{http://dx.doi.org/10.1016/j.cpc.2019.107058}{{\em Comput. Phys.
  Commun.} {\bfseries 248} (2020) 107058},
  \href{http://arxiv.org/abs/1908.02760}{{\ttfamily arXiv:1908.02760
  [hep-ph]}}.

\bibitem{Bjorken:1967qfa}
J.~Bjorken and S.~Drell, {\em Relativistic quantum fields}.
\newblock McGraw-Hill Book Company, New York, 1965.

\bibitem{Zwicky:2016lka}
R.~Zwicky, \href{http://dx.doi.org/10.3204/DESY-PROC-2016-04/Zwicky}{``{A brief
  Introduction to Dispersion Relations and Analyticity}''} in {\em {Quantum
  Field Theory at the Limits}: {from Strong Fields to Heavy Quarks}}.
\newblock 10, 2016.
\newblock \href{http://arxiv.org/abs/1610.06090}{{\ttfamily arXiv:1610.06090
  [hep-ph]}}.

\bibitem{Baulieu:2009ha}
L.~Baulieu, D.~Dudal, M.~S. Guimaraes, M.~Q. Huber, S.~P. Sorella,
  N.~Vandersickel, and D.~Zwanziger, ``{Gribov horizon and i-particles: about a
  toy model and the construction of physical operators}'',
\href{http://arxiv.org/abs/0912.5153}{{\ttfamily arXiv:0912.5153 [hep-th]}}.

\bibitem{Dudal:2010wn}
D.~Dudal and M.~S. Guimaraes, ``{On the computation of the spectral density of
  two-point functions: complex masses, cut rules and beyond}'',
  \href{http://dx.doi.org/10.1103/PhysRevD.83.045013}{{\em Phys. Rev.}
  {\bfseries D83} (2011) 045013},
\href{http://arxiv.org/abs/1012.1440}{{\ttfamily arXiv:1012.1440 [hep-th]}}.

\bibitem{Alkofer:2008dt}
R.~Alkofer, M.~Q. Huber, and K.~Schwenzer, ``{Infrared Behavior of Three-Point
  Functions in Landau Gauge Yang-Mills Theory}'',
  \href{http://dx.doi.org/10.1140/epjc/s10052-009-1066-3}{{\em Eur. Phys. J.}
  {\bfseries C62} (2009) 761--781},
\href{http://arxiv.org/abs/0812.4045}{{\ttfamily arXiv:0812.4045 [hep-ph]}}.

\bibitem{Eichmann:2014xya}
G.~Eichmann, R.~Williams, R.~Alkofer, and M.~Vujinovic, ``{The three-gluon
  vertex in Landau gauge}'',
  \href{http://dx.doi.org/10.1103/PhysRevD.89.105014}{{\em Phys.Rev.}
  {\bfseries D89} (2014) 105014},
\href{http://arxiv.org/abs/1402.1365}{{\ttfamily arXiv:1402.1365 [hep-ph]}}.

\bibitem{Huber:2011xc}
M.~Q. Huber and M.~Mitter, ``{CrasyDSE: A Framework for solving Dyson-Schwinger
  equations}'', \href{http://dx.doi.org/10.1016/j.cpc.2012.05.019}{{\em
  Comput.Phys.Commun.} {\bfseries 183} (2012) 2441--2457},
\href{http://arxiv.org/abs/1112.5622}{{\ttfamily arXiv:1112.5622 [hep-th]}}.

\bibitem{Blum:2014gna}
A.~Blum, M.~Q. Huber, M.~Mitter, and L.~von Smekal, ``{Gluonic three-point
  correlations in pure Landau gauge QCD}'',
  \href{http://dx.doi.org/10.1103/PhysRevD.89.061703}{{\em Phys. Rev. D}
  {\bfseries 89} (2014) 061703(R)},
\href{http://arxiv.org/abs/1401.0713}{{\ttfamily arXiv:1401.0713 [hep-ph]}}.

\bibitem{Eichmann:2021zuv}
G.~Eichmann, J.~M. Pawlowski, and J.~a.~M. Silva, ``{Mass generation in
  Landau-gauge Yang-Mills theory}'',
  \href{http://dx.doi.org/10.1103/PhysRevD.104.114016}{{\em Phys. Rev. D}
  {\bfseries 104} no.~11, (2021) 114016},
  \href{http://arxiv.org/abs/2107.05352}{{\ttfamily arXiv:2107.05352
  [hep-ph]}}.

\bibitem{Pinto-Gomez:2022brg}
F.~Pinto-G\'omez, F.~De~Soto, M.~N. Ferreira, J.~Papavassiliou, and
  J.~Rodr\'\i{}guez-Quintero, ``{Lattice three-gluon vertex in extended
  kinematics: planar degeneracy}'',
  \href{http://arxiv.org/abs/2208.01020}{{\ttfamily arXiv:2208.01020
  [hep-ph]}}.

\bibitem{Windisch:2021mem}
A.~Windisch, T.~Gallien, and C.~Schwarzlmueller, ``{A machine learning pipeline
  for autonomous numerical analytic continuation of Dyson-Schwinger
  equations}'', \href{http://dx.doi.org/10.1051/epjconf/202225809003}{{\em EPJ
  Web Conf.} {\bfseries 258} (2022) 09003},
  \href{http://arxiv.org/abs/2112.13011}{{\ttfamily arXiv:2112.13011
  [hep-ph]}}.

\bibitem{Windisch:2019byg}
A.~Windisch, T.~Gallien, and C.~Schwarzlm\"uller, ``{Deep reinforcement
  learning for complex evaluation of one-loop diagrams in quantum field
  theory}'', \href{http://dx.doi.org/10.1103/PhysRevE.101.033305}{{\em Phys.
  Rev. E} {\bfseries 101} no.~3, (2020) 033305},
  \href{http://arxiv.org/abs/1912.12322}{{\ttfamily arXiv:1912.12322
  [hep-ph]}}.

\bibitem{Kallen:1964epp}
G.~K\"all\'en, {\em Elementary Particle Physics}.
\newblock Addison-Wesley, 1964.

\bibitem{Cutkosky:1960sp}
R.~Cutkosky, ``{Singularities and discontinuities of Feynman amplitudes}'',
  \href{http://dx.doi.org/10.1063/1.1703676}{{\em J. Math. Phys.} {\bfseries 1}
  (1960) 429--433}.

\end{thebibliography}\endgroup

\end{document}